\documentclass[twocolumn]{aastex631}

\usepackage[caption=false]{subfig}
\usepackage{array}
\newcolumntype{H}{>{\setbox0=\hbox\bgroup}c<{\egroup}@{}}

\usepackage{comment}
\usepackage{color}
\usepackage{url}

\usepackage{amsmath}	
\usepackage{amssymb}	

\usepackage[percent]{overpic}

\newcommand{\OII}{[O\,{\sc ii}]}
\newcommand{\OIII}{[O\,{\sc iii}]}

\newcommand{\Ha}{H$\alpha$}
\newcommand{\Hb}{H$\beta$}

\newcommand{\HeII}{He{\sc ii}}

\newcommand{\NII}{[N\,{\sc ii}]}
\newcommand{\SII}{[S\,{\sc ii}]}

\newcommand{\kms}{km\,s$^{-1}$}
\newcommand{\Oabundance}{$12+\log ({\rm O/H})$}
\newcommand{\Zsun}{$Z_{\odot}$}

\newcommand{\Msun}{$M_{\odot}$}

\newcommand{\Msunyrkpc}{$M_{\odot}\,{\rm yr}^{-1}\,{\rm kpc}^{-2}$}
\newcommand{\Msunkpc}{$M_{\odot}\,{\rm kpc}^{-2}$}
\newcommand{\Msunpc}{$M_{\odot}\,{\rm pc}^{-2}$}

\submitjournal{ApJ on Dec 5, 2024}

\shorttitle{Resolved mass-metallicity relation in EMPGs}
\shortauthors{Nakajima et al.}

\begin{document}

\title{EMPRESS. X. Spatially resolved mass-metallicity relation in extremely metal-poor galaxies: \\
evidence of episodic star-formation fueled by a metal-poor gas infall}

\correspondingauthor{Kimihiko Nakajima}
\email{kimihiko.nakajima@nao.ac.jp}

\author[0000-0003-2965-5070]{Kimihiko Nakajima}
\affiliation{National Astronomical Observatory of Japan, 2-21-1 Osawa, Mitaka, Tokyo 181-8588, Japan}

\author[0000-0002-1049-6658]{Masami Ouchi}
\affiliation{National Astronomical Observatory of Japan, 2-21-1 Osawa, Mitaka, Tokyo 181-8588, Japan}
\affiliation{Institute for Cosmic Ray Research, The University of Tokyo, 5-1-5 Kashiwanoha, Kashiwa, Chiba 277-8582, Japan}
\affiliation{Department of Astronomical Science, SOKENDAI (The Graduate University for Advanced Studies), Osawa 2-21-1, Mitaka, Tokyo, 181-8588, Japan}
\affiliation{Kavli Institute for the Physics and Mathematics of the Universe (WPI), University of Tokyo, Kashiwa, Chiba 277-8583, Japan}

\author[0000-0001-7730-8634]{Yuki Isobe}
\affiliation{Kavli Institute for Cosmology, University of Cambridge, Madingley Road, Cambridge, CB3 0HA, UK}
\affiliation{Cavendish Laboratory, University of Cambridge, 19 JJ Thomson Avenue, Cambridge, CB3 0HE, UK}
\affiliation{Waseda Research Institute for Science and Engineering, Faculty of Science and Engineering, Waseda University, 3-4-1 Okubo, Shinjuku, Tokyo 169-8555, Japan}
\affiliation{Institute for Cosmic Ray Research, The University of Tokyo, 5-1-5 Kashiwanoha, Kashiwa, Chiba 277-8582, Japan}
\affiliation{Department of Physics, Graduate School of Science, The University of Tokyo, 7-3-1 Hongo, Bunkyo, Tokyo 113-0033, Japan}

\author[0000-0002-5768-8235]{Yi Xu}
\affiliation{Institute for Cosmic Ray Research, The University of Tokyo, 5-1-5 Kashiwanoha, Kashiwa, Chiba 277-8582, Japan}
\affiliation{Department of Astronomy, Graduate School of Science, The University of Tokyo, 7-3-1 Hongo, Bunkyo, Tokyo 113-0033, Japan}

\author[0000-0002-5443-0300]{Shinobu Ozaki}
\affiliation{National Astronomical Observatory of Japan, 2-21-1 Osawa, Mitaka, Tokyo 181-8588, Japan}

\author[0000-0002-7402-5441]{Tohru Nagao}
\affiliation{Research Center for Space and Cosmic Evolution, Ehime University, Bunkyo-cho 2-5, Matsuyama, Ehime 790-8577, Japan}

\author[0000-0002-7779-8677]{Akio K.~Inoue}
\affiliation{Waseda Research Institute for Science and Engineering, Faculty of Science and Engineering, Waseda University, 3-4-1 Okubo, Shinjuku, Tokyo 169-8555, Japan}
\affiliation{Department of Physics, School of Advanced Science and Engineering, Faculty of Science and Engineering, Waseda University, 3-4-1 Okubo, Shinjuku, Tokyo 169-8555, Japan}

\author{Michael Rauch}
\affiliation{Observatories of the Carnegie Institution for Science, 813 Santa Barbara St., Pasadena, CA 91101, USA}

\author[0000-0002-3801-434X]{Haruka Kusakabe}
\affiliation{Observatoire de Gen\`{e}ve, Universit\'e de Gen\`{e}ve, 51 Chemin de P\'egase, 1290 Versoix, Switzerland}
\affiliation{National Astronomical Observatory of Japan, 2-21-1 Osawa, Mitaka, Tokyo 181-8588, Japan}

\author[0000-0003-3228-7264]{Masato Onodera}
\affiliation{Subaru Telescope, National Astronomical Observatory of Japan, National Institutes of Natural Sciences (NINS), 650 North Aohoku Place, Hilo, HI 96720, USA}
\affiliation{Department of Astronomical Science, SOKENDAI (The Graduate University for Advanced Studies), Osawa 2-21-1, Mitaka, Tokyo, 181-8588, Japan}

\author[0000-0003-4321-0975]{Moka Nishigaki}
\affiliation{Department of Astronomical Science, SOKENDAI (The Graduate University for Advanced Studies), Osawa 2-21-1, Mitaka, Tokyo, 181-8588, Japan}
\affiliation{National Astronomical Observatory of Japan, 2-21-1 Osawa, Mitaka, Tokyo 181-8588, Japan}

\author[0000-0001-9011-7605]{Yoshiaki Ono}
\affiliation{Institute for Cosmic Ray Research, The University of Tokyo, 5-1-5 Kashiwanoha, Kashiwa, Chiba 277-8582, Japan}

\author[0000-0001-6958-7856]{Yuma Sugahara} 
\affiliation{Waseda Research Institute for Science and Engineering, Faculty of Science and Engineering, Waseda University, 3-4-1 Okubo, Shinjuku, Tokyo 169-8555, Japan}
\affiliation{Department of Physics, School of Advanced Science and Engineering, Faculty of Science and Engineering, Waseda University, 3-4-1 Okubo, Shinjuku, Tokyo 169-8555, Japan}

\author[0000-0002-8996-7562]{Takashi Hattori}
\affiliation{Subaru Telescope, National Astronomical Observatory of Japan, National Institutes of Natural Sciences (NINS), 650 North A'ohoku Place, Hilo, HI 96720, USA}

\author[0000-0002-5661-033X]{Yutaka Hirai}
\affiliation{Department of Physics and Astronomy, University of Notre Dame, 225 Nieuwland Science Hall, Notre Dame, IN 46556, USA}
\affiliation{Astronomical Institute, Tohoku University, 6-3 Aoba, Aramaki, Aoba-ku, Sendai, Miyagi 980-8578, Japan}

\author[0000-0002-0898-4038]{Takuya Hashimoto}
\affiliation{Division of Physics, Faculty of Pure and Applied Sciences, University of Tsukuba, Tsukuba, Ibaraki 305-8571, Japan}
\affiliation{Tomonaga Center for the History of the Universe (TCHoU), Faculty of Pure and Applied Sciences, University of Tsukuba, Tsukuba, Ibaraki 305-8571, Japan}

\author[0000-0002-1418-3309]{Ji Hoon Kim}
\affiliation{Astronomy Program, Department of Physics and Astronomy, Seoul National University, 1 Gwanak-ro, Gwanak-gu, Seoul 08826, Republic of Korea}
\affiliation{SNU Astronomy Research Center, Seoul National University, 1 Gwanak-ro, Gwanak-gu, Seoul 08826, Republic of Korea}

\author[0000-0003-1169-1954]{Takashi J.~Moriya}
\affiliation{National Astronomical Observatory of Japan, 2-21-1 Osawa, Mitaka, Tokyo 181-8588, Japan}
\affiliation{Department of Astronomical Science, SOKENDAI (The Graduate University for Advanced Studies), Osawa 2-21-1, Mitaka, Tokyo, 181-8588, Japan}
\affiliation{School of Physics and Astronomy, Faculty of Science, Monash University, Clayton, Victoria 3800, Australia}

\author[0009-0006-6763-4245]{Hiroto Yanagisawa}
\affiliation{Institute for Cosmic Ray Research, The University of Tokyo, 5-1-5 Kashiwanoha, Kashiwa, Chiba 277-8582, Japan}
\affiliation{Department of Physics, Graduate School of Science, The University of Tokyo, 7-3-1 Hongo, Bunkyo, Tokyo 113-0033, Japan}

\author[0000-0002-1005-4120]{Shohei Aoyama}
\affiliation{Institute of Management and Information Technologies, Chiba University, 1-33, Yayoi-cho, Inage-ward, Chiba, 263-8522, Japan}
\affiliation{Institute for Cosmic Ray Research, The University of Tokyo, 5-1-5 Kashiwanoha, Kashiwa, Chiba 277-8582, Japan}

\author[0000-0001-7201-5066]{Seiji Fujimoto} 
\affiliation{Department of Astronomy, The University of Texas at Austin, Austin, TX, USA}
\affiliation{Cosmic DAWN Center}
\affiliation{Niels Bohr Institute, University of Copenhagen, Lyngbyvej2, DK-2100, Copenhagen, Denmark}

\author[0000-0002-0547-3208]{Hajime Fukushima}
\affiliation{Center for Computational Sciences, University of Tsukuba, Ten-nodai, 1-1-1 Tsukuba, Ibaraki 305-8577, Japan}

\author{Keita Fukushima}
\affiliation{Theoretical Astrophysics, Department of Earth \& Space Science, Graduate School of Science, Osaka University, 
1-1 Machikaneyama, Toyonaka, Osaka 560-0043, Japan}

\author[0000-0002-6047-430X]{Yuichi Harikane}
\affiliation{Institute for Cosmic Ray Research, The University of Tokyo, 5-1-5 Kashiwanoha, Kashiwa, Chiba 277-8582, Japan}

\author{Shun Hatano}
\affiliation{Department of Astronomical Science, SOKENDAI (The Graduate University for Advanced Studies), Osawa 2-21-1, Mitaka, Tokyo, 181-8588, Japan}
\affiliation{National Astronomical Observatory of Japan, 2-21-1 Osawa, Mitaka, Tokyo 181-8588, Japan}

\author[0000-0002-8758-8139]{Kohei Hayashi}
\affiliation{National Institute of Technology, Sendai College, 48 Nodayama, Medeshima-Shiote, Natori, Miyagi 981-1239, Japan}
\affiliation{Astronomical Institute, Tohoku University, 6-3 Aoba, Aramaki, Aoba-ku, Sendai, Miyagi 980-8578, Japan}
\affiliation{Institute for Cosmic Ray Research, The University of Tokyo, 5-1-5 Kashiwanoha, Kashiwa, Chiba 277-8582, Japan}

\author{Tsuyoshi Ishigaki}
\affiliation{Department of Physical Science and Materials Engineering, Faculty of Science and Engineering, Iwate University \\
3-18-34 Ueda, Morioka, Iwate 020-8550, Japan}

\author{Masahiro Kawasaki}
\affiliation{Institute for Cosmic Ray Research, The University of Tokyo, 5-1-5 Kashiwanoha, Kashiwa, Chiba 277-8582, Japan}
\affiliation{Kavli Institute for the Physics and Mathematics of the Universe (WPI), University of Tokyo, Kashiwa, Chiba 277-8583, Japan}

\author[0000-0001-5780-1886]{Takashi Kojima}
\affiliation{Institute for Cosmic Ray Research, The University of Tokyo, 5-1-5 Kashiwanoha, Kashiwa, Chiba 277-8582, Japan}
\affiliation{Department of Physics, Graduate School of Science, The University of Tokyo, 7-3-1 Hongo, Bunkyo, Tokyo 113-0033, Japan}

\author[0000-0002-3852-6329]{Yutaka Komiyama} 
\affiliation{Department of Advanced Sciences, Faculty of Science and Engineering, Hosei University, 3-7-2 Kajino-cho, Koganei-shi, Tokyo 184-8584, Japan}

\author{Shuhei Koyama}
\affiliation{Institute of Astronomy, Graduate School of Science, The University of Tokyo, 2-21-1 Osawa, Mitaka, Tokyo 181-0015, Japan}

\author[0000-0002-0479-3699]{Yusei Koyama}
\affiliation{Subaru Telescope, National Astronomical Observatory of Japan, National Institutes of Natural Sciences (NINS), 650 North A'ohoku Place, Hilo, HI 96720, USA}
\affiliation{Department of Astronomical Science, SOKENDAI (The Graduate University for Advanced Studies), Osawa 2-21-1, Mitaka, Tokyo, 181-8588, Japan}

\author[0000-0003-1700-5740]{Chien-Hsiu Lee} 
\affiliation{W. M. Keck Observatory, Kamuela, HI 96743, USA}

\author{Akinori Matsumoto}
\affiliation{Institute for Cosmic Ray Research, The University of Tokyo, 5-1-5 Kashiwanoha, Kashiwa, Chiba 277-8582, Japan}
\affiliation{Department of Physics, Graduate School of Science, The University of Tokyo, 7-3-1 Hongo, Bunkyo, Tokyo 113-0033, Japan}

\author[0000-0003-4985-0201]{Ken Mawatari}
\affiliation{National Astronomical Observatory of Japan, 2-21-1 Osawa, Mitaka, Tokyo 181-8588, Japan}

\author{Kentaro Motohara}
\affiliation{National Astronomical Observatory of Japan, 2-21-1 Osawa, Mitaka, Tokyo 181-8588, Japan}
\affiliation{Institute of Astronomy, Graduate School of Science, The University of Tokyo, 2-21-1 Osawa, Mitaka, Tokyo 181-0015, Japan}

\author{Kai Murai}
\affiliation{Department of Physics, Tohoku University, 6-3 Aoba, Aramaki, Aoba-ku, Sendai, Miyagi 980-8578, Japan}

\author[0000-0001-7457-8487]{Kentaro Nagamine}
\affiliation{Theoretical Astrophysics, Department of Earth \& Space Science, Graduate School of Science, Osaka University, 
1-1 Machikaneyama, Toyonaka, Osaka 560-0043, Japan}
\affiliation{Kavli Institute for the Physics and Mathematics of the Universe (WPI), University of Tokyo, Kashiwa, Chiba 277-8583, Japan}
\affiliation{Department of Physics \& Astronomy, University of Nevada, Las Vegas, 4505 S. Maryland Pkwy, Las Vegas, NV 89154-4002, USA}

\author[0000-0002-5768-8235]{Minami Nakane}
\affiliation{Institute for Cosmic Ray Research, The University of Tokyo, 5-1-5 Kashiwanoha, Kashiwa, Chiba 277-8582, Japan}
\affiliation{Department of Physics, Graduate School of Science, The University of Tokyo, 7-3-1 Hongo, Bunkyo, Tokyo 113-0033, Japan}

\author{Tomoki Saito}
\affiliation{Nishi-Harima Astronomical Observatory, Centre for Astronomy, University of Hyogo, 407-2 Nishigaichi, Sayo, Sayo-gun, Hyogo 679-5313}

\author{Rin Sasaki}
\affiliation{Department of Physical Science and Materials Engineering, Faculty of Science and Engineering, Iwate University \\
3-18-34 Ueda, Morioka, Iwate 020-8550, Japan}

\author{Takatoshi Shibuya}
\affiliation{Kitami Institute of Technology, 165 Koen-cho, Kitami, Hokkaido 090-8507, Japan}

\author[0000-0002-7043-6112]{Akihiro Suzuki}
\affiliation{Research Center for the Early Universe, The University of Tokyo, 7-3-1 Hongo, Bunkyo, Tokyo 113-0033, Japan}

\author[0000-0001-8416-7673]{Tsutomu T.~Takeuchi}
\affiliation{Division of Particle and Astrophysical Science, Nagoya University, Furo-cho, Chikusa-ku, Nagoya 464--8602, Japan}
\affiliation{The Research Center for Statistical Machine Learning, the Institute of Statistical Mathematics, 10-3 Midori-cho, Tachikawa, Tokyo 190---8562, Japan}

\author{Hiroya Umeda}
\affiliation{Institute for Cosmic Ray Research, The University of Tokyo, 5-1-5 Kashiwanoha, Kashiwa, Chiba 277-8582, Japan}
\affiliation{Department of Physics, Graduate School of Science, The University of Tokyo, 7-3-1 Hongo, Bunkyo, Tokyo 113-0033, Japan}

\author{Masayuki Umemura}
\affiliation{Center for Computational Sciences, University of Tsukuba, Ten-nodai, 1-1-1 Tsukuba, Ibaraki 305-8577, Japan}

\author[0000-0002-2740-3403]{Kuria Watanabe}
\affiliation{Department of Astronomical Science, SOKENDAI (The Graduate University for Advanced Studies), Osawa 2-21-1, Mitaka, Tokyo, 181-8588, Japan}
\affiliation{National Astronomical Observatory of Japan, 2-21-1 Osawa, Mitaka, Tokyo 181-8588, Japan}

\author[0000-0001-6229-4858]{Kiyoto Yabe}
\affiliation{Kavli Institute for the Physics and Mathematics of the Universe (WPI), University of Tokyo, Kashiwa, Chiba 277-8583, Japan}
\affiliation{Subaru Telescope, National Astronomical Observatory of Japan, National Institutes of Natural Sciences (NINS), 650 North Aohoku Place, Hilo, HI 96720, USA}

\author[0000-0002-1319-3433]{Hidenobu Yajima}
\affiliation{Center for Computational Sciences, University of Tsukuba, Ten-nodai, 1-1-1 Tsukuba, Ibaraki 305-8577, Japan}

\author[0000-0003-3817-8739]{Yechi Zhang}
\affiliation{Institute for Cosmic Ray Research, The University of Tokyo, 5-1-5 Kashiwanoha, Kashiwa, Chiba 277-8582, Japan}
\affiliation{Department of Physics, Graduate School of Science, The University of Tokyo, 7-3-1 Hongo, Bunkyo, Tokyo 113-0033, Japan}



\begin{abstract}
Using the Subaru/FOCAS IFU capability, we examine the spatially resolved relationships between gas-phase metallicity, stellar mass, and star-formation rate surface densities ($\Sigma_{\star}$ and $\Sigma_{\rm SFR}$, respectively) in extremely metal-poor galaxies (EMPGs) in the local universe. Our analysis includes 24 EMPGs, comprising 9,177 spaxels, which span a unique parameter space of local metallicity (\Oabundance\ $\simeq 6.9$ to $7.9$) and stellar mass surface density ($\Sigma_{\star} \sim 10^5$ to $10^7$\,\Msunkpc), extending beyond the range of existing large integral-field spectroscopic surveys. Through spatially resolved emission line diagnostics based on the \NII\ BPT-diagram, we verify the absence of evolved active galactic nuclei in these EMPGs. Our findings reveal that, while the resolved mass--metallicity relation exhibits significant scatter in the low-mass regime, this scatter is closely correlated with local star-formation surface density. Specifically, metallicity decreases as $\Sigma_{\rm SFR}$ increases for a given $\Sigma_{\star}$. Notably, half of the EMPGs show a distinct metal-poor horizontal branch on the resolved mass--metallicity relation. This feature typically appears at the peak clump with the highest $\Sigma_{\star}$ and $\Sigma_{\rm SFR}$ and is surrounded by a relatively metal-enriched ambient region. These findings support a scenario in which metal-poor gas infall fuels episodic star formation in EMPGs, consistent with the kinematic properties observed in these systems. In addition, we identify four EMPGs with exceptionally low central metallicities (\Oabundance\ $\lesssim 7.2$), which display only a metal-poor clump without a surrounding metal-rich region. This suggests that such ultra-low metallicity EMPGs, at less than a few percent of the solar metallicity, may serve as valuable analogs for galaxies in the early stages of galaxy evolution.
\end{abstract}

\keywords{Chemical abundances(224) --- Galaxy chemical evolution(580) --- Galaxy evolution(594) --- Dwarf galaxies(416)}

\section{Introduction} \label{sec:introduction}

The study of extremely metal-poor galaxies (EMPGs) provides a unique window into the fundamental astrophysical processes that occur during the early stages of galaxy evolution. These galaxies, characterized by metallicities below 10\% of the solar value ($Z<0.1$\,\Zsun), are thought to possess inherent similarities to systems in the early universe, where the first episodes of star formation occurred from gas that had experienced minimal chemical enrichment. Investigating EMPGs allows us to probe the mechanisms that drive the formation of stars and the build-up of metals, as well as how such galaxies evolve over time (e.g., \citealt{izotov1990,izotov2012,izotov2018_lowZ,izotov2019_lowZspec, izotov2021_J2229, KO2000, kniazev2003, TI2005, pustilnik2005, berg2012, skillman2013, hirschauer2016}). Crucially, understanding the interplay between star formation, metal enrichment, and gas inflows or outflows is key to constructing a comprehensive picture of galaxy evolution across cosmic history (e.g., \citealt{lilly2013, dekel2014, somerville2015, MM2019}).

One of the fundamental questions in this field is whether EMPGs in the local universe can serve as analogs of galaxies in the early universe. Early-universe galaxies are expected to experience episodic bursts of star formation, fueled by infalling gas from the intergalactic medium (IGM; e.g., \citealt{keres2005, dekel2009_nature, schaye2010}), which drives star formation in chemically pristine environments (e.g., \citealt{wise2012,yajima2023}). However, local EMPGs may not universally adhere to this idealized scenario. Chemical enrichment through successive episodes of star formation or gas mixing processes could result in diverse metallicity distributions within individual systems (e.g., \citealt{mott2013, gibson2013, ma2017_gradZ, molla2019, sun2024_rMZR_FIRE2}). To unravel these complexities, spatially-resolved spectroscopic studies are crucial. By mapping metallicity variations across EMPGs, we can distinguish between different evolutionary stages and better understand the chemical enrichment processes shaping these galaxies.

The spatially-resolved mass-metallicity relation (referred to as rMZR throughout this paper), which connects the local stellar mass surface density ($\Sigma_{\star}$) with metallicity, provides an essential framework for investigating these processes. At higher metallicities ($Z>0.1$\,\Zsun) in the local universe, large optical integral field unit (IFU) surveys, such as MaNGA and CALIFA, along with observations using IFUs on 8m-class telescopes like VLT/MUSE, have established that the rMZR is relatively tight, with metallicity decreasing toward lower $\Sigma_{\star}$ values (e.g., \citealt{gonzalez-delgado2014, barrera-ballesteros2016, gao2018, erroz-ferrer2019, yao2022}). This trend is similar to those observed in the global stellar mass-metallicity relation (e.g., \citealt{tremonti2004, AM2013, curti2020}). Interestingly, some studies have also suggested a dependence on star formation rate (SFR) surface density ($\Sigma_{\rm SFR}$) and global SFR, indicating a potential link between star formation activity and metallicity.
Moreover, investigations of metallicity gradients in local galaxies has been routinely examined to probe past episodes of star formation and feedback (e.g., \citealt{sanchez2014, ho2015, sanchez-menguiano2016, belfiore2017, khoram2024}).
These investigations often reveal lower metallicity in the outer regions of galaxies, i.e., negative metallicity gradients that become steeper with increasing global stellar mass, consistent with an inside-out growth scenario. 
These large surveys predominantly cover galaxies with global stellar mass in the range of $10^9-10^{11}$\,\Msun, local stellar mass densities around $\Sigma_{\star}=10^{6.5}-10^{9}$\,\Msunkpc, and metallicities spanning \Oabundance\ $=8.-8.7$.

At high redshift, near-infrared IFU spectroscopy with instruments such as Keck/OSIRIS, VLT/SINFONI, and KMOS has extended studies of resolved metallicity to $z=1-3$ (e.g., \citealt{jones2010_gradZ, jones2013_gradZ, cresci2010, swinbank2012, troncoso2014, wuyts2016, curti2020_klever}).
More recently, JWST/NIRSpec observations have begun to reveal resolved metallicity distributions in galaxies at even higher redshifts, up to $z=3-10$ (e.g., \citealt{venturi2024, tripodi2024, marconcini2024}).
The studies generally reveal flat or mildly negative metallicity gradients in galaxies with comparable masses and metallicities as explored in the local universe, suggesting only moderate cosmological evolution in gradient trends compared to the local universe, albeit with large scatters. Comparisons with cosmological simulations (e.g., \citealt{mott2013, gibson2013, ma2017_gradZ, molla2019, sun2024_rMZR_FIRE2}) indicate that factors like enhanced feedbacks from star formation, galaxy interactions, and/or inflows of metal-poor gas driven by high accretion rates from the IGM play a key role at high redshift, leading to greater mixing of metals and, in some cases, ``inverted'' positive metallicity gradients.

Extending the rMZR to the extremely metal-poor regime in the local universe is essential for understanding the low-mass end of galaxy evolution, where EMPGs can serve as nearby analogs of high-redshift star-forming systems.
However, probing the rMZR at the extremely metal-poor regime remains challenging due to the observational limitations and the rarity of EMPGs with spatially-resolved data. In this study, we address this gap by exploring the rMZR in EMPGs, providing new insights into how gas inflows, star formation, and metal enrichment interact in low-metallicity environments.

In this paper, we present the first systematic, spatially resolved spectroscopic analysis of 24 EMPGs observed with the FOCAS IFU instrument on Subaru as part of the EMPRESS 3D collaboration (PI: M.~Ouchi).
The EMPRESS project (Extremely Metal-Poor Representatives Explored by the Subaru Survey) represents a cutting-edge initiative aimed at uncovering metal-poor systems and exploring their fundamental properties. The project encompasses diverse investigations, including the search for rare metal-poor systems and their global properties \citep{kojima2020, nakajima2022_empressV, nishigaki2023}, the study of chemical abundances to trace past massive star formation and explosion \citep{kojima2021, isobe2022_fe, watanabe2024}, and the examination of ionizing spectra to infer the presence of intermediate-mass black holes \citep{umeda2022, hatano2023, hatano2024}. Furthermore, EMPRESS examines the dynamics and outflows of EMPGs to explore gas properties and feedback mechanisms \citep{isobe2021_tail, isobe2023_focasifu, xu2022, xu2024_focasifu} and delves into primordial abundances as a direct test of Big Bang nucleosynthesis models (\citealt{matsumoto2022}; see also \citealt{kawasaki2022}). 
In this study, we aim to map the metallicity distribution within these galaxies by utilizing the 3D datacubes of EMPGs, focusing on how stellar mass and star formation activity shape the local chemical enrichment processes. We pay particular attention to the presence of metal-poor branches on the rMZR, which may indicate ongoing gas inflows and localized star-formation bursts. A key question is the diversity of evolutionary stages among EMPGs, i.e., whether such galaxies are experiencing their first episode of star formation and chemical enrichment, or whether they represent systems with more complex star-formation histories and metal-mixing processes.

This paper is structured as follows. In Section \ref{sec:data}, we describe the observational setup and data reduction procedures of the EMPRESS 3D. Section \ref{sec:results} presents the key results, including the rMZR analysis and metallicity gradients. In Section \ref{sec:discussions}, we interpret the results, categorize the EMPGs into evolutionary groups, and discuss their relevance to early-universe analogs. Finally, we summarize our conclusions in Section \ref{sec:summary}.
Throughout the paper we assume a solar chemical composition following \citet{asplund2009}, and adopt a standard $\Lambda$CDM cosmology with $\Omega_{\Lambda}=0.7$, $\Omega_{m}=0.3$, and $H_0=70$\,\kms\,Mpc$^{-1}$.

\begin{deluxetable*}{lcccccccc}
\tablecaption{EMPRESS 3D objects studied in this work
\label{tbl:objects}}
\tablewidth{0.99\columnwidth}
\tablehead{
\colhead{ID}
& \colhead{R.A.}
& \colhead{Decl.}
& \colhead{Redshift}
& \colhead{12+log(O/H)}
& \colhead{Date of Obs.}
& \colhead{Airmass}
& \colhead{Exposure}
& \colhead{Category} \\ 
 & \colhead{(J2000)}
& \colhead{(J2000)}
& \colhead{}
& \colhead{(literature)}
& \colhead{(UT)}
& \colhead{}
& \colhead{(sec)} 
& \colhead{($\P$)} 
} 
\startdata
J1631$+$4426 & 16:31:14.24 & $+$44:26:04.43 & $0.031$ & $6.90\pm 0.03$ & Jun 18, 2021 & $1.28$ & 1200 & B \\ 
J1234$+$3901 & 12:34:15.70 & $+$39:01:16.41 & $0.133$ & $7.04\pm 0.03$ & Jul 13, 2021 & $1.60$ & 1200 & D \\ 
DDO68-\#2-\#3 & 09:56:46.05 & $+$28:49:43.78 & $0.002$ & $7.08\pm 0.09$ & Dec 14, 2021 & $1.34$ & 1200 & B \\ 
J0228$-$0210 & 02:28:02.59 & $-$02:10:55.55 & $0.042$ & $7.08\pm 0.07$ & Nov 25, 2021 & $1.24$ & 1200 & C \\ 
LeoP & 10:21:45.10 & $+$18:05:17.20 & 4e-04 & $7.17\pm 0.05$ & Dec 14, 2021 & $1.14$ & 1200 & B \\ 
IZw18-NW-SE & 09:34:02.03 & $+$55:14:28.07 & $0.003$ & $7.17\pm 0.01$ & Dec 14, 2021 & $1.36$ & 1200 & C \\ 
J0935$-$0115 & 09:35:39.20 & $-$01:15:41.41 & $0.016$ & $7.17\pm 0.07$ & Dec 14, 2021 & $1.08$ & 1200 & B \\ 
J1452$+$0241 & 14:52:55.28 & $+$02:41:01.31 & $0.006$ & $7.21\pm 0.12$ & Jul 13, 2021 & $1.33$ & 1200 & C \\ 
SBS0335-052E & 03:37:44.06 & $-$05:02:40.19 & $0.012$ & $7.23\pm 0.01$ & Nov 25, 2021 & $1.40$ & 1200 & A \\ 
J2104$-$0035 & 21:04:55.30 & $-$00:35:22.00 & $0.004$ & $7.26\pm 0.03$ & Jul 13, 2021 & $1.35$ & 1200 & A \\ 
J0036$+$0052 & 00:36:30.40 & $+$00:52:34.71 & $0.028$ & $7.39\pm 0.08$ & Aug 14, 2021 & $1.64$ & 1200 & C \\ 
J2136$+$0414 & 21:36:58.81 & $+$04:14:04.31 & $0.017$ & $7.42\pm 0.02$ & Nov 25, 2021 & $1.26$ & 1200 & A \\ 
J0057$-$0941 & 00:57:57.32 & $-$09:41:19.20 & $0.015$ & $7.43\pm 0.11$ & Nov 25, 2021 & $1.53$ & 1920 & C \\ 
HS0822+3542 & 08:25:55.44 & $+$35:32:31.92 & $0.002$ & $7.45\pm 0.02$ & Dec 14, 2021 & $1.24$ & 1200 & A \\ 
J1044$+$0353 & 10:44:57.79 & $+$03:53:13.15 & $0.013$ & $7.45\pm 0.04$ & Dec 14, 2021 & $1.17$ & 1200 & A \\ 
J1702$+$2120 & 17:02:39.88 & $+$21:20:08.91 & $0.025$ & $7.48\pm 0.11$ & Aug 14, 2021 & $1.18$ & 1200 & C \\ 
J2302$+$0049 & 23:02:10.00 & $+$00:49:38.78 & $0.033$ & $7.55\pm 0.07$ & Aug 14, 2021 & $1.32$ & 1200 & A \\ 
J1423$+$2257 & 14:23:42.88 & $+$22:57:28.80 & $0.033$ & $7.56\pm 0.04$ & Aug 14, 2021 & $1.61$ & 1200 & C \\ 
J1355$+$4651 & 13:55:25.64 & $+$46:51:51.34 & $0.028$ & $7.56\pm 0.04$ & Aug 14, 2021 & $1.62$ & 1200 & A \\ 
J1418$+$2102 & 14:18:51.12 & $+$21:02:39.74 & $0.009$ & $7.64\pm 0.04$ & Jul 13, 2021 & $1.50$ & 1200 & A \\ 
J1016$+$3754 & 10:16:24.53 & $+$37:54:45.97 & $0.004$ & $7.64\pm 0.01$ & Dec 14, 2021 & $1.05$ & 1200 & A \\ 
J0125$+$0759 & 01:25:34.19 & $+$07:59:24.69 & $0.010$ & $7.65\pm 0.02$ & Nov 25, 2021 & $1.21$ & 1200 & A \\ 
J2115$-$1734 & 21:15:58.33 & $-$17:34:45.09 & $0.023$ & $7.68\pm 0.01$ & Jul 13, 2021 & $1.57$ & 900 & A \\ 
J1323$-$0132$^{\ddag}$ & 13:23:47.46 & $-$01:32:51.94 & $0.022$ & $7.74\pm 0.05$ & Jul 13, 2021 & $1.74$ & 1200 & $\cdots$ \\ 
J1253$-$0312$^{\ddag}$ & 12:53:05.96 & $-$03:12:58.94 & $0.023$ & $8.01\pm 0.01$ & Jul 13, 2021 & $1.44$ & 600 & $\cdots$ \\ 
J2314$+$0154 & 23:14:37.55 & $+$01:54:14.27 & $0.033$ & $7.04$ -- $7.25^{\dag}$ & Jun 18, 2021 & $1.29$ & 1200 & D \\ 
\enddata
\tablecomments{%
($\dag$) The metallicity is based on this study.
($\ddag$) These objects are not used in the analyses in this paper.
($\P$) These categories are based on the patterns found on the resolved mass-metallicity relation (see \S\ref{ssec:results_rMZR_individual}).
}
\end{deluxetable*}


\section{EMPRESS 3D data} \label{sec:data}

\subsection{Sample} \label{ssec:data_sample}

As presented in part by \citet{isobe2023_focasifu} and \citet{xu2024_focasifu} (see also \citealt{kashiwagi2021} as a pilot study), our target sample for the FOCAS IFU spectroscopy is derived from the compilation of local metal-poor galaxies provided in \citet{nakajima2022_empressV}. This comprehensive catalog includes 103 EMPGs with \Oabundance\ $\leq 7.69$, alongside 82 moderately metal-poor galaxies with $7.69 <$ \Oabundance\ $\lesssim 8.1$. The EMPGs identified by the EMPRESS survey using HSC are also incorporated into this compilation.
For the EMPRESS 3D survey, targets are primarily selected based on their low metallicities, typically measured at the central region of each object via slit or fiber spectroscopy. Additionally, we consider the \Hb\ flux to ensure the robust detection of multiple key emission lines with high significance, allowing us to investigate the spatial distribution of these emission lines.

\subsection{FOCAS IFU observations} \label{ssec:data_observations}

The FOCAS IFU observations were conducted on the Subaru Telescope across multiple nights in June, July, August, November, and December 2021, under varying sky conditions with seeing values of $0\farcs6$--$0\farcs8$, $0\farcs6$--$0\farcs8$, $\sim0\farcs8$, $\sim 1\farcs0$, and $0\farcs7$--$1\farcs2$, respectively. Each object was observed with a single FOCAS IFU pointing, providing a field of view of $13\farcs5 \times 10\farcs0$ and a slice width of $0\farcs435$ \citep{ozaki2020}. Table \ref{tbl:objects} summarizes the basic properties of the EMPRESS 3D objects and their observations.
The position angle for each target was adjusted based on the $r$-band image to capture the full extent of the structure, as the $r$-band traces both the strong \Ha\ emission and the stellar continuum. Observations presented in this paper were performed in low-resolution mode using the 300B grism without a filter, offering wavelength coverage from $\sim 3700$\,\AA\ to $7200$\,\AA\ at an average spectral resolution of $R \sim 1000$. 
It is worth noting that the sensitivity drops significantly at both the shorter and longer ends of the spectrum by a few hundred \AA, limiting the detection of faint emission lines such as \OII$\lambda 3727$ from faint sources.

Our observations consisted of one to six observing block(s) per target, with on-source integration times ranging from $120$ to $1200$\,sec. For cases where the integration was repeated, we applied a $\pm 1^{\prime\prime}$ dithering along the long axis of the detector to improve spatial sampling and mitigate detector artifacts.
Atmospheric dispersion corrector was used in all observations. 
Wavelength calibration data of ThAr were acquired immediately following each sequence of observing blocks, using the same telescope pointing as the science observations to ensure consistency. Dome flat data were obtained during twilight.
Spectrophotometric standard stars (Feige 67, HZ 44, Feige 110, and Feige 34) were observed either at the beginning or the end of each observing run to enable accurate flux calibration.

The FOCAS IFU data was reduced using the dedicated reduction software%
\footnote{
\url{https://www2.nao.ac.jp/~shinobuozaki/focasifu/index.html}
}.
We used version 20210210 for data collected in June and July, and the version 20210818 for the remaining observations. 
The data reduction process for each observing block included bias subtraction, flat-field correction, cosmic ray removal, wavelength calibration, flux calibration, and data-cube production. 
Sky subtraction was carried out using two complementary methods. The default method utilized sky spectra simultaneously obtained with a dedicated sky slit in the IFU, while the alternative approach employed the external package \verb+ZAP+ \citep{soto2016_zap}, which performs sky subtraction using the science data-cubes. 
As part of the reduction strategy, we initially created preliminary \Ha\ intensity maps using the data-cubes produced by the default sky subtraction method. These \Ha\ maps were then used to mask the source regions, allowing us to reprocess the sky subtraction with \verb+ZAP+. 
We adopted the default sky subtraction for objects where the \Ha\ emission occupied more than 75\% of the field of view (5/26 objects) and used the \verb+ZAP+-processed data-cubes for the remaining 21/26 objects. 
For sources observed in multiple blocks, we aligned the spatial zero-points of the flux-calibrated, sky-subtracted data-cubes and combined them to produce the final data-cube.
Error data-cubes were generated by accounting for readout noise, photon noise, and error propagation throughout the reduction process, including the combination of multiple observing blocks for each object.

\subsection{Emission line measures in 3D} \label{ssec:data_maps}

Each of the EMPRESS 3D objects has a known spectroscopic redshift based on previous spectroscopic observations of its brightest core region. We use this redshift as an initial prior to search for key emission lines, such as the Balmer lines, \OII$\lambda\lambda 3726,3729$, \OIII$\lambda 4363$, \HeII$\lambda 4686$, \OIII$\lambda\lambda 5007, 4959$, \NII$\lambda 6584$, and \SII$\lambda\lambda 6716, 6731$, which are essential for constraining the physical conditions of the gas.
For each spaxel, we extract a 1D spectrum by applying a 2-pixel Gaussian kernel and fit a single Gaussian profile with a flat continuum within a wavelength window of $\Delta \lambda \sim 50$\,\AA\ around each key emission line. The corresponding error spectrum is extracted in the same manner from the error data-cube, and used during the fitting procedure. 
These processes provide measurements of flux, equivalent width (EW), velocity information and their corresponding uncertainties for the detected lines. For closely spaced emission lines, we perform simultaneous fitting with multiple Gaussian components, sharing the same continuum.
We impose a detection threshold of $5\sigma$ for both emission lines and continuum to ensure reliable measurements, and only spaxel positions meeting this criterion are included in the subsequent analysis.

Note that we do not apply slit-loss corrections to the spaxel flux measurements as a function of wavelength, as the primary metallicity measurement method described in Sect.~\ref{ssec:results_metallicity_measurement} relies on emission lines and the continuum at similar wavelengths, minimizing the impact on the results. To evaluate the effect of wavelength-dependent changes in the point spread function, we used standard stars and confirmed that these changes are typically within 5\% across the wavelength range of $\sim4500$--$7000$\,\AA. However, point spread functions at shorter wavelengths can be larger, potentially contributing to flux losses of \OII\ emission, although this does not affect our primary results, as \OII\ is not used for the main analysis (Sect.~\ref{ssec:results_metallicity_measurement}).

\begin{figure}[t]
    \begin{center}
     \includegraphics[bb=0 0 543 527, width=0.99\columnwidth]{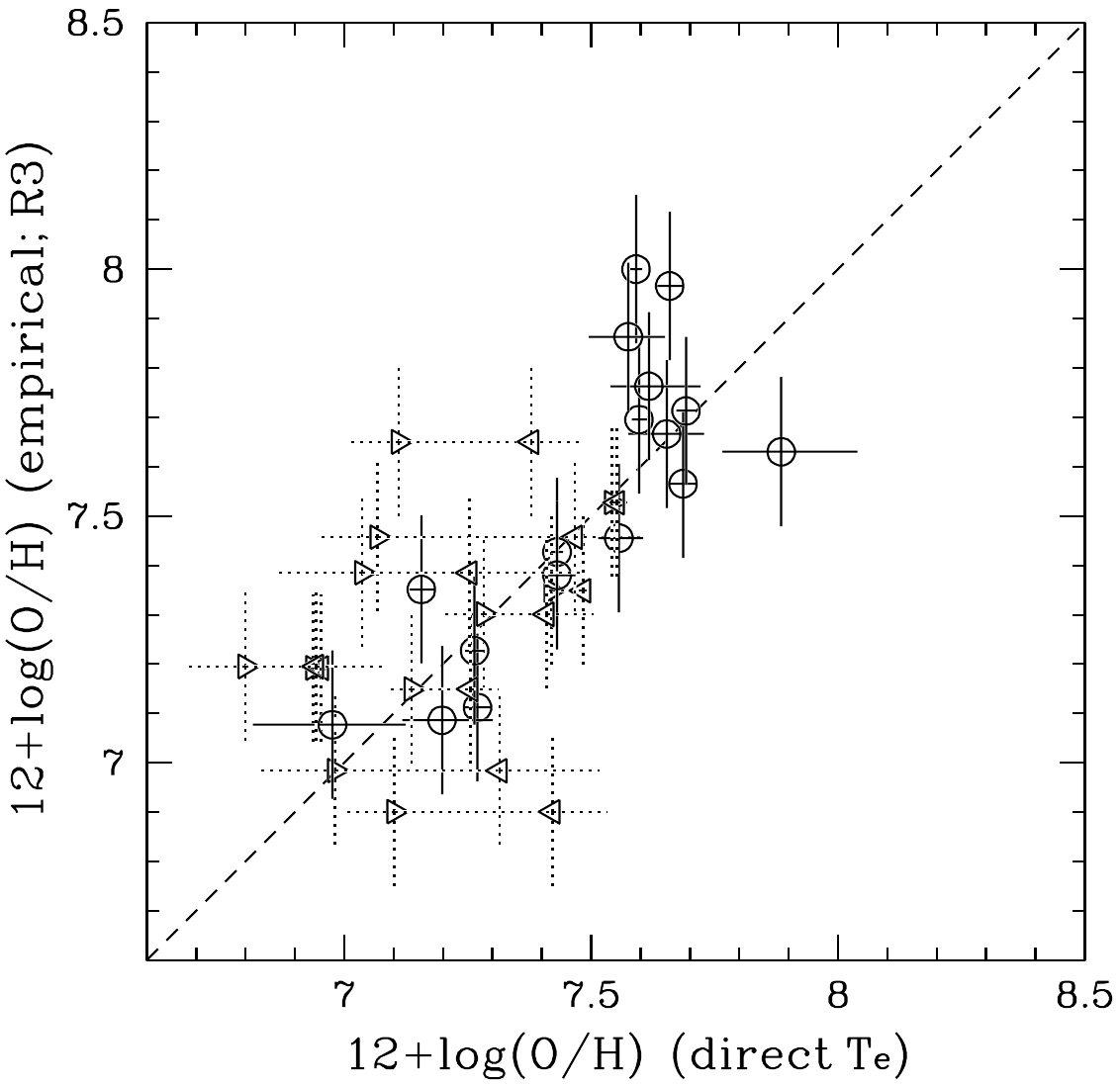}
    \caption{%
    		Comparison of metallicities derived from the direct temperature method and the empirical R3-index, using the integrated spectrum around the central brightest region of each galaxy. Black circles represent EMPGs with well-determined temperature-based metallicities, where the \OII$\lambda 3727$ line is detected. For EMPGs lacking \OII$\lambda 3727$ (and thus missing a robust measurement of O$^{+}$/H$^{+}$), the T${e}$-based oxygen abundance is constrained between two values: the lower limit given by O$^{2+}$/H$^{+}$ and the upper limit defined as O$^{2+}$/H$^{+}$ plus the $3\sigma$ upper limit of O$^{+}$/H$^{+}$. These limits are depicted as two black triangles connected by a dotted line. The dashed line represents the 1:1 correspondence between the metallicities obtained from the two methods, showing overall agreement.
			}
    \label{fig:compare_Z}
    \end{center} 
\end{figure}

\section{Results} \label{sec:results}

\subsection{Metallicity measurements} \label{ssec:results_metallicity_measurement}

Using the emission line flux maps obtained from the FOCAS IFU data, we now examine the metallicity distribution within galaxies in the extremely metal-poor regime. 
Given that the temperature-sensitive line \OIII$\lambda 4363$ and the singly ionized oxygen line \OII$\lambda 3727$ are not consistently detected across the regions where extended \Ha\ emission is observed, we instead utilize strong-line metallicity indicators to construct metallicity maps. We adopt the indicators presented by \citet{nakajima2022_empressV}, which are calibrated based on the largest EMPG sample to date and are well-suited for metal-poor objects with metallicities below $0.1$\,\Zsun.
Among the available indicators, we use the R3-index, defined as R3 = \OIII$\lambda 5007$/\Hb, throughout this study. Although the R23-index, given by R23 = (\OIII$\lambda\lambda 5007,4959$ + \OII$\lambda\lambda 3726, 3729$)/\Hb, is generally recommended for low-metallicity regimes, we refrain from using it due to the unavailable \OII\ in all but the brightest targets or clumps, owing to poor sensitivity below 4000\,\AA. For most of the objects in this study, the metallicities derived from the R3-index exhibit smaller uncertainties, even when accounting for systematic errors, compared to those from the R23-index, which relies on upper limits for \OII\ to constrain the metallicity range.
To account for variations in ionization, we use the equivalent width of \Hb, EW(\Hb), as a proxy for ionization-sensitive line ratios, such as \OIII/\OII, following the method proposed by \citet{nakajima2022_empressV}.

\begin{figure*}
  \centering
    \begin{tabular}{c}      %
      \begin{minipage}{0.304\hsize}
        \begin{center}
         \includegraphics[bb=0 0 402 274, width=1.0\textwidth]{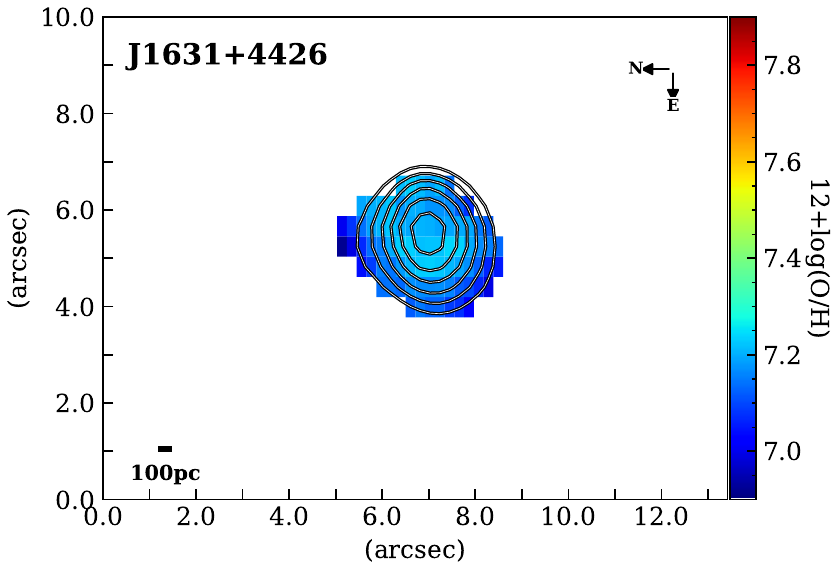}
        \end{center}
      \end{minipage}      %
      \begin{minipage}{0.304\hsize}
        \begin{center}
         \includegraphics[bb=0 0 402 274, width=1.0\textwidth]{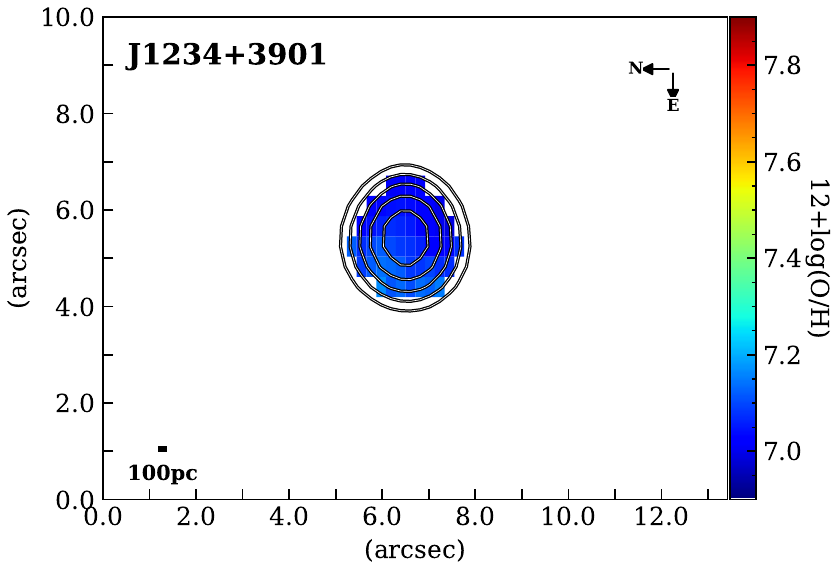}
        \end{center}
      \end{minipage}      %
      \begin{minipage}{0.304\hsize}
        \begin{center}
         \includegraphics[bb=0 0 402 274, width=1.0\textwidth]{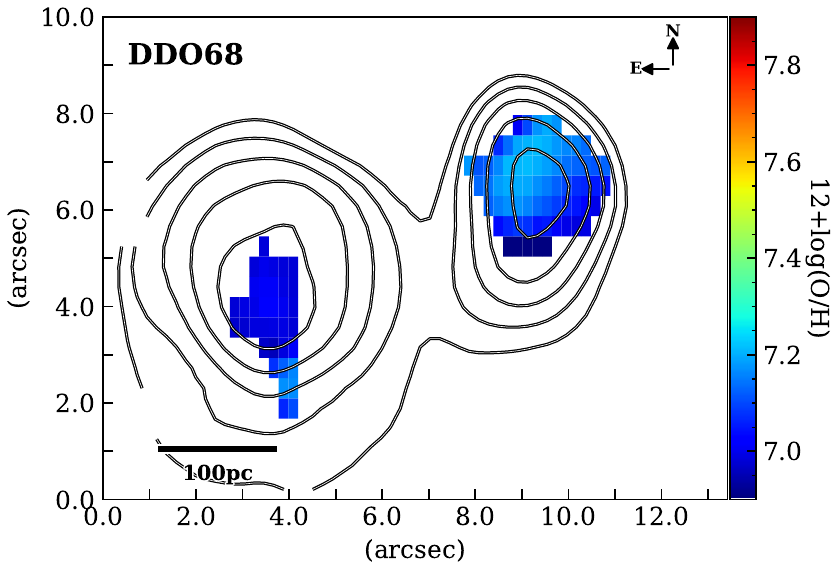}
        \end{center}
      \end{minipage}      \\
      \begin{minipage}{0.304\hsize}
        \begin{center}
         \includegraphics[bb=0 0 402 274, width=1.0\textwidth]{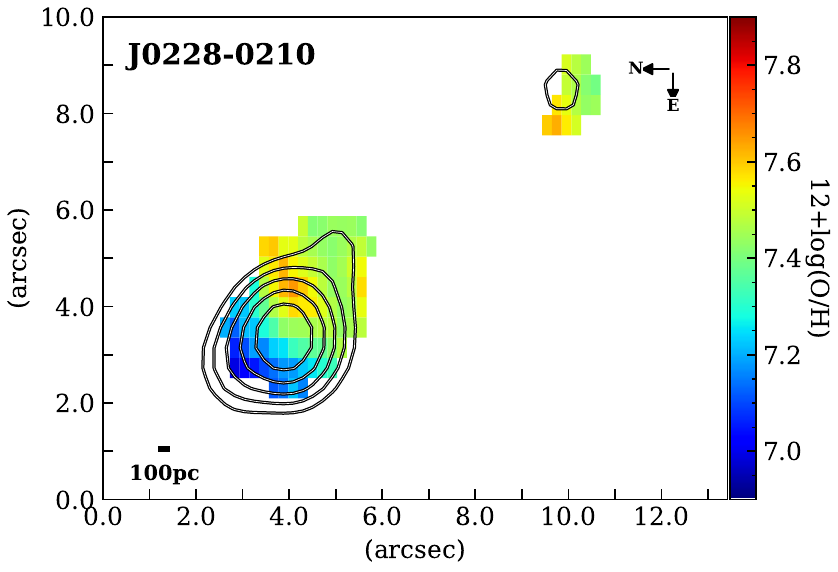}
        \end{center}
      \end{minipage}      %
      \begin{minipage}{0.304\hsize}
        \begin{center}
         \includegraphics[bb=0 0 402 274, width=1.0\textwidth]{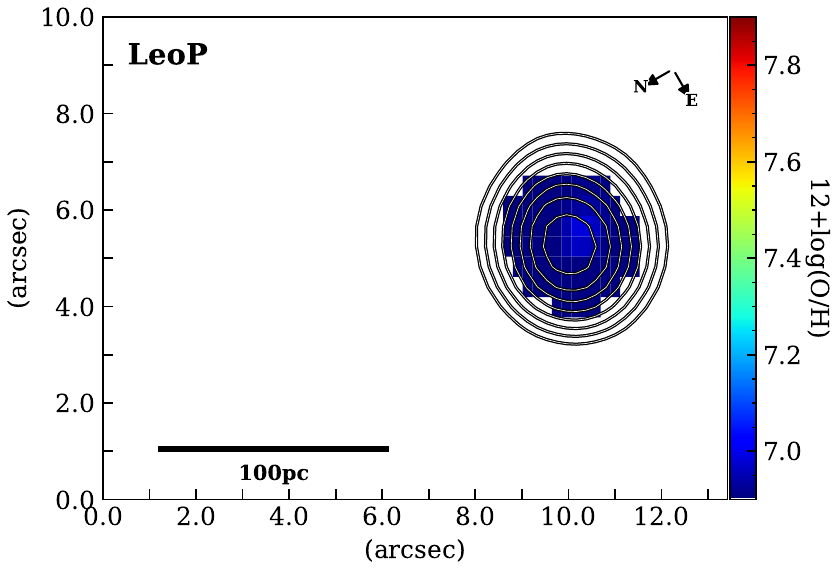}
        \end{center}
      \end{minipage}      %
      \begin{minipage}{0.304\hsize}
        \begin{center}
         \includegraphics[bb=0 0 402 274, width=1.0\textwidth]{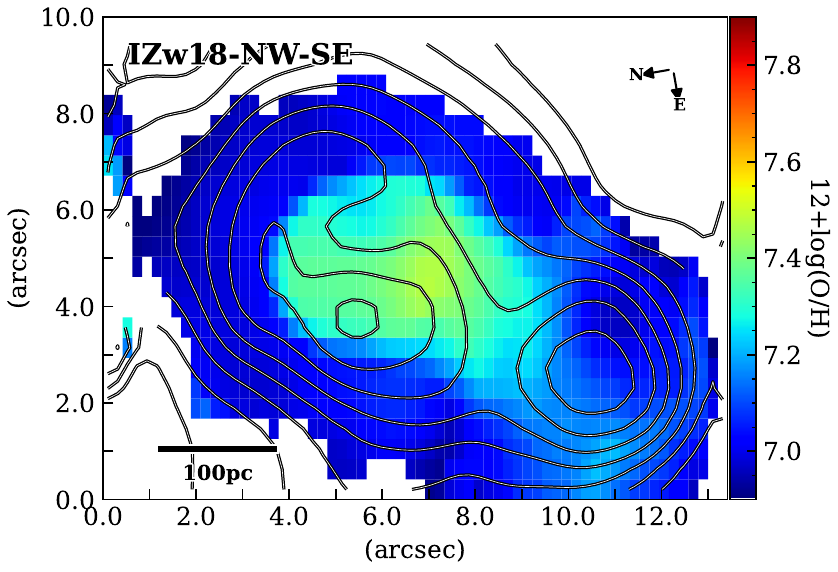}
        \end{center}
      \end{minipage}      \\
      \begin{minipage}{0.304\hsize}
        \begin{center}
         \includegraphics[bb=0 0 402 274, width=1.0\textwidth]{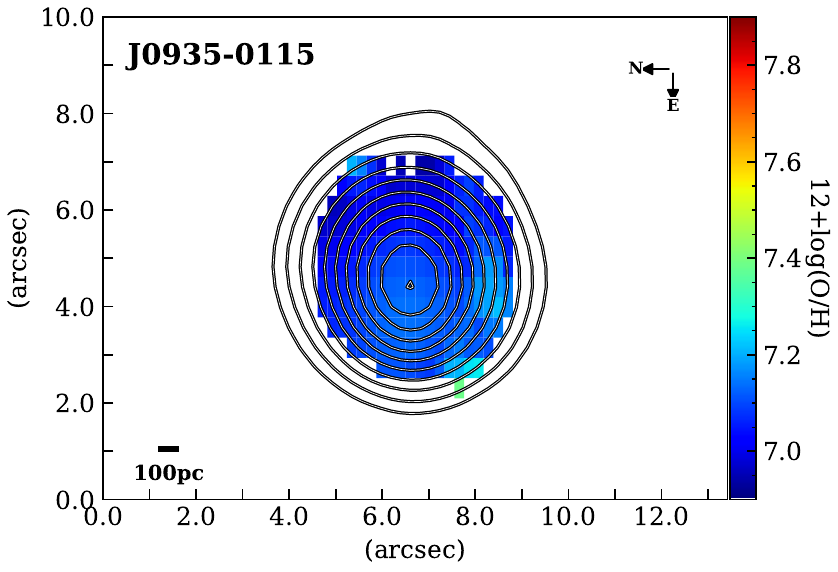}
        \end{center}
      \end{minipage}      %
      \begin{minipage}{0.304\hsize}
        \begin{center}
         \includegraphics[bb=0 0 402 274, width=1.0\textwidth]{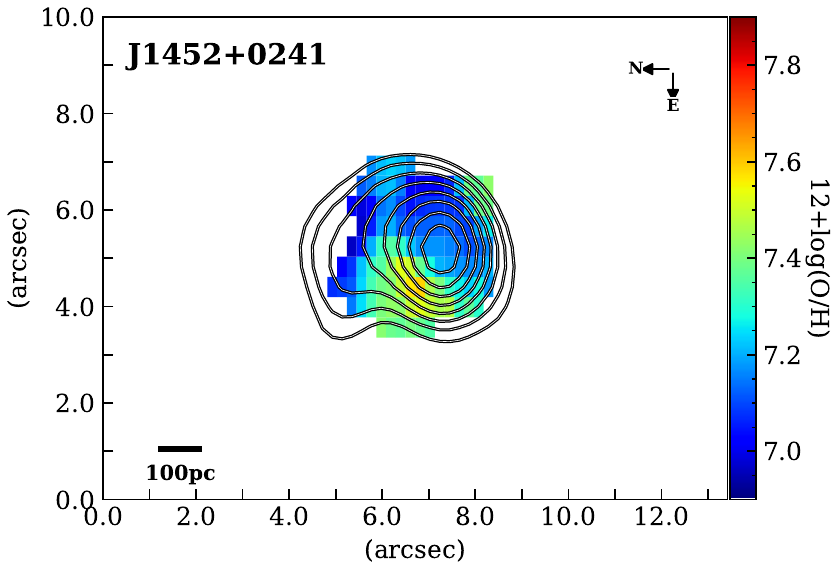}
        \end{center}
      \end{minipage}      %
      \begin{minipage}{0.304\hsize}
        \begin{center}
         \includegraphics[bb=0 0 402 274, width=1.0\textwidth]{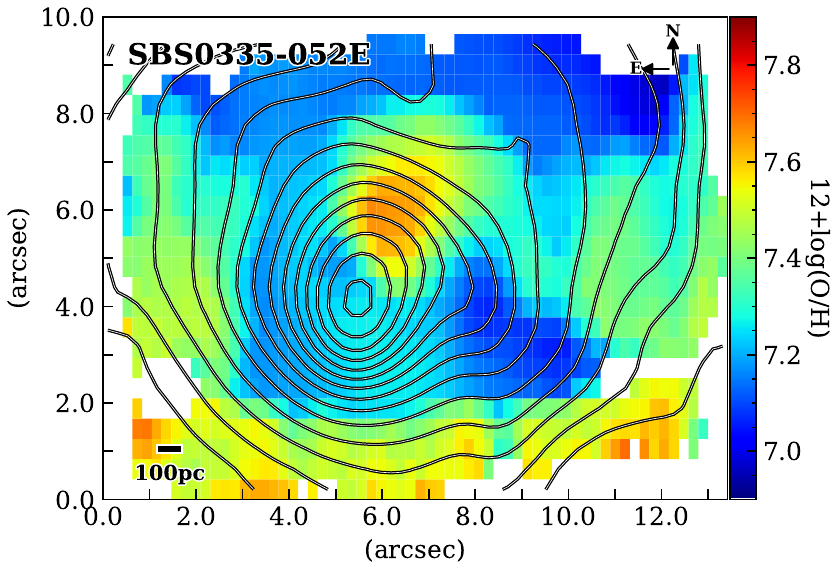}
        \end{center}
      \end{minipage}      \\
      \begin{minipage}{0.304\hsize}
        \begin{center}
         \includegraphics[bb=0 0 402 274, width=1.0\textwidth]{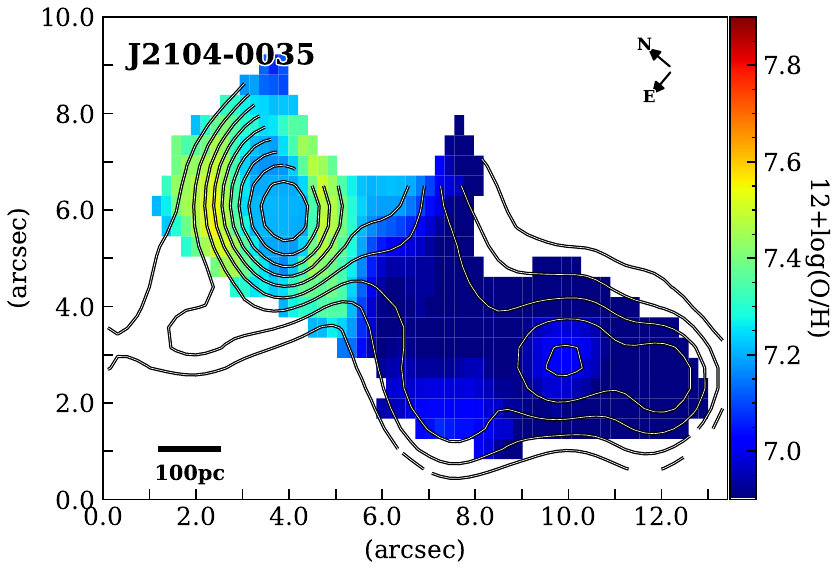}
        \end{center}
      \end{minipage}      %
      \begin{minipage}{0.304\hsize}
        \begin{center}
         \includegraphics[bb=0 0 402 274, width=1.0\textwidth]{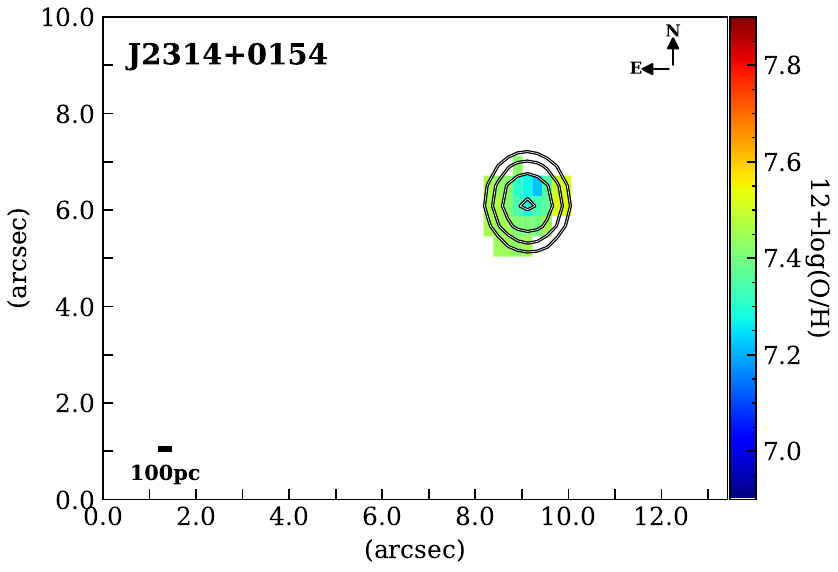}
        \end{center}
      \end{minipage}      %
      \begin{minipage}{0.304\hsize}
        \begin{center}
         \includegraphics[bb=0 0 402 274, width=1.0\textwidth]{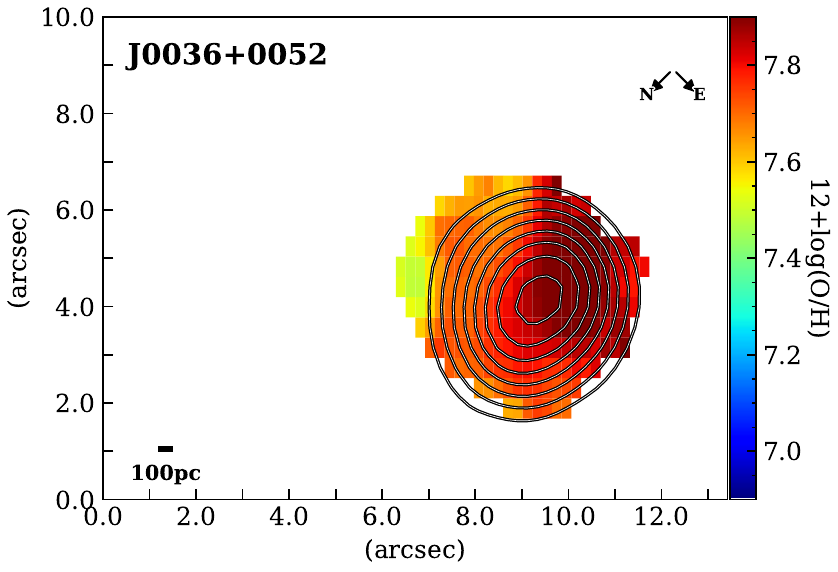}
        \end{center}
      \end{minipage}      \\
    \end{tabular}
    \caption{%
      Gas-phase metallicity maps for the 24 EMPGs. Spaxels with measured metallicities are shown (see Sect.~\ref{ssec:results_metallicity_measurement}). Lower metallicities are represented by bluer colors, as indicated in the legend. Contours trace the \Ha\ intensity distribution, outlining the spatial structure of each galaxy. Each panel includes a 100\,pc scale bar in the bottom left corner and directional markers for North and East in the top right corner for reference.
    }
    \label{fig:map_Z}
\end{figure*}


\begin{figure*}
  \addtocounter{figure}{-1}
  \centering
    \begin{tabular}{c}      %
      \begin{minipage}{0.304\hsize}
        \begin{center}
         \includegraphics[bb=0 0 402 274, width=1.0\textwidth]{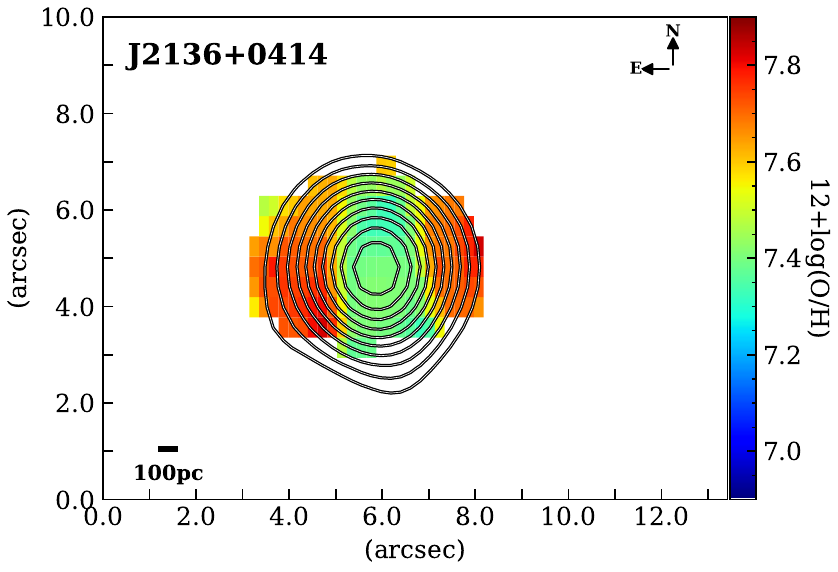}
        \end{center}
      \end{minipage}      %
      \begin{minipage}{0.304\hsize}
        \begin{center}
         \includegraphics[bb=0 0 402 274, width=1.0\textwidth]{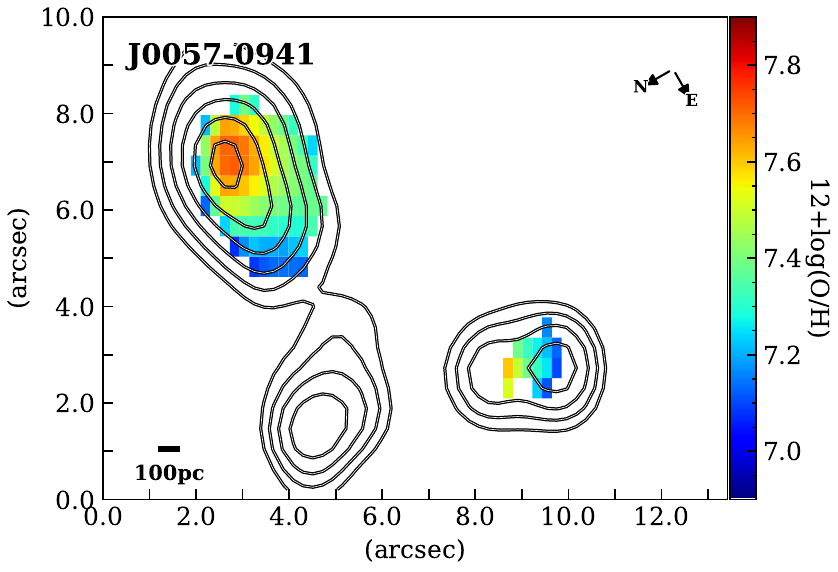}
        \end{center}
      \end{minipage}      %
      \begin{minipage}{0.304\hsize}
        \begin{center}
         \includegraphics[bb=0 0 402 274, width=1.0\textwidth]{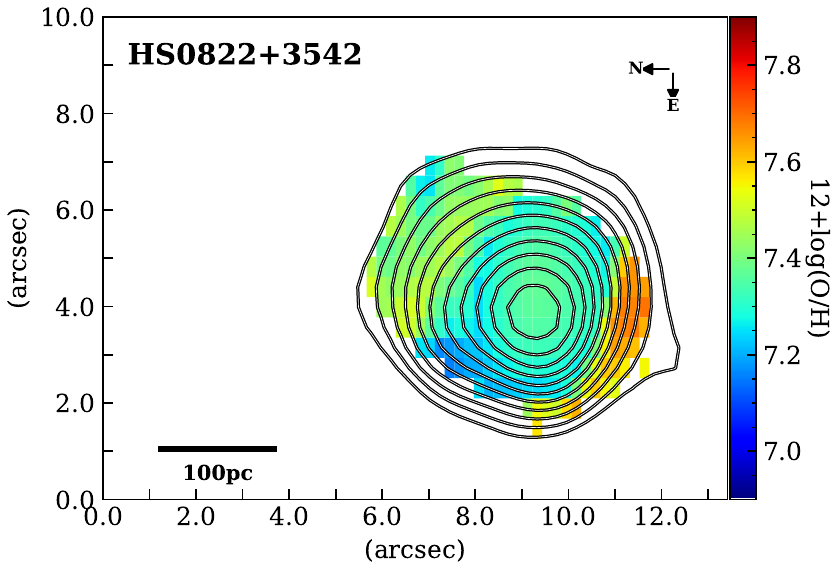}
        \end{center}
      \end{minipage}      \\
      \begin{minipage}{0.304\hsize}
        \begin{center}
         \includegraphics[bb=0 0 402 274, width=1.0\textwidth]{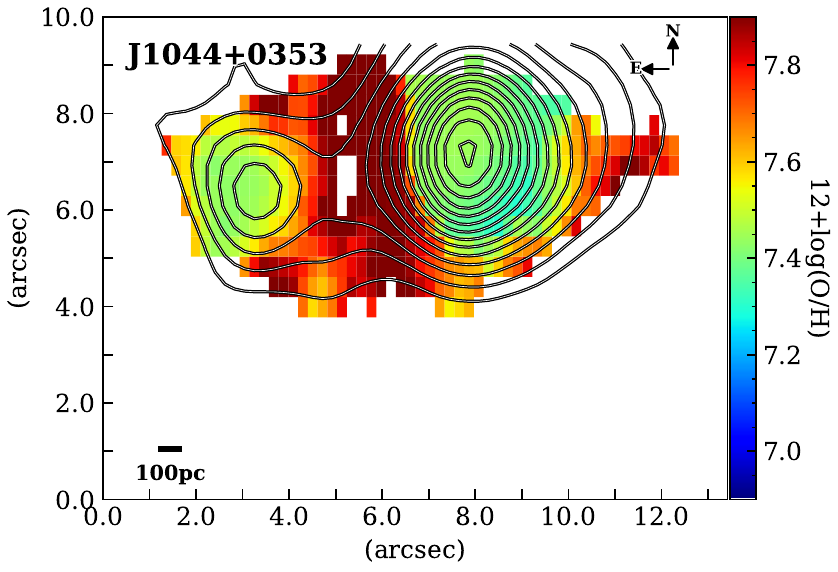}
        \end{center}
      \end{minipage}      %
      \begin{minipage}{0.304\hsize}
        \begin{center}
         \includegraphics[bb=0 0 402 274, width=1.0\textwidth]{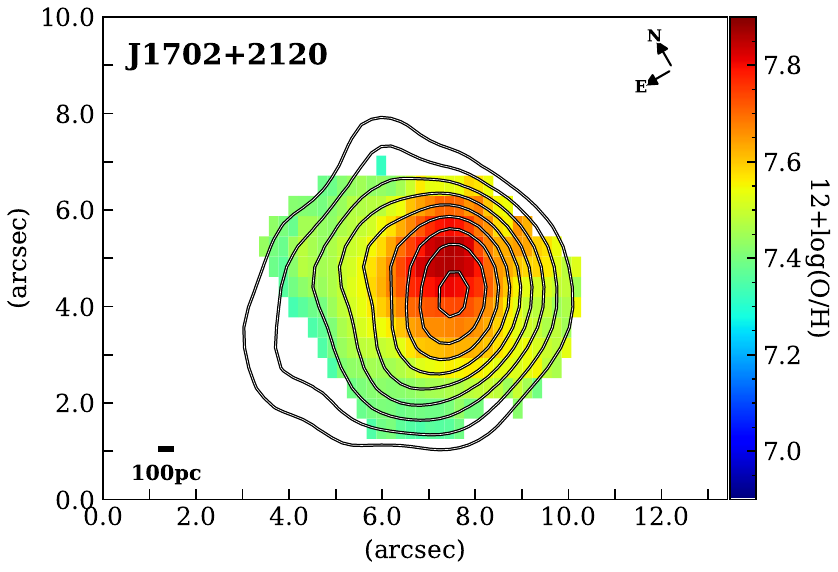}
        \end{center}
      \end{minipage}      %
      \begin{minipage}{0.304\hsize}
        \begin{center}
         \includegraphics[bb=0 0 402 274, width=1.0\textwidth]{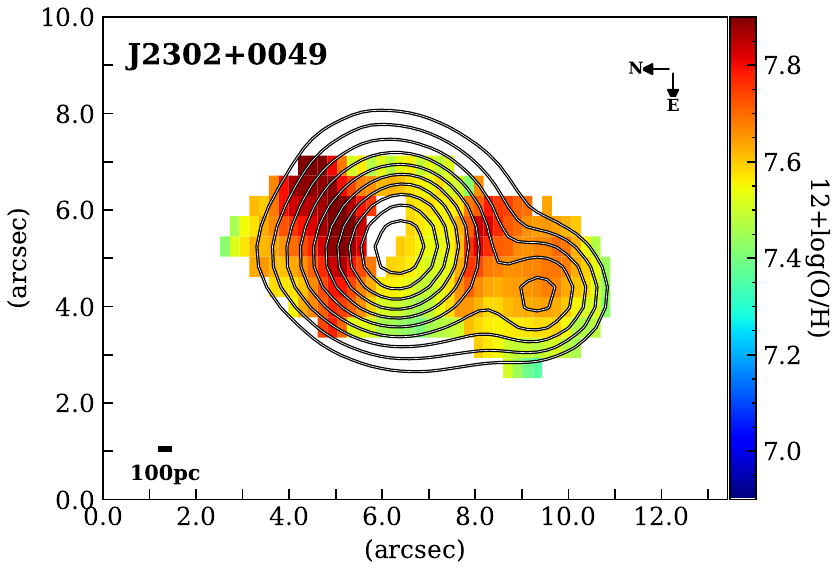}
        \end{center}
      \end{minipage}      \\
      \begin{minipage}{0.304\hsize}
        \begin{center}
         \includegraphics[bb=0 0 402 274, width=1.0\textwidth]{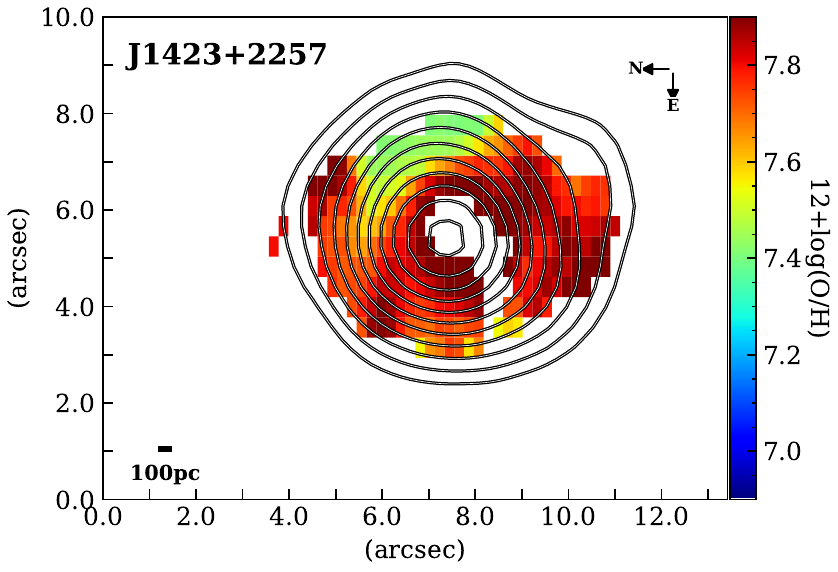}
        \end{center}
      \end{minipage}      %
      \begin{minipage}{0.304\hsize}
        \begin{center}
         \includegraphics[bb=0 0 402 274, width=1.0\textwidth]{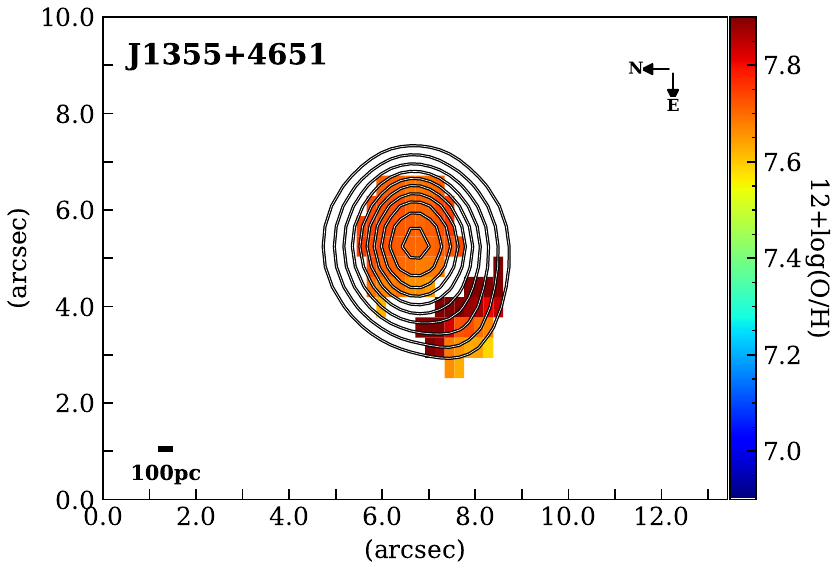}
        \end{center}
      \end{minipage}      %
      \begin{minipage}{0.304\hsize}
        \begin{center}
         \includegraphics[bb=0 0 402 274, width=1.0\textwidth]{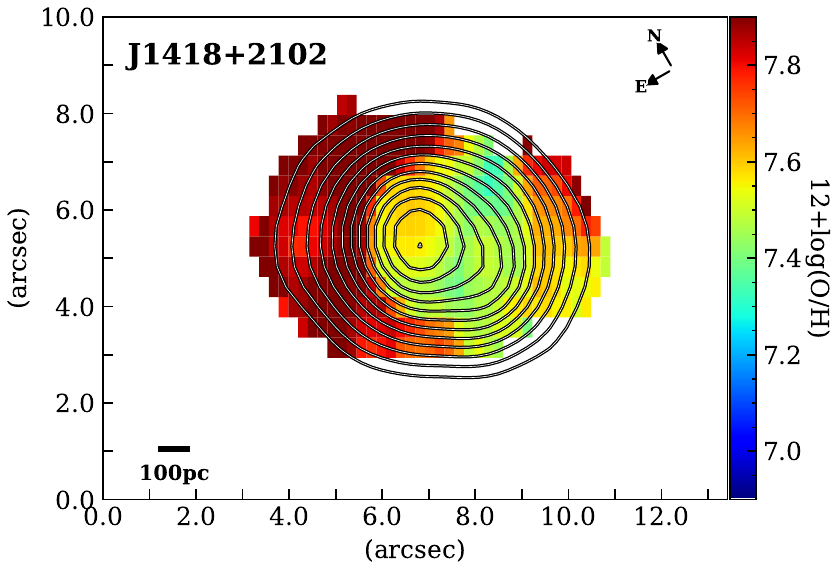}
        \end{center}
      \end{minipage}      \\
      \begin{minipage}{0.304\hsize}
        \begin{center}
         \includegraphics[bb=0 0 402 274, width=1.0\textwidth]{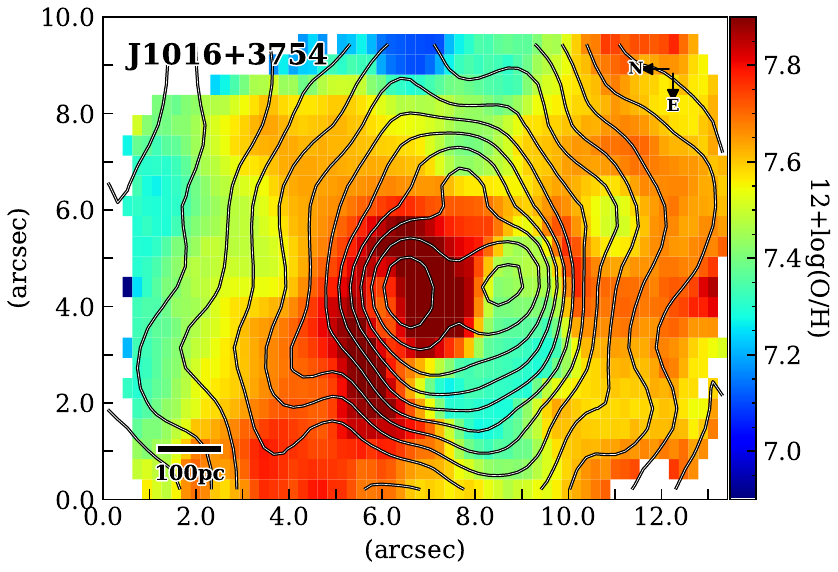}
        \end{center}
      \end{minipage}      %
      \begin{minipage}{0.304\hsize}
        \begin{center}
         \includegraphics[bb=0 0 402 274, width=1.0\textwidth]{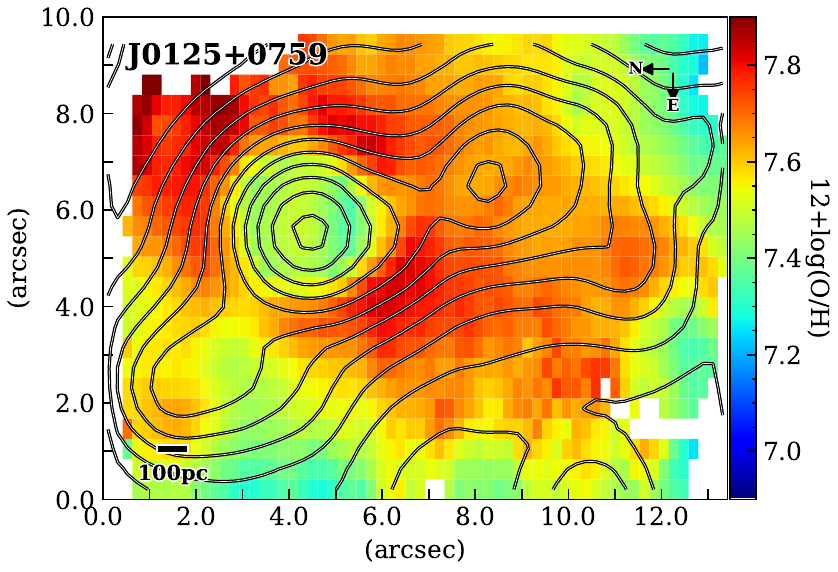}
        \end{center}
      \end{minipage}      %
      \begin{minipage}{0.304\hsize}
        \begin{center}
         \includegraphics[bb=0 0 402 274, width=1.0\textwidth]{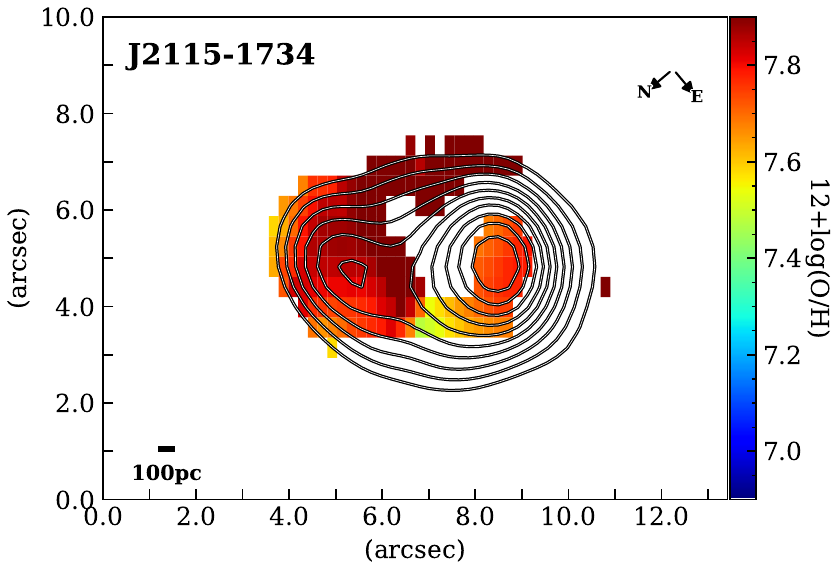}
        \end{center}
      \end{minipage}      \\
    \end{tabular}
    \caption{%
      (Continued.) 
    }
\end{figure*}



In Figure \ref{fig:compare_Z}, we assess the reliability of the R3-index by comparing metallicities derived from the direct temperature method with those obtained using the R3-index. For this comparison, we use spectra summed from the central spaxels around the brightest core of each galaxy.
For the direct temperature method, we apply the dust attenuation correction following the law of \citet{calzetti2000}. We then determine the gas temperature using \OIII$\lambda 4363$ and derive the oxygen abundance from the \OIII/\Hb\ and \OII/\Hb\ ratios (e.g., \citealt{izotov2009,nakajima2022_empressV}). For those without a significant detection of \OII$\lambda 3727$, the oxygen abundance is bracketed by O$^{2+}$/H$^{+}$ and (O$^{2+}$/H$^{+}$ $+$ $3\sigma$ upper-limit of O$^{+}$/H$^{+}$).

Figure \ref{fig:compare_Z} shows that metallicities derived using the R3-index generally align well with those obtained through the direct-temperature method, supporting the use of the R3-index for our resolved metallicity distribution analysis. Notably, for the most metal-poor galaxy in our sample, J1631$+$4426, our direct-temperature metallicity closely matches the value initially reported in \citet{kojima2020} based on their direct-temperature method, while the R3-based metallicity is approximately 0.3\,dex higher. This discrepancy may be attributed to the fact that this galaxy lies at the extreme low-metallicity end, where empirical metallicity relations are less robustly calibrated. Interestingly, \citet{thuan2022} re-evaluated the electron temperature for this galaxy based on independently taken new observations and reported a 0.24 dex higher metallicity compared to the original value. This ongoing discussion highlights the need for further refinement and calibration of metallicity indicators, particularly at the lowest metallicity regime.
Another consideration is that the contribution of O$^{+}$ to the direct-temperature metallicity may be underestimated due to potential slit losses of \OII\ at shorter wavelengths (Sect.~\ref{ssec:data_maps}). This effect could be more pronounced in metal-rich systems, where lower ionization states (i.e., lower \OIII$/$\OII\ ratios) are typically observed (e.g., \citealt{nagao2006_metallicity, maiolino2008, curti2017, nakajima2022_empressV}). Such a bias may explain the data-points with \Oabundance\ $\gtrsim 7.5$ in Figure \ref{fig:compare_Z}, where metallicities derived using the direct temperature method appear lower than those estimated with the R3-index.

A key limitation of the R3-index is that it exhibits a two-valued behavior, diminishing its diagnostic power around \Oabundance\ $=8.0$. For our objects, we consistently select the low-metallicity solution, following the procedure outlined in \citet{nakajima2022_empressV}, as the central metallicities are already known to be \Oabundance\ $<7.8$. This choice is further supported by the low \NII$/$\Ha\ ratios observed in the spaxels of EMPGs (see Sect.~\ref{ssec:results_BPT}), ensuring that the derived metallicity distributions remain smooth across the systems. 
However, exceptions arise with the two highest metallicity galaxies in our sample, J1323$-$0132 and J1253$-$0312, with central metallicities around \Oabundance\ $\sim7.8-8.0$. In these cases, the outer regions may fall on either the low- or high-metallicity branch of the R3-index. To avoid potential inconsistencies, we exclude these two galaxies from further analyses in this paper. Consequently, the final sample thus comprises 24 objects, totaling 9,177 spaxels with reliable metallicity measurements, defined by $>5\sigma$ detections of \OIII, \Hb, and the underlying continuum (Sect.~\ref{ssec:data_maps}).

Using the 9,177 spaxels with reliable metallicity measurements, Figure \ref{fig:map_Z} presents the metallicity maps of the 24 EMPGs, arranged in descending order based on the metallicity measured at the central region of each object. The center is defined as the peak of the \Ha\ intensity. The maps reveal a diverse range of metallicity distributions within the extremely metal-poor clumps. In the following sections, we explore how the local metallicity varies as a function of parameters such as local stellar mass, SFR, distance from the center, and other relevant factors.

\subsection{BPT diagram} \label{ssec:results_BPT}

\begin{figure}[!t]
    \begin{center}
     \includegraphics[bb=0 0 510 407, width=0.99\columnwidth]{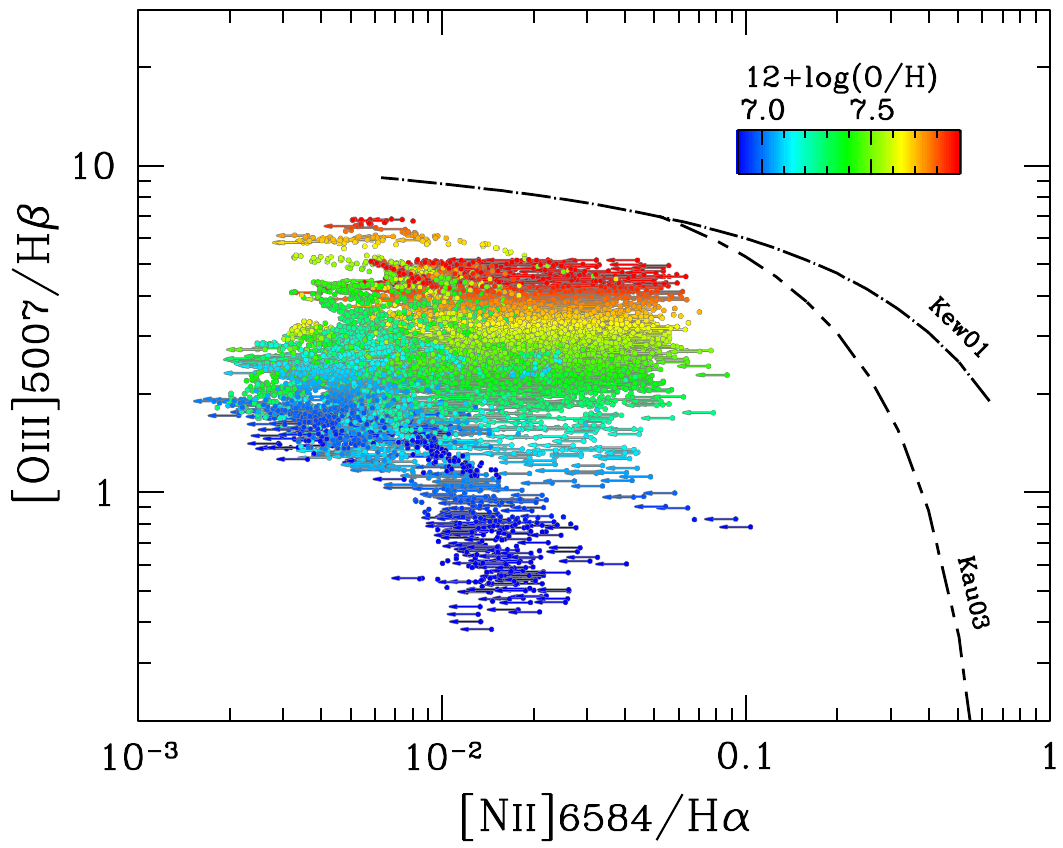}
    \caption{%
   	Resolved \NII\ BPT diagram for the EMPGs. Each spaxel with a metallicity measurement is plotted, color-coded by metallicity as indicated in the legend. For spaxels lacking a \NII\ detection, $3\sigma$ upper limits are shown with leftward arrows. The conventional demarcation curves between AGNs and galaxies on the diagrams are depicted as introduced by \citet{kewley2001} (dot-long dashed) and \citet{kauffmann2003} (short dash-long dashed).
	}
    \label{fig:rBPT}
    \end{center} 
\end{figure}

Before moving in detail to the metallicity discussion, we examine the placement of our sources and spaxels on the \citet{BPT1981} diagram to identify any clear evidence of active galactic nuclei (AGNs) within our sample. Figure \ref{fig:rBPT} displays the \OIII/\Hb\ and \NII/\Ha\ line ratios (referred to as the \NII\ BPT diagram) for all spaxels with reliable metallicity measurements (i.e., spaxels with robust detections of \OIII\ and \Hb\ and reliable EW(\Hb); see Sect.~\ref{ssec:results_metallicity_measurement}). The two demarcation curves, from \citet{kewley2001} and \citet{kauffmann2003}, distinguish between AGN-powered and star-formation-dominated systems, with spaxels lying above the curves classified as AGN-dominated.

Figure \ref{fig:rBPT} shows that all spaxels, including those with upper limits on \NII, fall below the demarcation curves, confirming that the emission is consistent with star-formation-dominated processes. This result aligns with previous studies of EMPGs based on integrated emission-line analyses (e.g., \citealt{izotov2018_lowZ,kojima2020,isobe2022_fe}). However, it is worth noting that low-metallicity AGNs with $Z\lesssim0.5$\,\Zsun\ are theoretically predicted to overlap with star-forming regions on the BPT diagram (e.g., \citealt{kewley2013_theory,NM2022}). While we cannot fully exclude the possibility of low-metallicity AGNs within our sample, our results indicate an absence of evolved AGN-dominated regions within these EMPGs.

Evidence from other studies suggests that black hole-dominated components, such as high-mass X-ray binaries, ultraluminous X-ray sources, and intermediate-mass black holes, may be present in metal-poor dwarf galaxies (e.g., \citealt{schaerer2019,umeda2022,hatano2024}), although some studies present counterarguments \citep{senchyna2020,kehrig2021}. A detailed analysis of the \HeII$\lambda 4686$ line \citep{TI2005,kehrig2015,kehrig2018,rickardsvaught2021} and/or the \OIII$\lambda 4363$ auroral line \citep{mazzolari2024}, based on the EMPRESS 3D data, will be presented in a forthcoming paper to further investigate the presence of high-energy sources in EMPGs.

For the purposes of this paper, we assume that all spaxels are dominated by star formation. Thus, we apply the metallicity calibration from \citet{nakajima2022_empressV}, which is appropriate for star-forming regions.

\begin{figure*}[t]
  \centering
    \begin{tabular}{c}
      \begin{minipage}{0.99\hsize}
        \begin{center}
         \includegraphics[bb=0 0 557 402, width=0.8\textwidth]{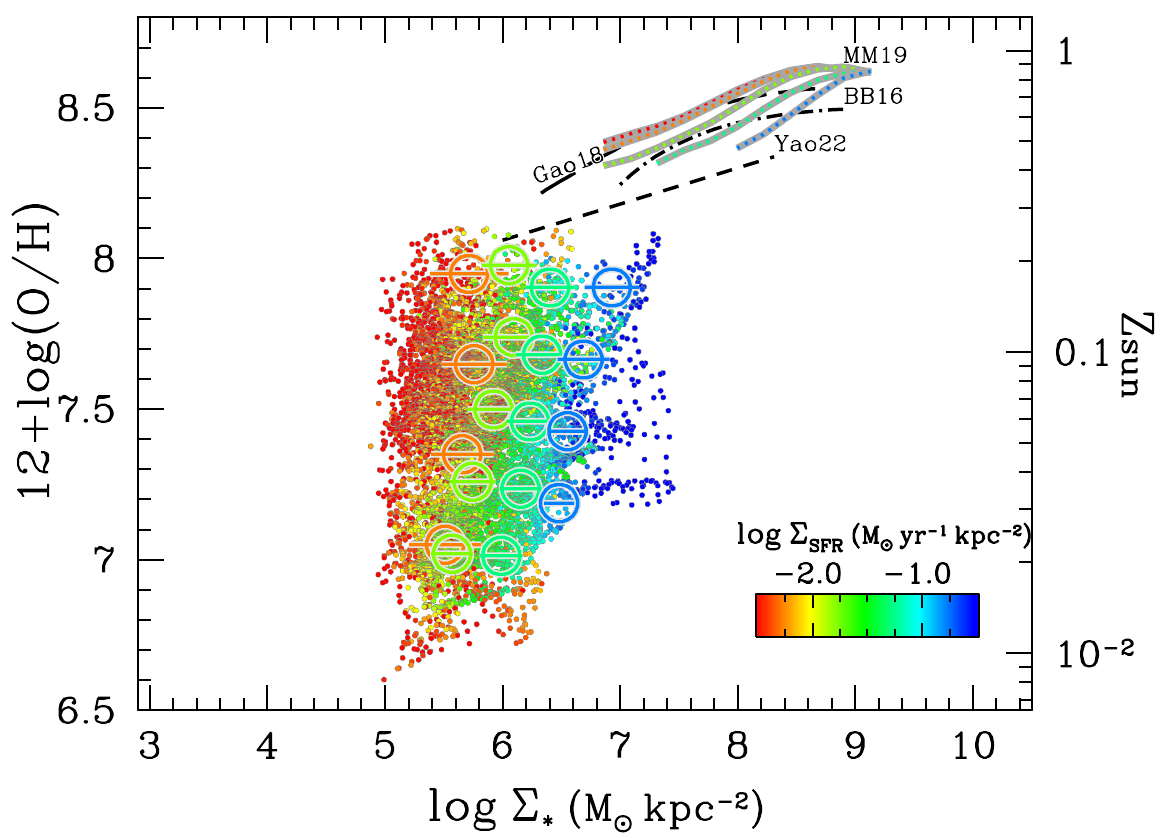}
        \end{center}
      \end{minipage}
      \\
      \\
      \begin{minipage}{0.49\hsize}
        \begin{center}
         \includegraphics[bb=0 0 557 402, width=0.9\textwidth]{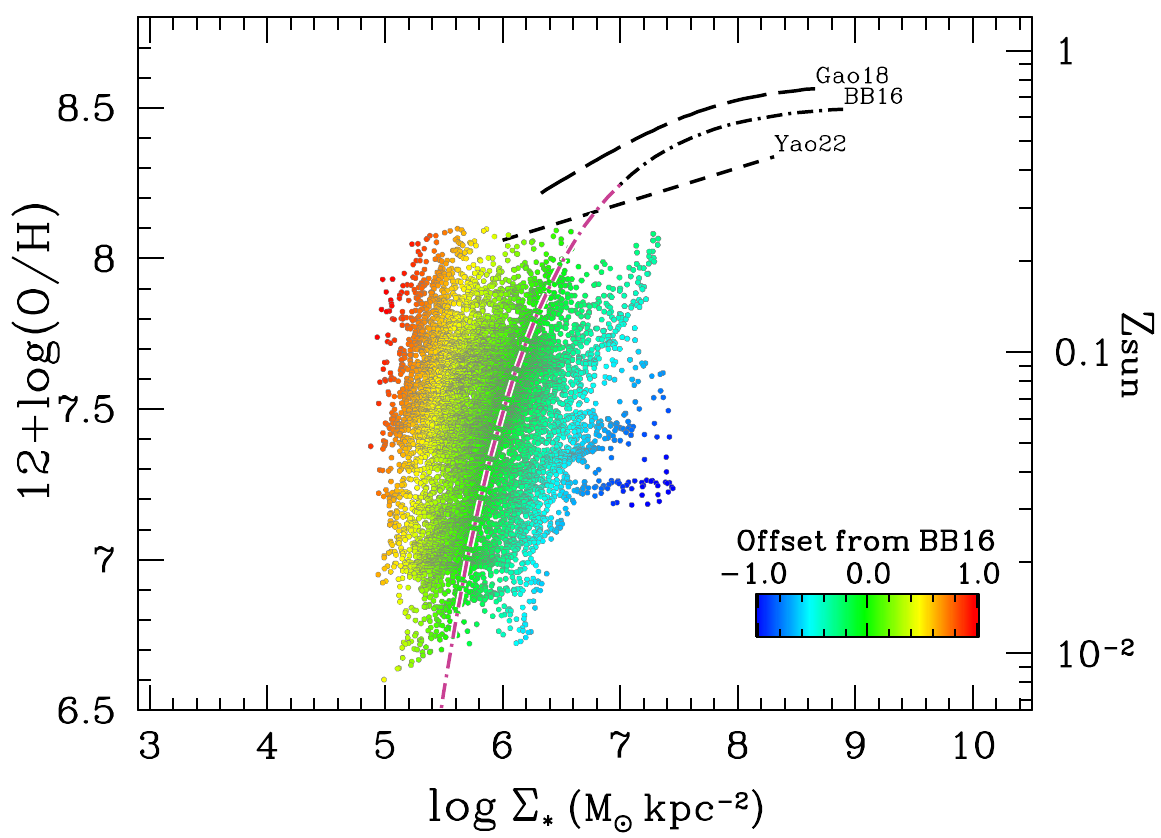}
        \end{center}
      \end{minipage}
      \begin{minipage}{0.49\hsize}
        \begin{center}
         \includegraphics[bb=0 0 557 402, width=0.9\textwidth]{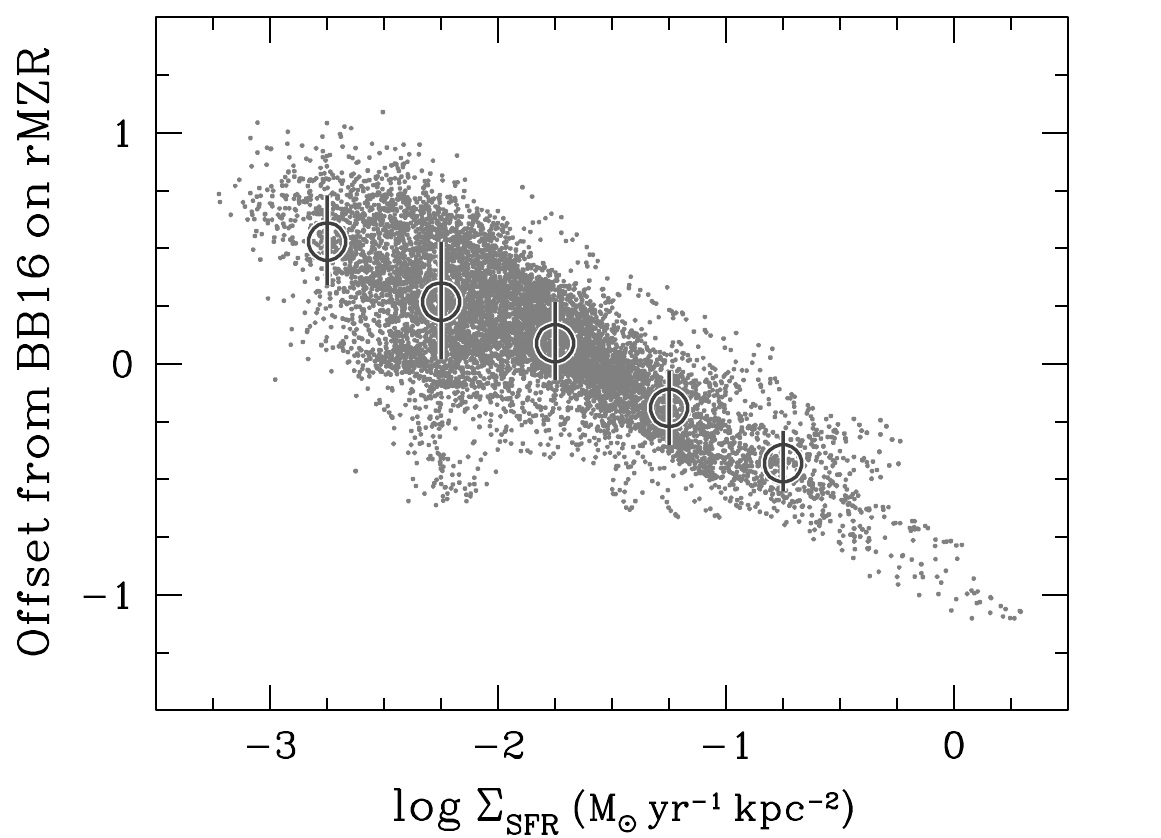}
        \end{center}
      \end{minipage}
    \end{tabular}
    \caption{%
      		(a) The relationship between local stellar mass surface density ($\Sigma_{\star}$) and metallicity for the spaxels of EMPGs observed in this study. The spaxels are represented by colored dots, with colors indicating the local SFR surface density ($\Sigma_{\rm SFR}$) as shown in the color bar. Large open circles indicate the average relationships for EMPG spaxels across four subsamples of $ \log \Sigma_{\rm SFR}$ values from $-2.5$ (orange) to $-0.5$ (blue). For comparison, high-metallicity studies are shown with colored curves (Belfiore et al., in prep.; data from \citealt{MM2019}), and previously explored resolved mass-metallicity relations in the high-metallicity regime are shown with black curves: dot-dashed for \citet{barrera-ballesteros2016}, long-dashed for \citet{gao2018}, and dashed for \citet{yao2022}. These comparisons are illustrated only within the $\Sigma_{\star}$ ranges originally explored in their respective studies.
		(b) Same as in Panel (a), but each spaxel is color-coded by its offset from the extrapolated relationship of \citet{barrera-ballesteros2016} (magenta line), where a negative offset indicates a spaxel falling below the relation.
		(c) The relationship between the offset from the \citet{barrera-ballesteros2016} relation and $\Sigma_{\rm SFR}$ for the spaxels of EMPGs. Individual spaxels are shown as dots, and large open circles represent average values across $\Sigma_{\rm SFR}$ bins.
    }
    \label{fig:rMZR}
\end{figure*}

\subsection{Resolved mass-metallicity relation} \label{ssec:results_rMZR}

We now shift our focus back to the metallicity, investigating its correlation with stellar mass and SFR. 
The $y$-axis values in Figure \ref{fig:rBPT} represent the R3-index used for metallicity measurements (Sect.~\ref{ssec:results_metallicity_measurement}). Notably, the extremely low R3-index values below $1$--$2$, combined with low \NII$/$\Ha\ ratios ($\ll 0.1$) are rarely explored in prior studies (e.g., \citealt{nagao2006_metallicity, maiolino2008, curti2017}). Some regions exhibit R3-index values as low as $\sim 0.4$, corresponding to metallicities below 1\% of the solar value, an exceptional finding even among EMPG samples \citep{nakajima2022_empressV}. By spatially resolving the metallicity distributions within EMPGs, we anticipate uncovering additional, even more metal-deficient regions.

To investigate the correlation between metallicity and stellar mass on local scales, we estimate the stellar mass for each spaxel using the stellar continuum within the wavelength range $\lambda_0 = 6500$--$7000$\,\AA\ and a mass-to-optical-luminosity relation typical of EMPGs \citep{kojima2020, isobe2021_tail, xu2022}. In these studies, they estimate stellar masses of local EMPGs by fitting SEDs to the broadband photometry using \verb+BEAGLE+ \citep{chevallard2016}, assuming a constant star formation and a \citet{chabrier2003} initial mass function (IMF). The typical best-fit SEDs has a stellar age of $\sim 5$\,Myr, one tenth of metallicity, and dust-free conditions. 
The mass-to-optical-luminosity relation is then obtained by linear fitting to the stellar masses and $i$-band luminosities. 
The SFR for each spaxel is obtained from the dust-corrected \Ha\ flux following the \citet{kennicutt1998} relation, also using the Chabrier IMF. To compute the surface densities of mass ($\Sigma_{\star}$) and SFR ($\Sigma_{\rm SFR}$), we divide the mass and SFR by the projection-corrected area of each spaxel. For projection corrections, we adopt the best-fit inclination angles derived from the morphokinematic analysis of \citet{isobe2023_focasifu} for the six objects studied therein using high-resolution 3D data. For the remaining objects, we use the projected axis ratio, $b/a$, calculated from the stellar light distribution, with values ranging between $0.61$ and $0.93$.

Figure \ref{fig:rMZR}(a) presents the spatially resolved mass-metallicity relation (rMZR), plotting all spaxels from the EMPGs along with previous high-metallicity results. Large IFU surveys, such as CALIFA and MaNGA, have established a tight rMZR in the high-metallicity regime (\Oabundance\ $\gtrsim 8.3$) in the local universe \citep{barrera-ballesteros2016, gao2018, sanchez2013}. Notably, the metallicity indicators used in these surveys are calibrated consistently using the direct temperature method.
By targeting EMPGs with the high-sensitivity spectrograph on the 8m Subaru Telescope, the EMPRESS 3D survey extends the rMZR into previously unexplored parameter spaces, probing extremely low metallicity (\Oabundance\ ranging from $6.9$ to $7.9$) and low stellar mass surface densities ($\Sigma_{\star} = 10^5$--$10^7$\,\Msunkpc), as shown in Figure \ref{fig:rMZR}(a). 
Interestingly, the lowest observed stellar mass surface density, $\Sigma_{\star}\sim 10^5$\,\Msunkpc, aligns with the star formation threshold density of $\Sigma_{\rm gas} \sim 1$\,\Msunpc\ \citep{kennicutt1989}, assuming the typical $10:1$ gas-to-stellar mass ratio found in EMPGs \citep{isobe2023_focasifu, xu2024_focasifu}. While this limit may partly reflect the observational detection constraints, our observations approach the critical threshold density for gravitational instability, potentially explaining the absence of regions below $\Sigma_{\star} \sim 10^5$\,\Msunkpc, where massive star formation may be entirely suppressed. In any case, these observations offer a unique opportunity to examine the metal enrichment processes at the low-metallicity and low-$\Sigma_{\star}$ regime, providing insights into the early evolutionary phases of galaxies in the local universe.

The spaxels of the 24 EMPGs reveal a substantial scatter in the low $\Sigma_{\star}$ regime on the rMZR plot. A similar trend has also been observed at the low $\Sigma_{\star}$ end of the MaNGA sample (e.g., \citealt{gao2018}), and our EMPRESS 3D data provides statistically significant confirmation of this behavior.

Despite the large scatter, the EMPG data points align, on average, with the extrapolation of the rMZR from the high-metallicity regime. For instance, \citet{barrera-ballesteros2016} adopted the following functional form, used previously by \citet{moustakas2011} and \citet{sanchez2013}, to describe the rMZR:
\begin{equation}
	y = a + b (x-c) e^{-(x-c)}.
\label{eq:rMZR}
\end{equation}
In this equation, $y$ represents \Oabundance\ and $x$ denotes the logarithm of $\Sigma_{\star}$ in units of \Msunkpc. Among the three free parameters, $a$ corresponds to the asymptotic oxygen abundance at high mass densities, while $b$ and $c$ determine the rate of metallicity increase with mass density.
In Figure \ref{fig:rMZR}(b), we show the best fit of Equation (\ref{eq:rMZR}) to the MaNGA galaxies from \citet{barrera-ballesteros2016} (black line) and its simple extrapolation toward lower $\Sigma_{\star}$ values (magenta line). The resolved components of our EMPGs tend to fall along this extrapolated relation, extending the rMZR to lower metallicities and mass densities. Notably, the rapidly declining component of the rMZR emerges at $\Sigma_{\star} \sim 10^7$\,\Msunkpc, likely contributing to the increased scatter observed at the low-mass end, which has not been fully explored in previous studies (e.g., \citealt{gao2018,barrera-ballesteros2016}).
Although the scatter in the low-$\Sigma_{\star}$ regime is substantial, understanding the factors driving this variability is crucial. A detailed investigation of the causes behind this scatter, as well as the differences in rMZR trends between our results and prior work \citep{gao2018,yao2022}, will offer valuable insights into the mass-metallicity relationship at the low-metallicity, low-mass end of galaxy evolution.

\begin{figure}[t]
    \begin{center}
     \includegraphics[bb=0 0 557 402, width=0.99\columnwidth]{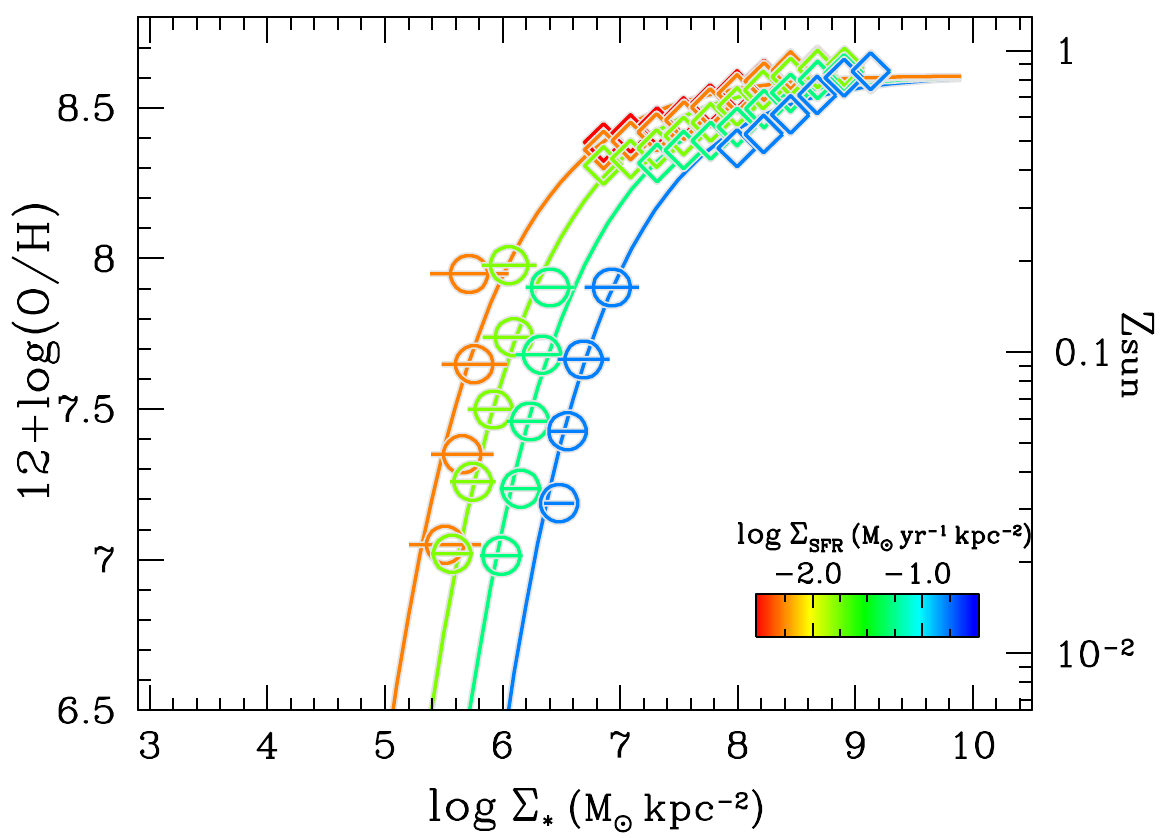}
    \caption{%
    		The best-fit relation for the $\Sigma_{\star}$--metallicity--$\Sigma_{\rm SFR}$ relation (Equation~\ref{eq:rMZR_modified}). The four colored curves represent the best-fit relations for $ \log \Sigma_{\rm SFR}$ values of $-2.25$ (orange), $-1.75$ (yellow), $-1.25$ (green), and $-0.75$ (blue). Open circles correspond to the average subsample of EMPGs (this study), while open diamonds represent the high-metallicity sample (Belfiore et al., in prep.; data from \citealt{MM2019}). Both datasets, spanning $ \log \Sigma_{\rm SFR}$ values from $-2.5$ to $-2.0$ (orange), $-2.0$ to $-1.5$ (yellow), $-1.5$ to $-1.0$ (green), and $-1.0$ to $-0.5$ (blue), are combined and used to get the best-fit.
		}
    \label{fig:rMZR_sfr_bestfit}
    \end{center} 
\end{figure}

Motivated by the well-established global relationship between stellar mass, metallicity, and SFR (e.g., \citealt{mannucci2010,lara-lopez2010,AM2013}), we color-code the individual data points in Figure \ref{fig:rMZR}(a) according to the local SFR surface density, $\Sigma_{\rm SFR}$. This approach allows us to explore potential dependencies on the local mass-metallicity relation. 
Figure \ref{fig:rMZR}(a) reveals a clear trend: regions with higher local $\Sigma_{\rm SFR}$ tend to lie below the average rMZR. This suggests a strong connection between elevated star formation activity and local metal deficiency. To quantify this relationship, we examine the correlation between $\Sigma_{\rm SFR}$ and the offset (i.e., the orthogonal distance) from the extrapolated relation of \citet{barrera-ballesteros2016}. As shown in panel (c) of Figure \ref{fig:rMZR}, the correlation between the two is remarkably tight.
A similar dependence of metallicity on star formation activity has also been reported in the high-metallicity regime (Belfiore et al., in prep.), as depicted with colored dotted curves (extracted from Fig.\,34 in \citealt{MM2019} as to be presented in Belfiore et al., in prep.). Our results extend this trend to the low-metallicity regime, providing further evidence for the influence of star formation on the mass-metallicity relation on local scales.

\begin{figure*}
  \centering
    \begin{tabular}{c}      %
      \begin{minipage}{0.304\hsize}
        \begin{center}
         \includegraphics[bb=0 0 533 352, width=1.0\textwidth]{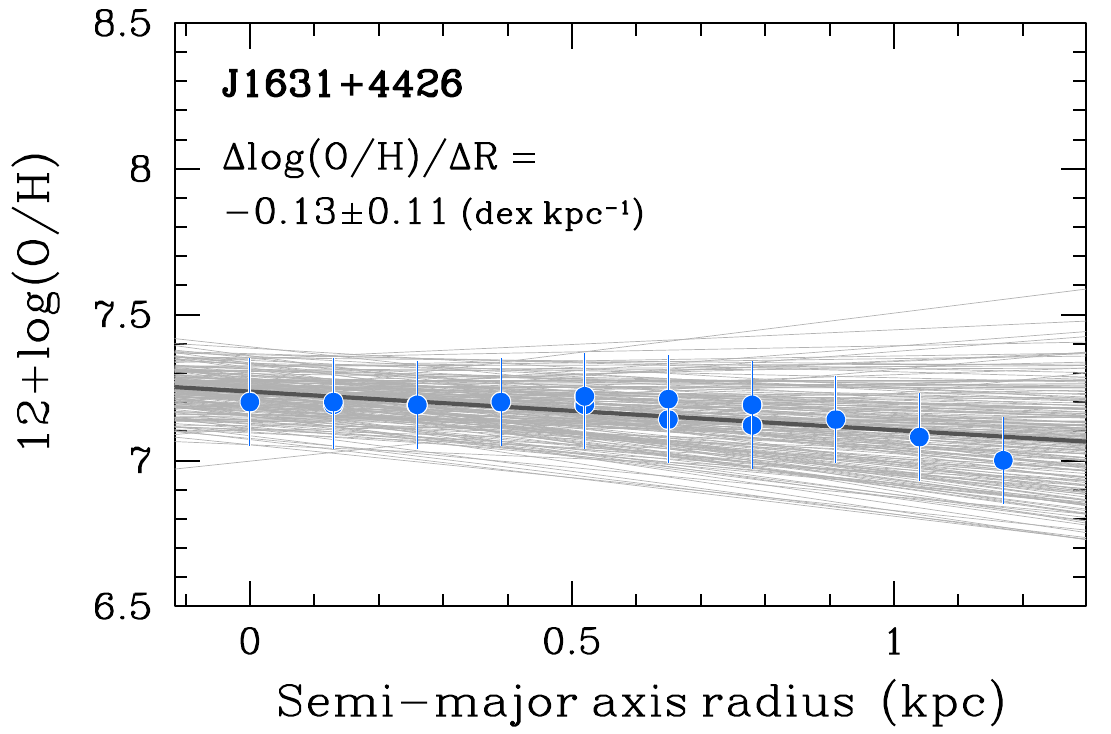}
        \end{center}
      \end{minipage}      %
      \begin{minipage}{0.304\hsize}
        \begin{center}
         \includegraphics[bb=0 0 533 352, width=1.0\textwidth]{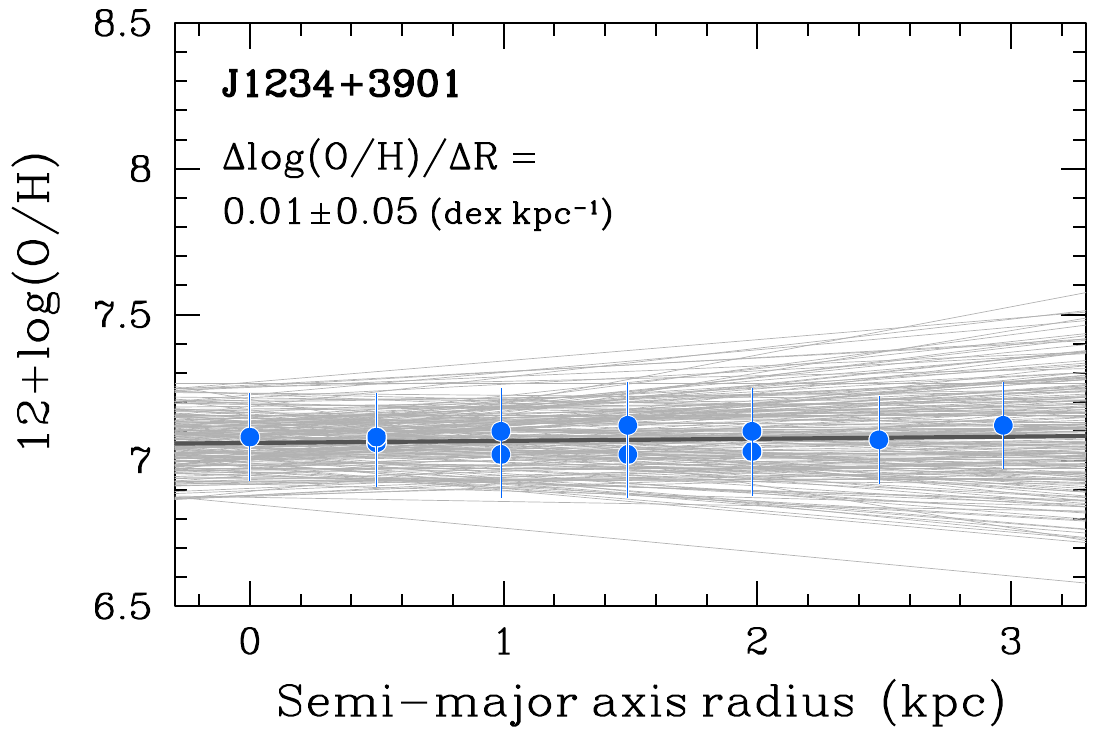}
        \end{center}
      \end{minipage}      %
      \begin{minipage}{0.304\hsize}
        \begin{center}
         \includegraphics[bb=0 0 533 352, width=1.0\textwidth]{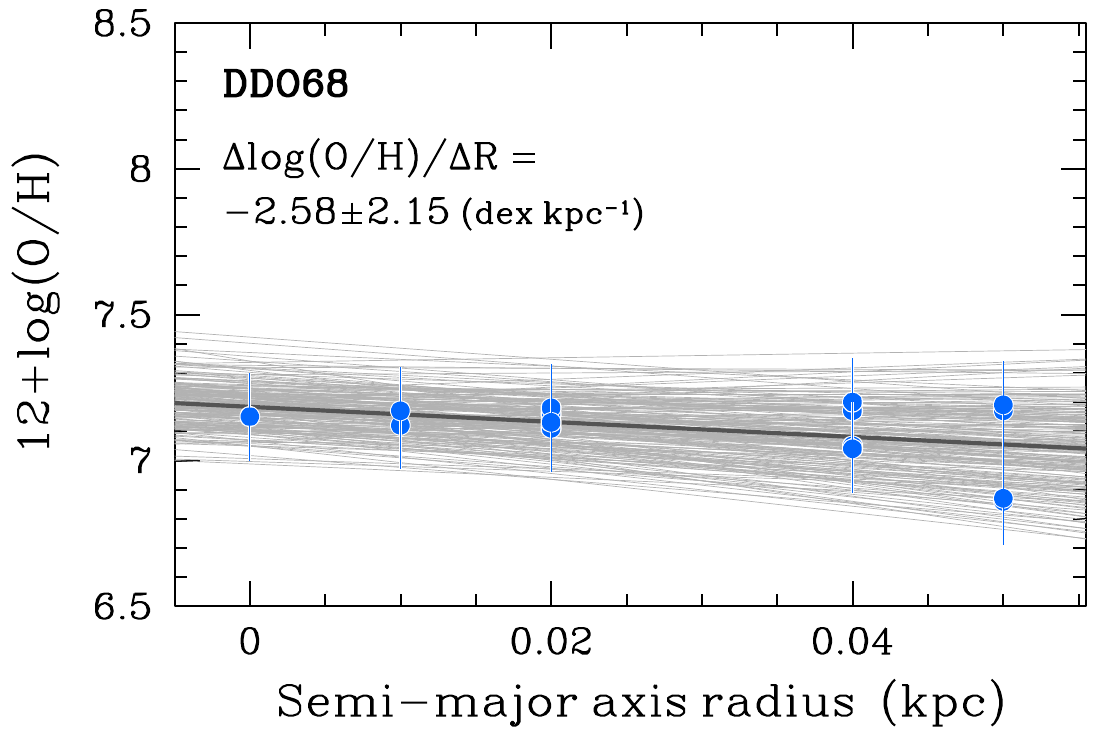}
        \end{center}
      \end{minipage}      \\
      \begin{minipage}{0.304\hsize}
        \begin{center}
         \includegraphics[bb=0 0 533 352, width=1.0\textwidth]{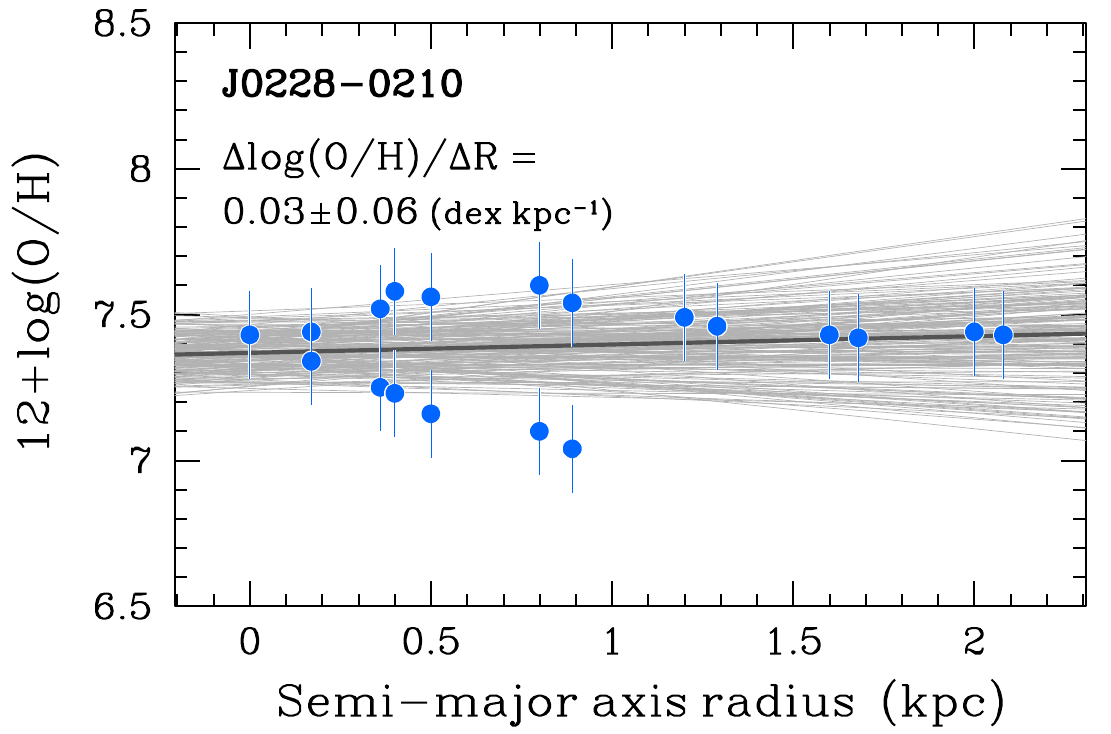}
        \end{center}
      \end{minipage}      %
      \begin{minipage}{0.304\hsize}
        \begin{center}
         \includegraphics[bb=0 0 533 352, width=1.0\textwidth]{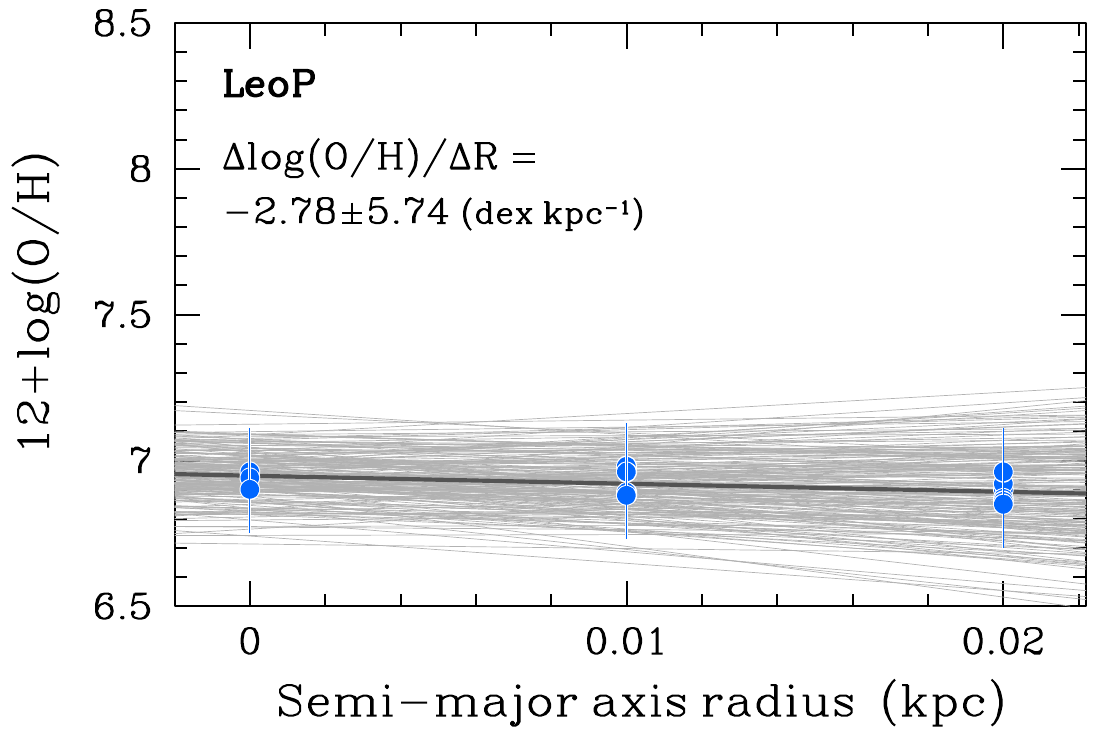}
        \end{center}
      \end{minipage}      %
      \begin{minipage}{0.304\hsize}
        \begin{center}
         \includegraphics[bb=0 0 533 352, width=1.0\textwidth]{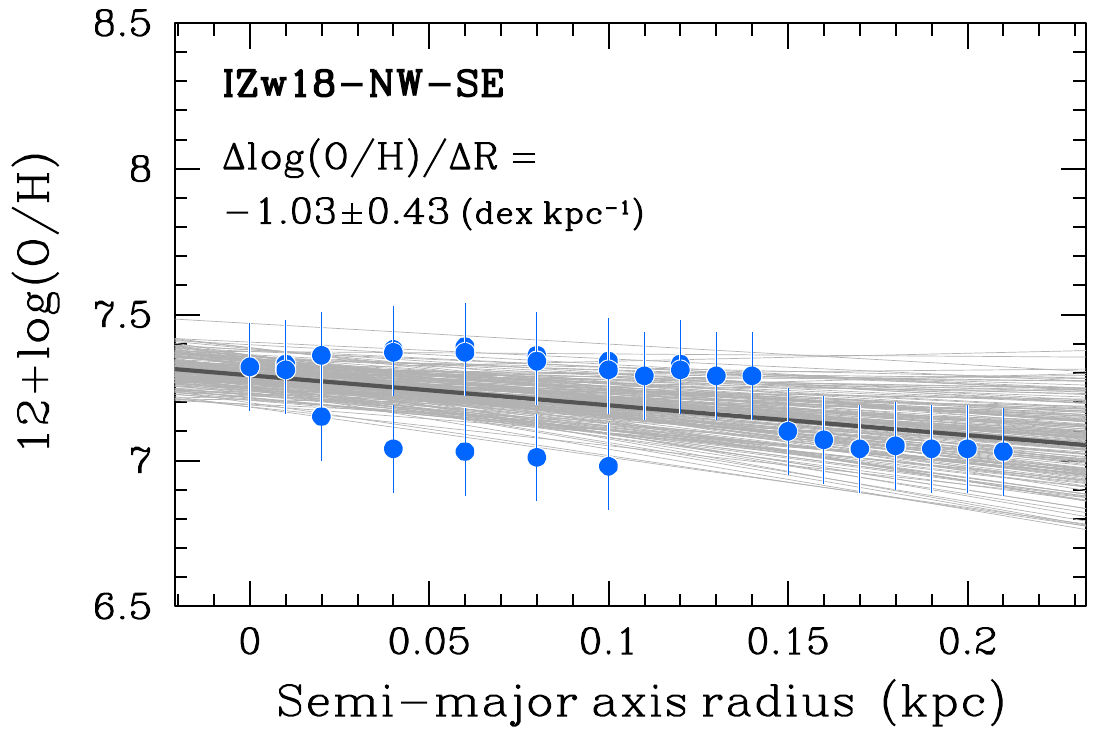}
        \end{center}
      \end{minipage}      \\
      \begin{minipage}{0.304\hsize}
        \begin{center}
         \includegraphics[bb=0 0 533 352, width=1.0\textwidth]{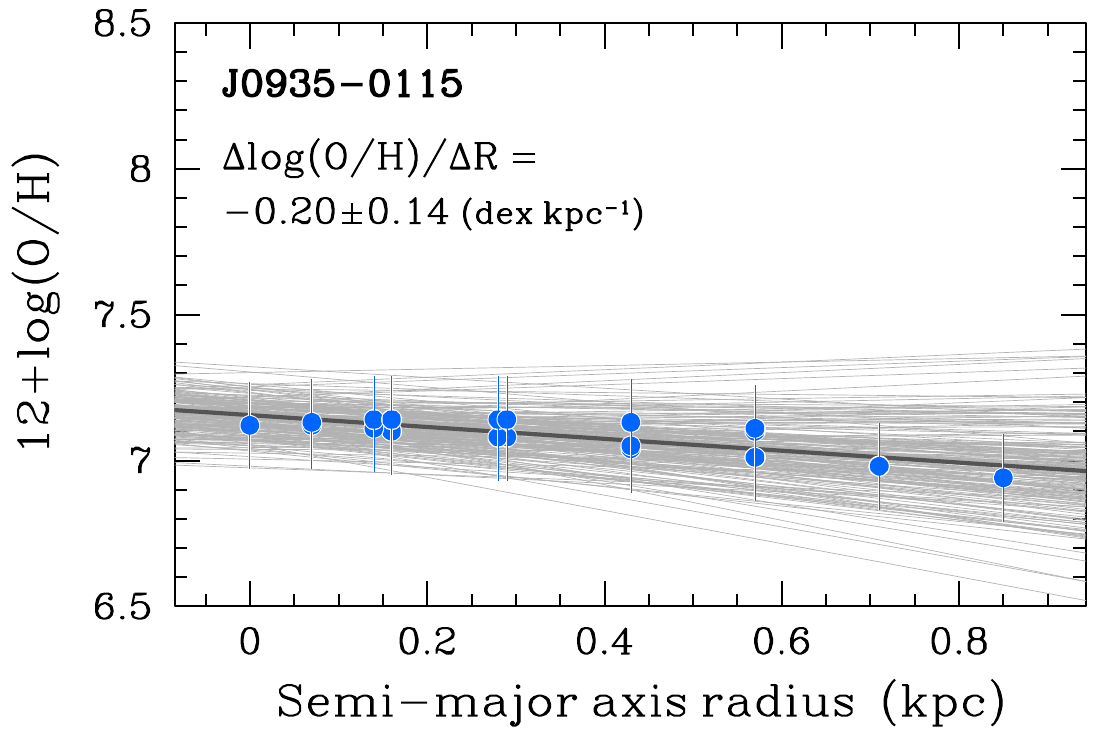}
        \end{center}
      \end{minipage}      %
      \begin{minipage}{0.304\hsize}
        \begin{center}
         \includegraphics[bb=0 0 533 352, width=1.0\textwidth]{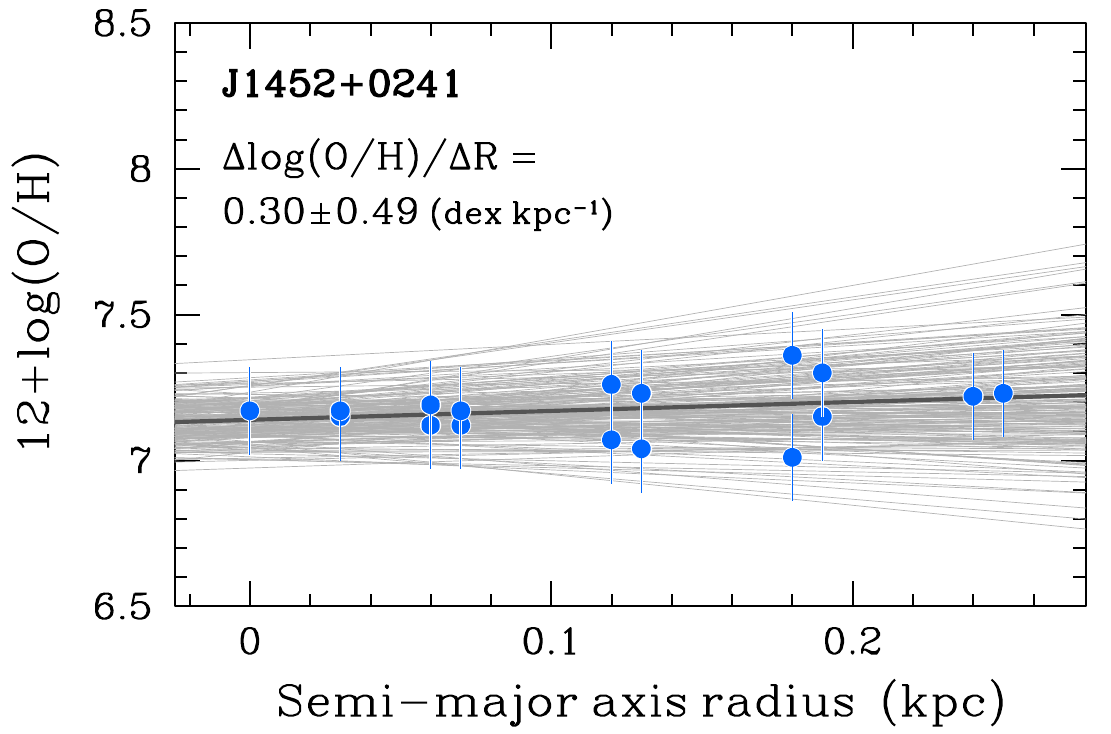}
        \end{center}
      \end{minipage}      %
      \begin{minipage}{0.304\hsize}
        \begin{center}
         \includegraphics[bb=0 0 533 352, width=1.0\textwidth]{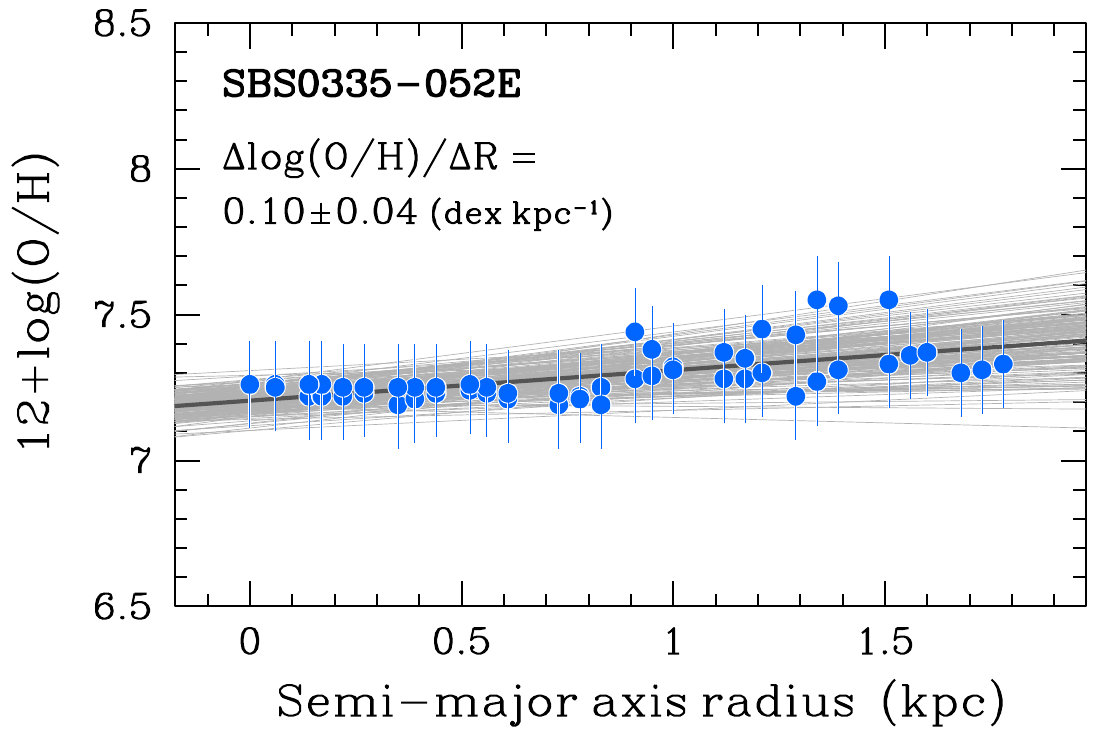}
        \end{center}
      \end{minipage}      \\
      \begin{minipage}{0.304\hsize}
        \begin{center}
         \includegraphics[bb=0 0 533 352, width=1.0\textwidth]{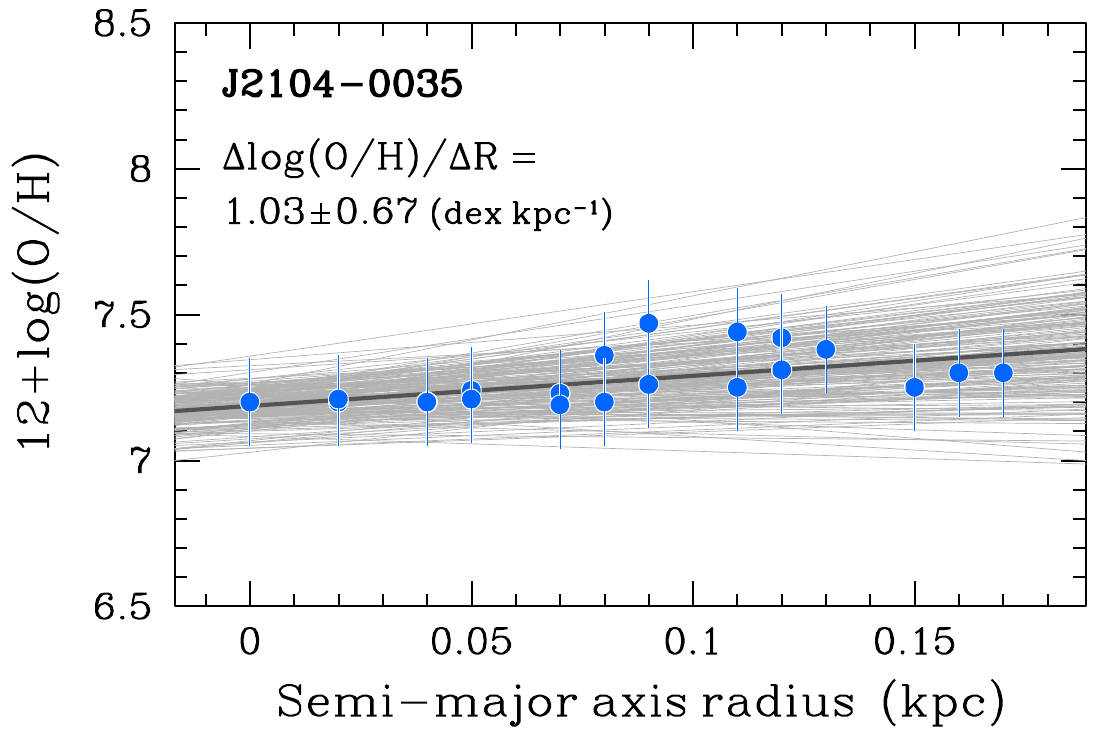}
        \end{center}
      \end{minipage}      %
      \begin{minipage}{0.304\hsize}
        \begin{center}
         \includegraphics[bb=0 0 533 352, width=1.0\textwidth]{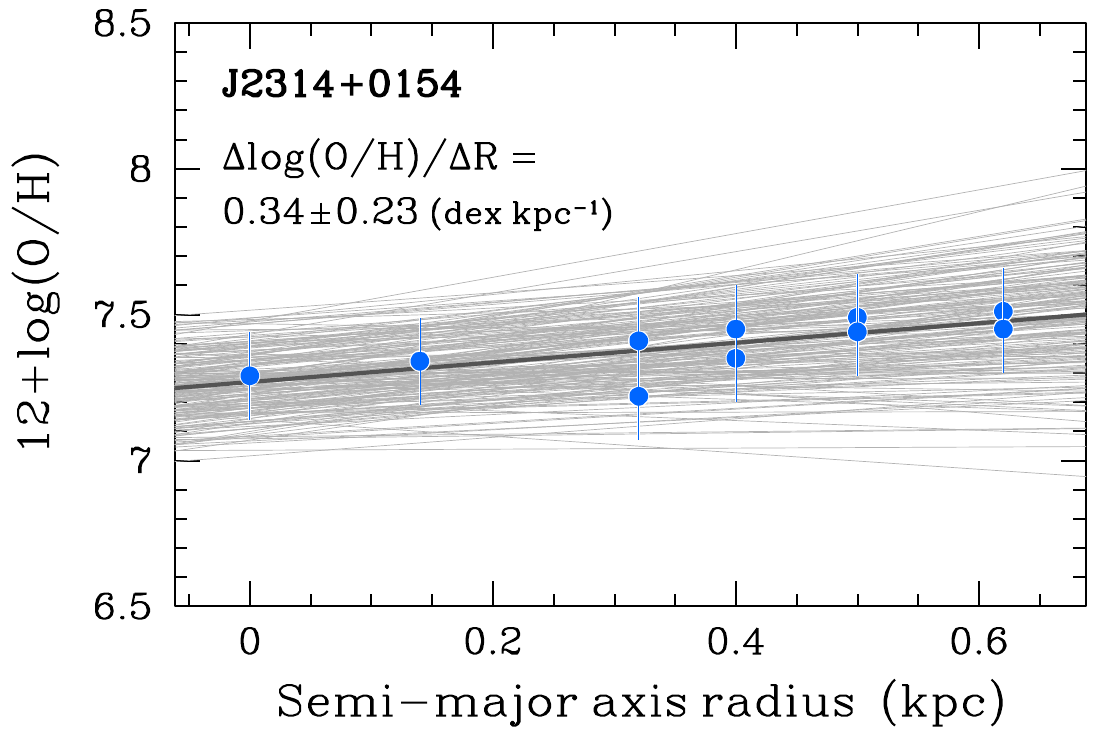}
        \end{center}
      \end{minipage}      %
      \begin{minipage}{0.304\hsize}
        \begin{center}
         \includegraphics[bb=0 0 533 352, width=1.0\textwidth]{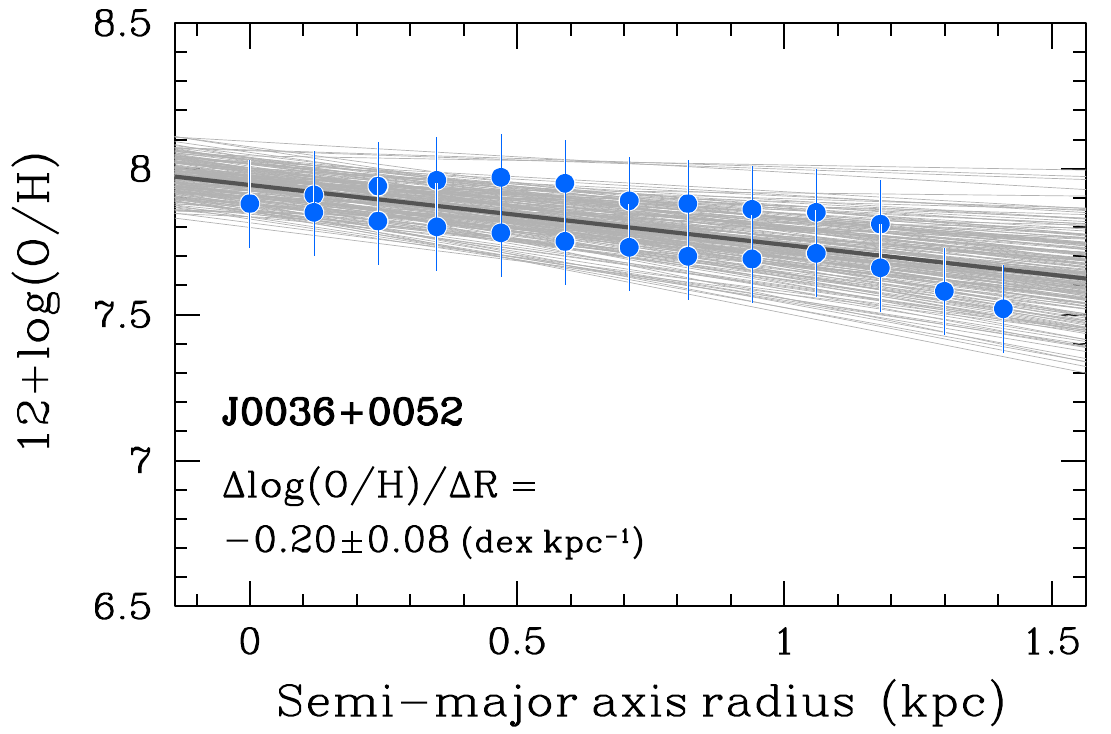}
        \end{center}
      \end{minipage}      \\
    \end{tabular}
    \caption{%
      Metallicity gradients for the 24 EMPGs. Metallicities are plotted as blue points as a function of radius along the semi-major axis, with the peak of $\Sigma_{\rm SFR}$ used as the center. A best-fit linear function (black solid line) is derived by perturbing each blue point within its $1\sigma$ uncertainty, performing a linear regression (shown in light gray lines), and repeating this process 1000 times to determine the best fit. The resulting best-fit metallicity gradient and its uncertainty are displayed in each panel for each EMPG.
    }
    \label{fig:gradZ}
\end{figure*}


\begin{figure*}
  \addtocounter{figure}{-1}
  \centering
    \begin{tabular}{c}      %
      \begin{minipage}{0.304\hsize}
        \begin{center}
         \includegraphics[bb=0 0 533 352, width=1.0\textwidth]{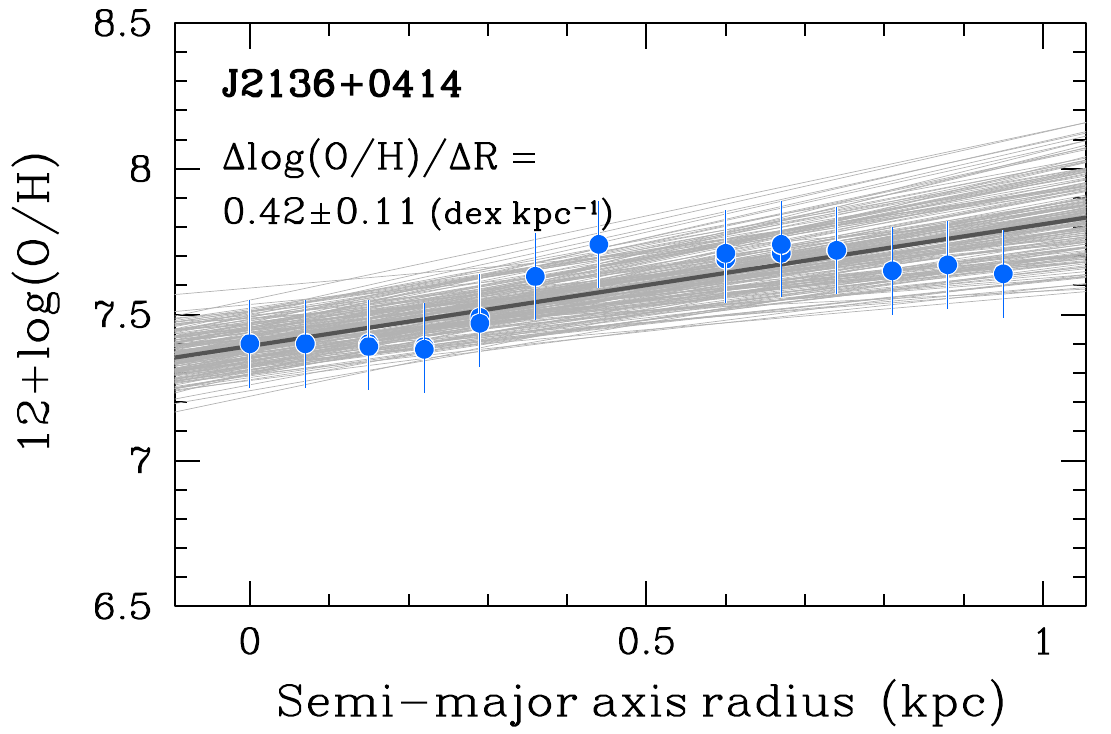}
        \end{center}
      \end{minipage}      %
      \begin{minipage}{0.304\hsize}
        \begin{center}
         \includegraphics[bb=0 0 533 352, width=1.0\textwidth]{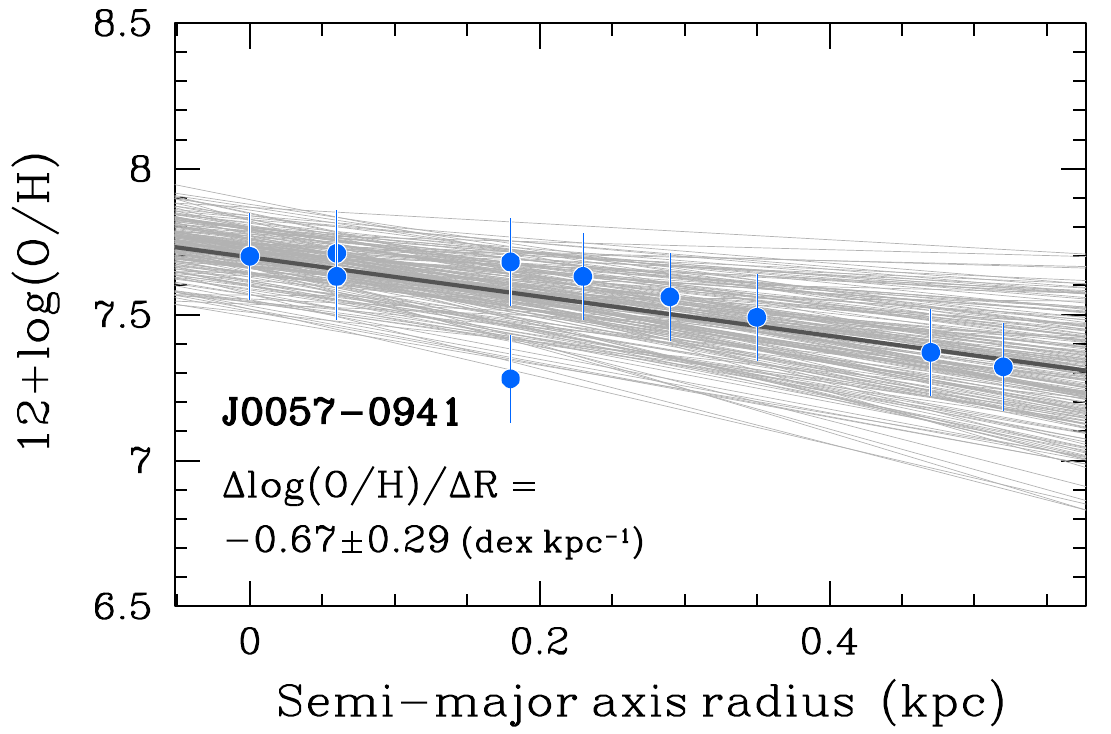}
        \end{center}
      \end{minipage}      %
      \begin{minipage}{0.304\hsize}
        \begin{center}
         \includegraphics[bb=0 0 533 352, width=1.0\textwidth]{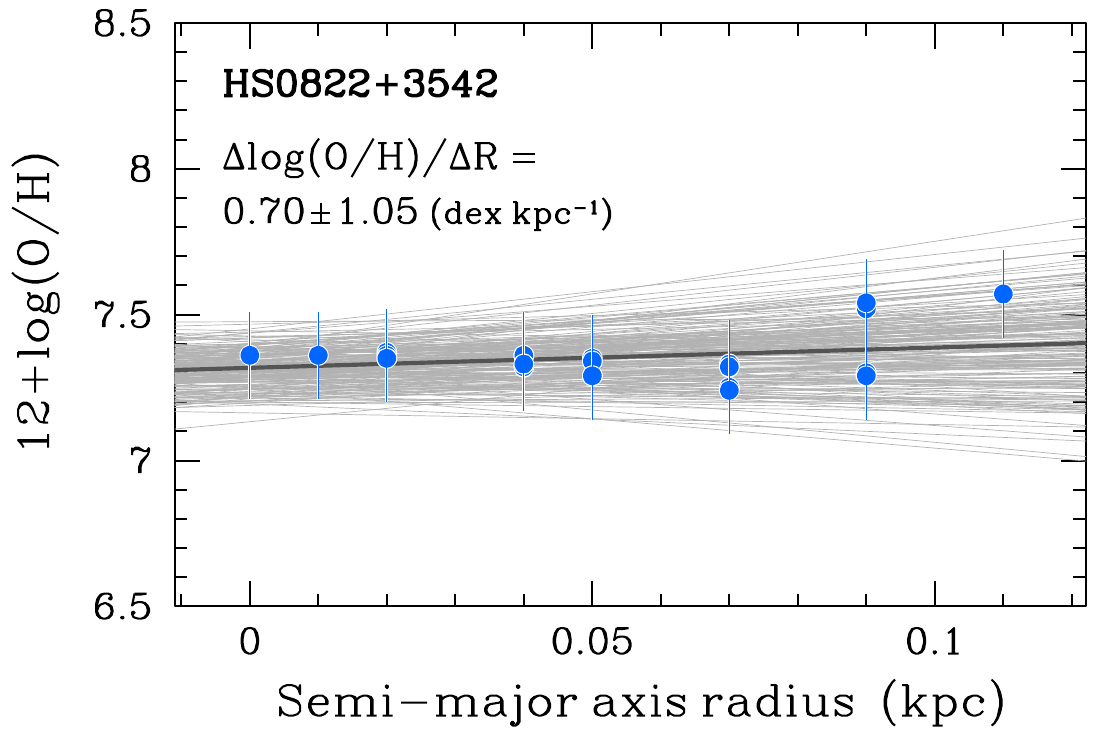}
        \end{center}
      \end{minipage}      \\
      \begin{minipage}{0.304\hsize}
        \begin{center}
         \includegraphics[bb=0 0 533 352, width=1.0\textwidth]{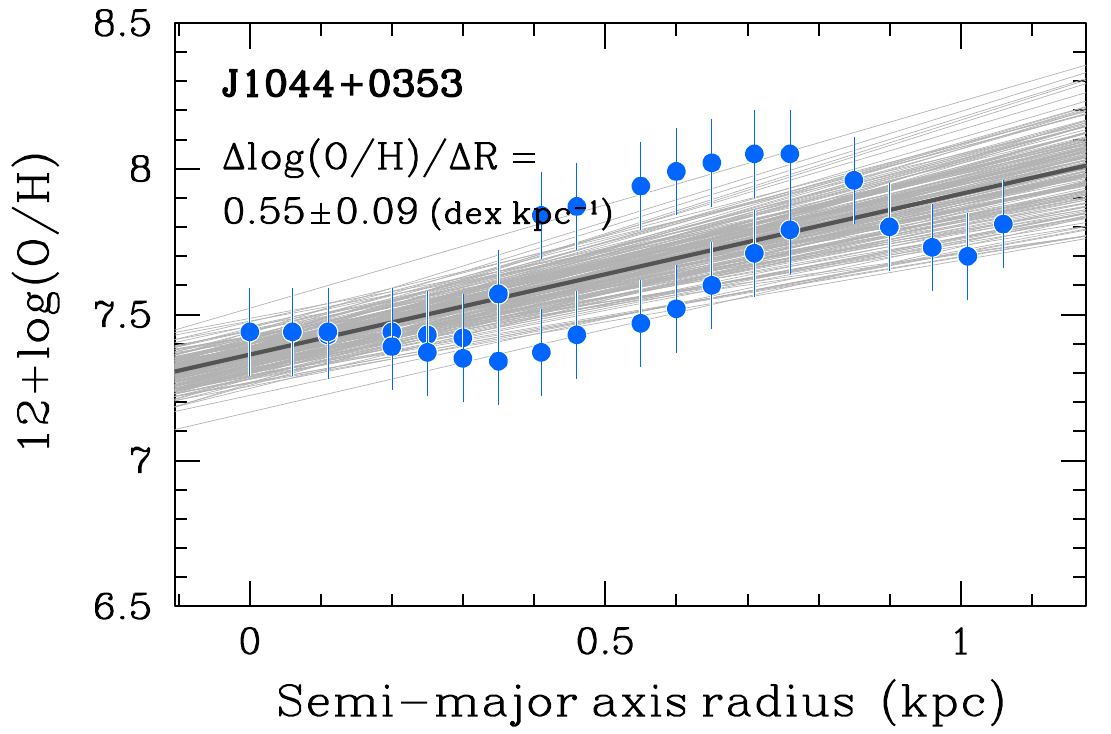}
        \end{center}
      \end{minipage}      %
      \begin{minipage}{0.304\hsize}
        \begin{center}
         \includegraphics[bb=0 0 533 352, width=1.0\textwidth]{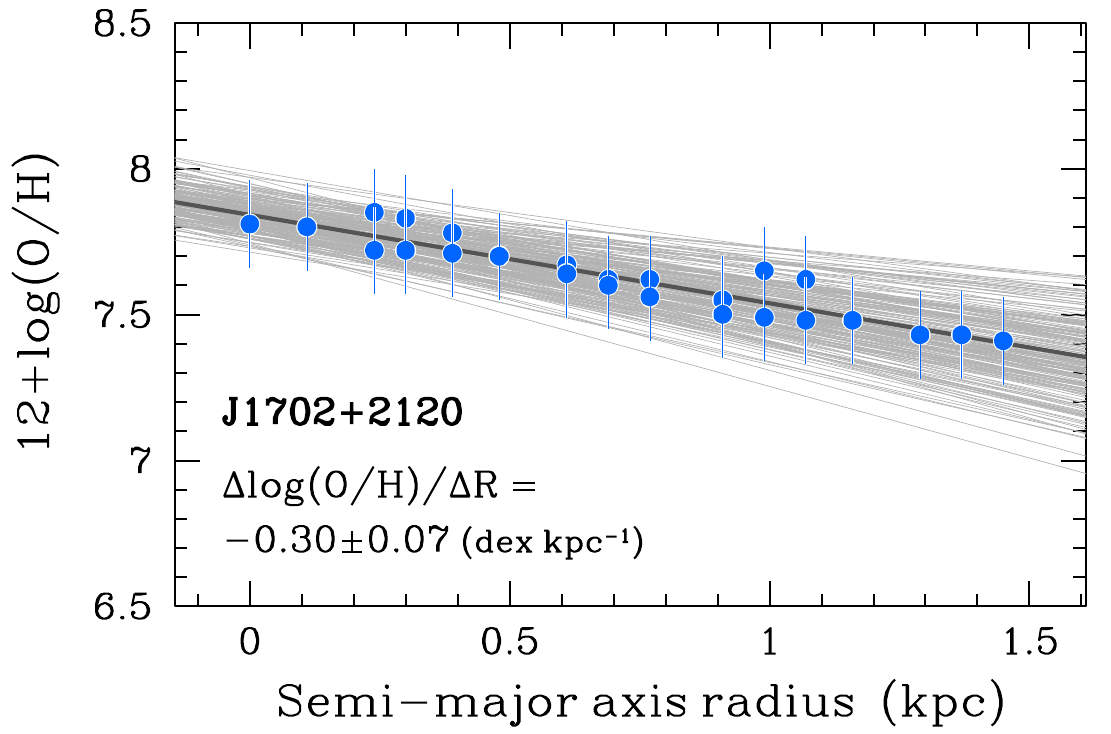}
        \end{center}
      \end{minipage}      %
      \begin{minipage}{0.304\hsize}
        \begin{center}
         \includegraphics[bb=0 0 533 352, width=1.0\textwidth]{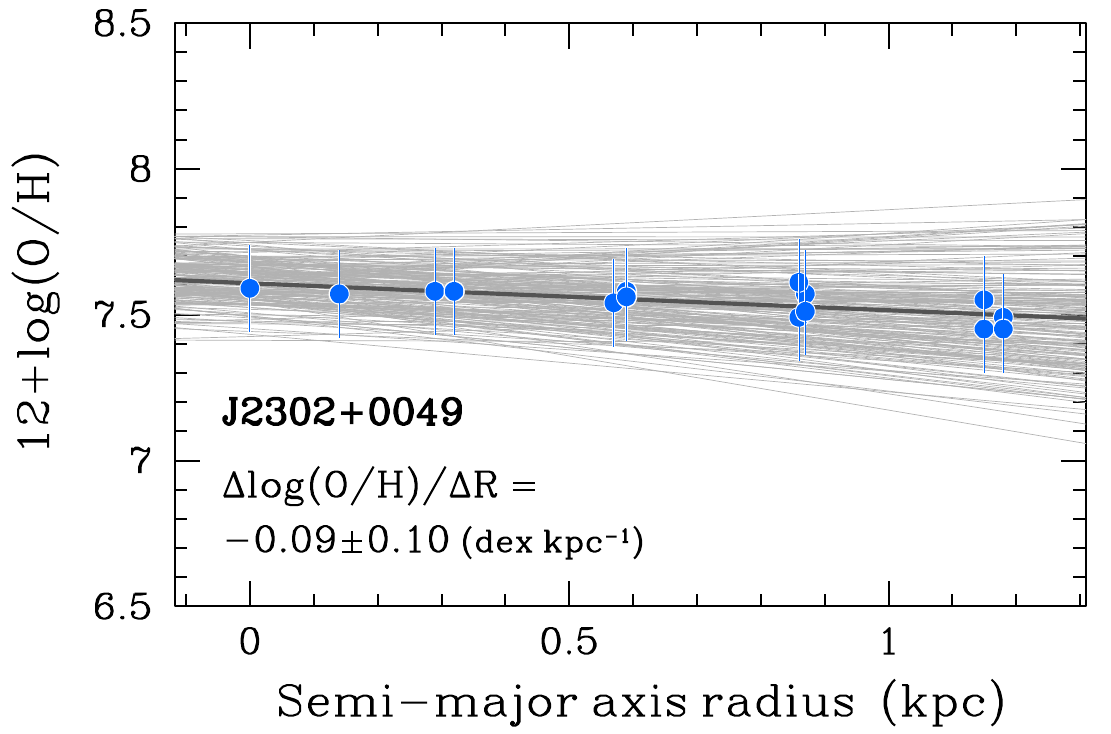}
        \end{center}
      \end{minipage}      \\
      \begin{minipage}{0.304\hsize}
        \begin{center}
         \includegraphics[bb=0 0 533 352, width=1.0\textwidth]{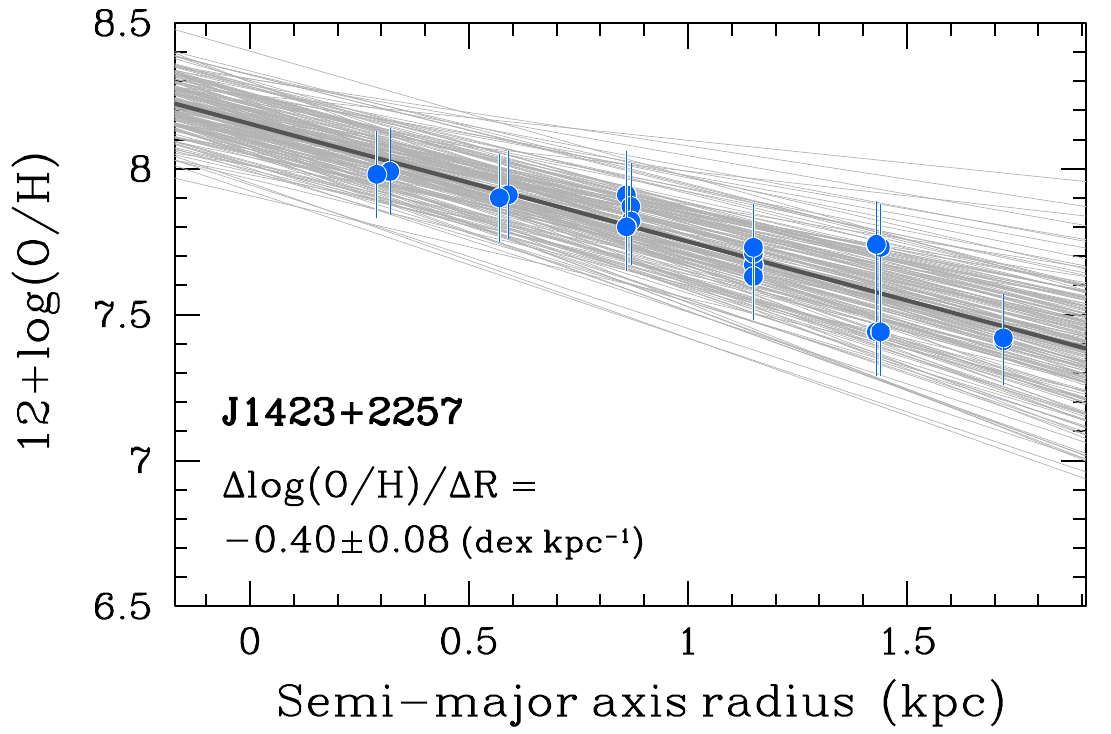}
        \end{center}
      \end{minipage}      %
      \begin{minipage}{0.304\hsize}
        \begin{center}
         \includegraphics[bb=0 0 533 352, width=1.0\textwidth]{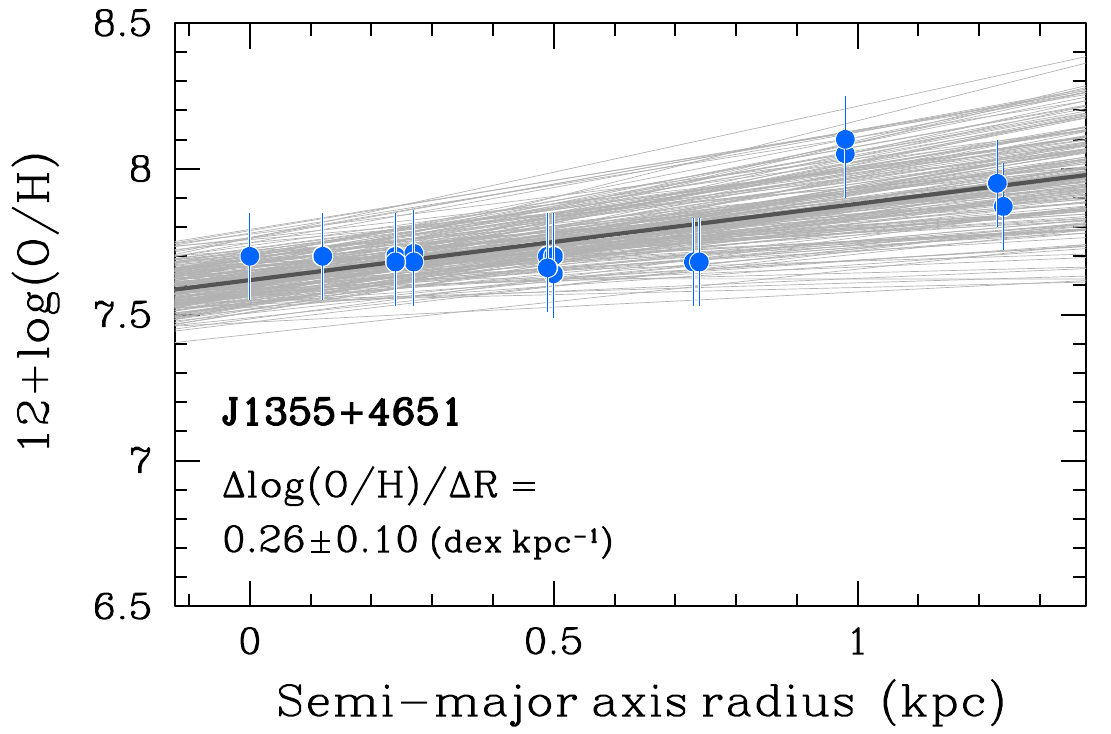}
        \end{center}
      \end{minipage}      %
      \begin{minipage}{0.304\hsize}
        \begin{center}
         \includegraphics[bb=0 0 533 352, width=1.0\textwidth]{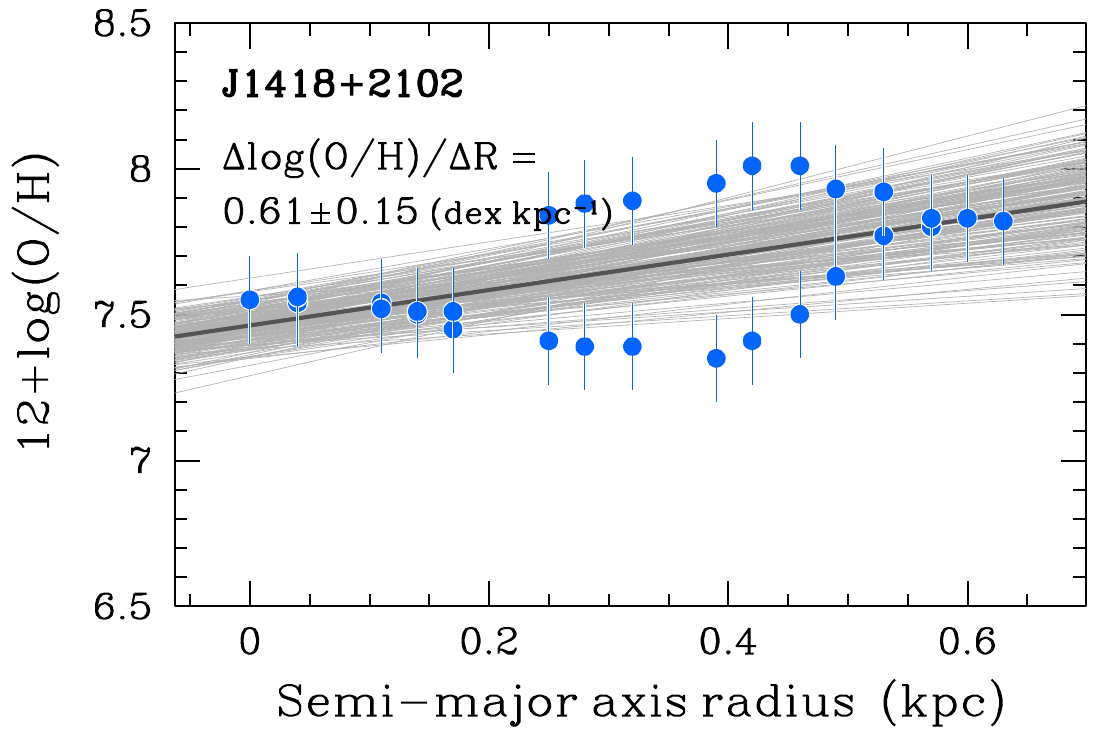}
        \end{center}
      \end{minipage}      \\
      \begin{minipage}{0.304\hsize}
        \begin{center}
         \includegraphics[bb=0 0 533 352, width=1.0\textwidth]{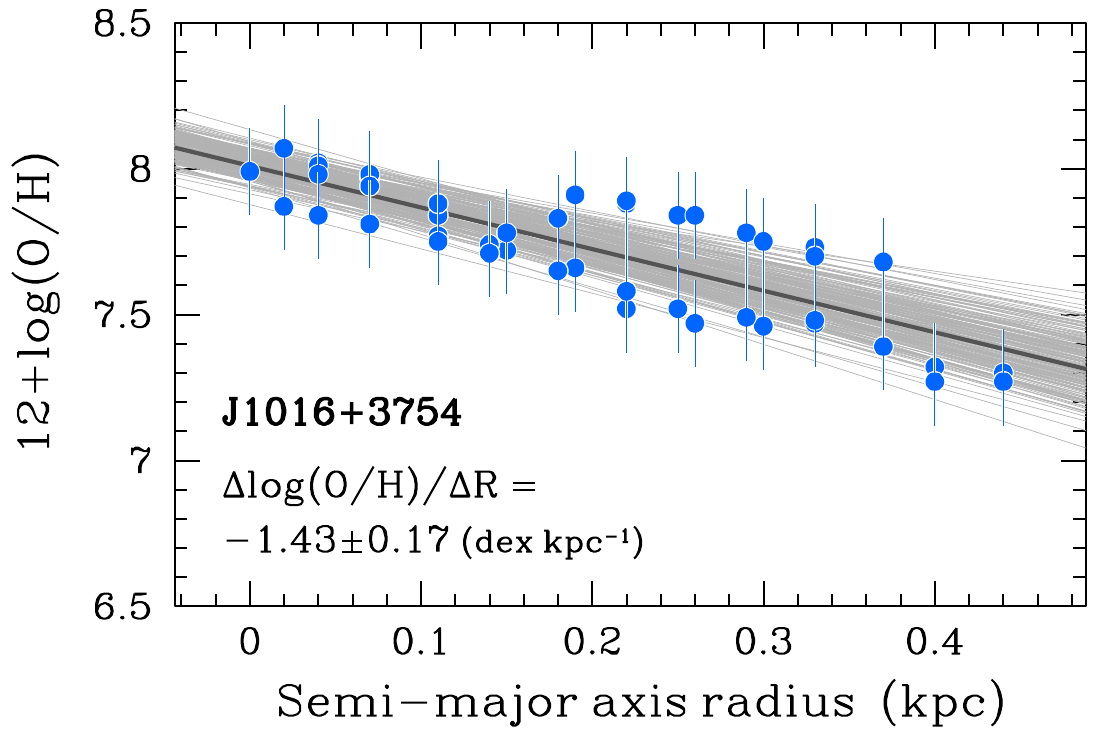}
        \end{center}
      \end{minipage}      %
      \begin{minipage}{0.304\hsize}
        \begin{center}
         \includegraphics[bb=0 0 533 352, width=1.0\textwidth]{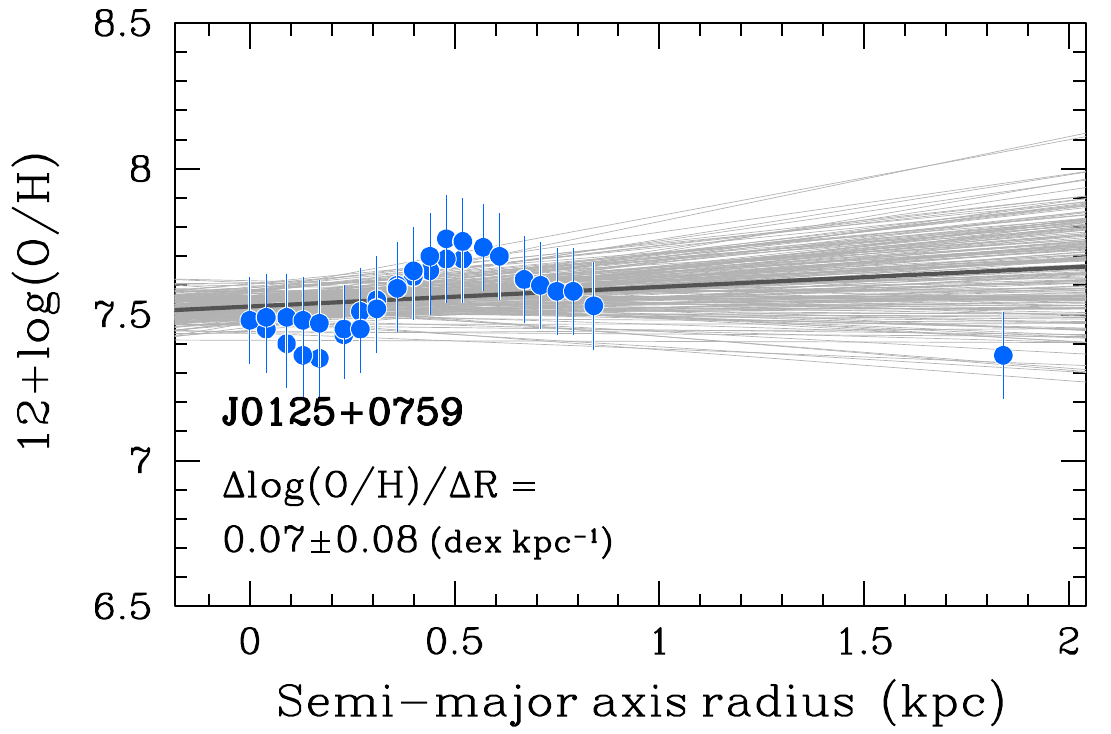}
        \end{center}
      \end{minipage}      %
      \begin{minipage}{0.304\hsize}
        \begin{center}
         \includegraphics[bb=0 0 533 352, width=1.0\textwidth]{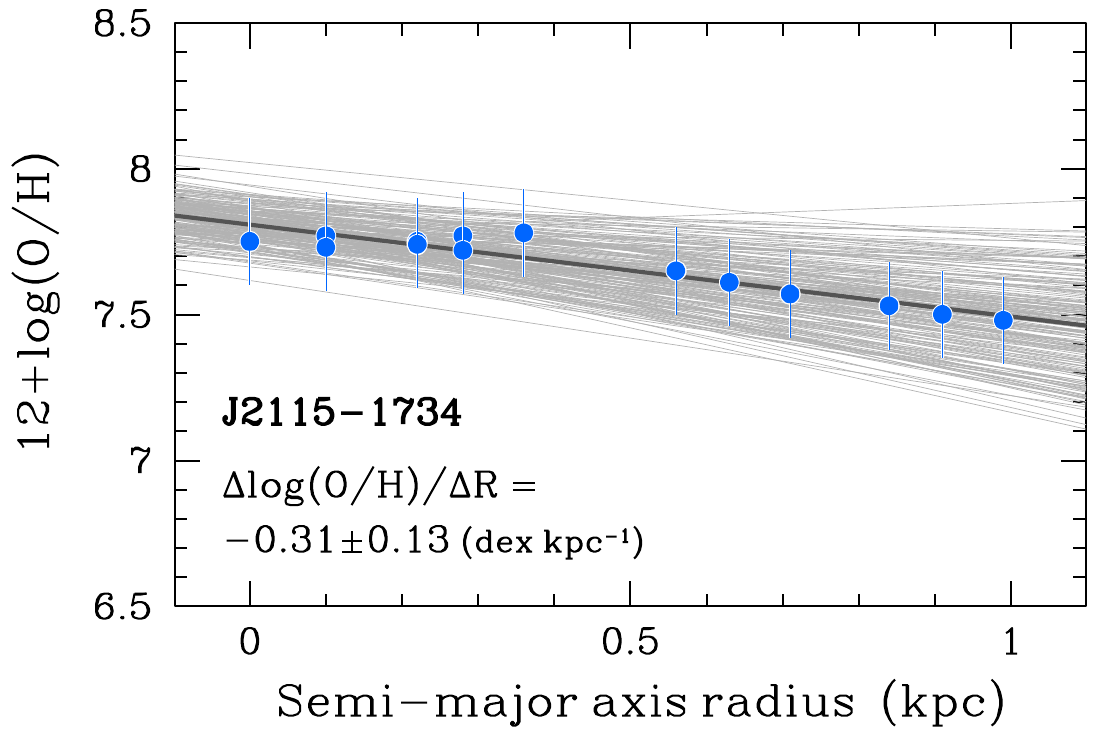}
        \end{center}
      \end{minipage}      \\
    \end{tabular}
    \caption{%
      (Continued.)
    }
\end{figure*}



Following the observed dependence of $\Sigma_{\rm SFR}$ on the rMZR, we modify the functional form of the relation as:
\begin{equation}
    y = a + b (\mu_{\alpha} - c) e^{-(\mu_{\alpha} - c)}, 
\label{eq:rMZR_modified}
\end{equation}
where $\mu_{\alpha} = \log \Sigma_{\star} - \alpha \log \Sigma_{\rm SFR}$, with $\Sigma_{\star}$ and $\Sigma_{\rm SFR}$ given in units of \Msunkpc\ and \Msunyrkpc, respectively. The additional free parameter $\alpha$ gauges the influence of SFR on the mass-metallicity relation, as frequently applied in studies of the global mass-metallicity relation (e.g., \citealt{mannucci2010, AM2013}).
For the fitting, we use the sample of EMPG spaxels spanning $\log \Sigma_{\rm SFR} = -2.5$ to $-0.5$, divided into four subsamples based on $\Sigma_{\rm SFR}$, each with a bin size of $\Delta \log \Sigma_{\rm SFR} = 0.5$. This ensures consistency with the $\Sigma_{\rm SFR}$ ranges used in high-metallicity studies (Belfiore et al., in prep.; \citealt{MM2019}). To remain within the observational limits, we restrict our sample to spaxels with $\log \Sigma_{\rm SFR} \geq -2.5$, corresponding to a minimum $\Sigma_{\star}$ of $10^5$\,\Msunkpc\ for regions with a typical specific SFR of $10^{1.5}$\,Gyr$^{-1}$ seen in our EMPGs.
For each subsample, we bin the spaxels to obtain the average $\Sigma_{\star}$ along with its uncertainty for a given metallicity range, shown as large open circles in Figure \ref{fig:rMZR}(a). To account for systematic uncertainties in metallicity measurements, we add an uncertainty of $\sim 0.15$\,dex. The binned EMPG data points, combined with those from the high-metallicity regime (Belfiore et al., in prep.; \citealt{MM2019}), are used for the least-squares fitting. We do not apply weighting to any of the binned data points during the fitting procedure.

We obtain the best-fit parameters as $a = 8.61 \pm 0.06$, $b = 0.004 \pm 0.013$, $c = 11.3 \pm 2.7$, and $\alpha = 0.66 \pm 0.04$. Figure \ref{fig:rMZR_sfr_bestfit} presents the fitting results across the four binned $\Sigma_{\rm SFR}$ values, demonstrating a good reproduction of the observed data. Notably, the best-fit value of $\alpha$ aligns with that reported in the global mass-metallicity-SFR relation for metal-poor galaxies in the local universe \citep{AM2013}, at $z=2-3$ \citep{sanders2021}, as well as at higher redshifts up to $z\sim 8$ (\citealt{nakajima2023_jwst}; cf. \citealt{curti2024_MZR}). This consistency suggests that the interplay between metallicity, $\Sigma_{\star}$, and $\Sigma_{\rm SFR}$ at local scales governs the global scaling relations observed in low-mass, metal-poor galaxies. Furthermore, it implies that the local mass and SFR surface densities play a fundamental role in shaping the metallicity distribution, even in early-stage galaxies. This finding is consistent with previous studies that explored these relationships in more evolved systems (e.g., \citealt{barrera-ballesteros2016}). 
Moreover, apparent differences between our results and some previous studies on the rMZR (e.g., \citealt{gao2018,yao2022}) can be attributed to variations in star formation activity across samples. For instance, the \citet{gao2018} sample includes regular star-forming MaNGA galaxies in the local universe with a median $\log \Sigma_{\rm SFR}=-2.2$, whereas \citet{yao2022} focuses on regions with clear \OIII$\lambda 4363$ detections, yielding a higher median $\log \Sigma_{\rm SFR}$ of $-1.5$. When comparing our results at similar $\Sigma_{\rm SFR}$ level in Figure \ref{fig:rMZR}(a), we observe a smoother and more consistent agreement with these prior studies on the $\Sigma_{\star}$--metallicity--$\Sigma_{\rm SFR}$ diagram.

\begin{figure*}[!t]
	\centering   
     	\includegraphics[bb=0 0 329 184, width=0.75\textwidth]{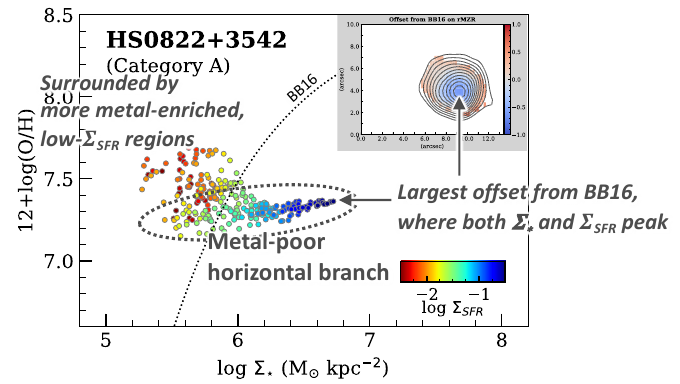}
     	\caption{%
		An example of an EMPG in Category A, HS0822$+$3542, is presented on the resolved mass-metallicity relation (main panel). Symbols and color coding follow Figure \ref{fig:rMZR}(a), but only spaxels from this single EMPG are shown. (Subpanel:) A map of the spatial distribution of the offset from the \citet{barrera-ballesteros2016} relation is provided, where bluer colors indicate regions farther below the relation, highlighting the metallicity variation across the galaxy. In this category, a distinct ``metal-poor horizontal branch'' (indicated with a dotted oval) is evident on the rMZR, extending toward the peak of $\Sigma_{\star}$ and $\Sigma_{\rm SFR}$, while being surrounded by more metal-enriched regions.
		}
    	\label{fig:rMZR_individual_a}
\end{figure*}

\begin{figure*}[!t]
  \centering
    \begin{tabular}{c}      %
      \begin{minipage}{0.48\hsize}
        \begin{center}
	 \begin{overpic}[bb=0 0 325 239, width=1.0\textwidth]{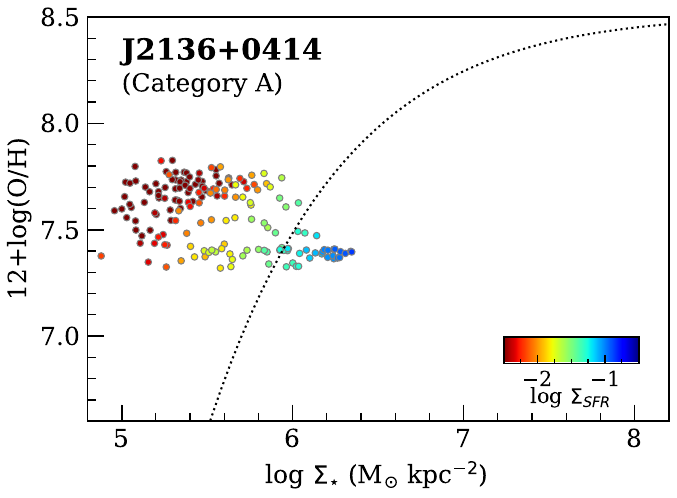}
           \put(61,44){\includegraphics[bb=0 0 399 285, scale=0.23]{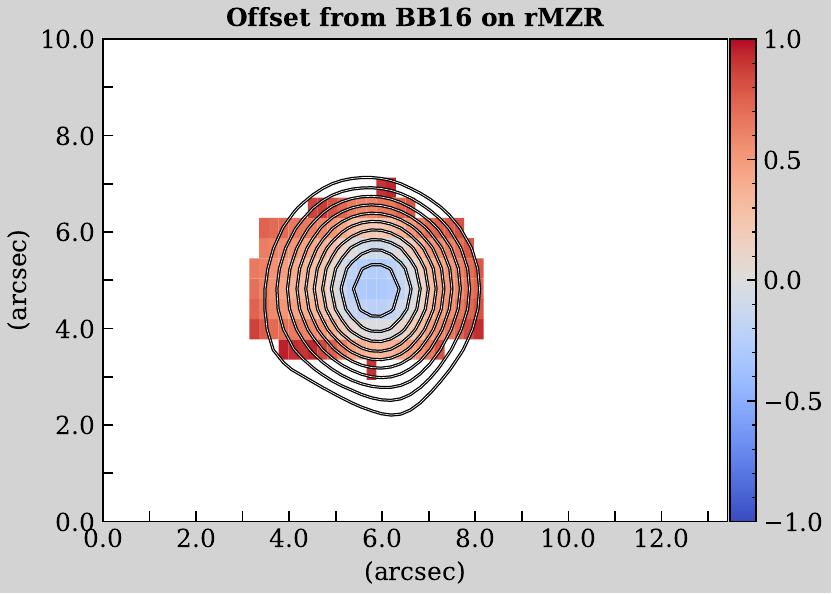}}  
         \end{overpic}
        \end{center}
      \end{minipage}      %
      \begin{minipage}{0.48\hsize}
        \begin{center}
	 \begin{overpic}[bb=0 0 325 239, width=1.0\textwidth]{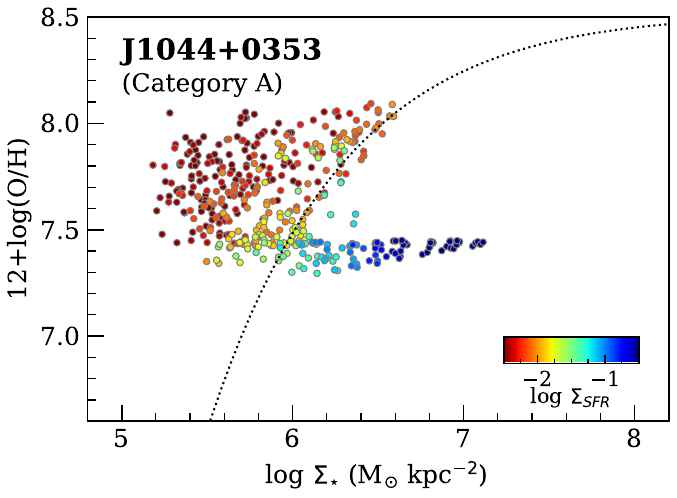}
           \put(61,44){\includegraphics[bb=0 0 399 285, scale=0.23]{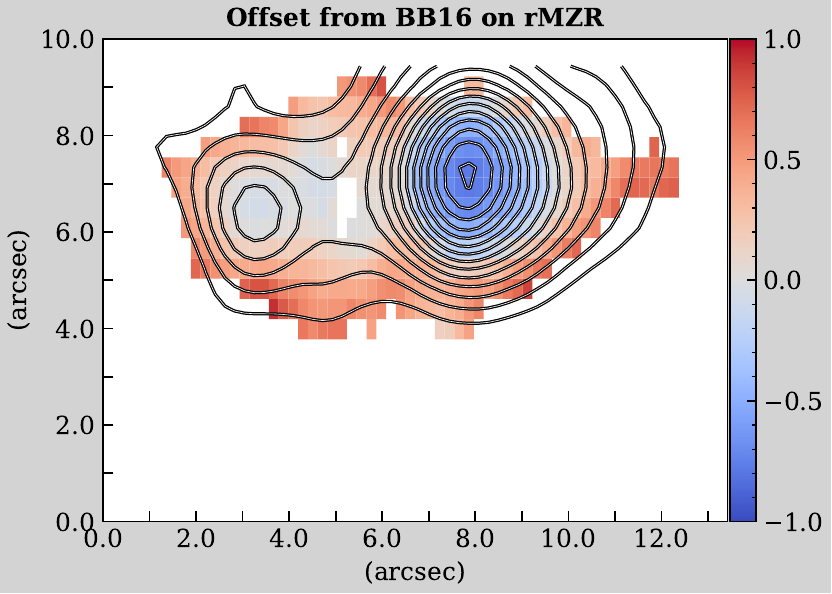}}  
         \end{overpic}
        \end{center}
      \end{minipage}      \\
      \begin{minipage}{0.48\hsize}
        \begin{center}
	 \begin{overpic}[bb=0 0 325 239, width=1.0\textwidth]{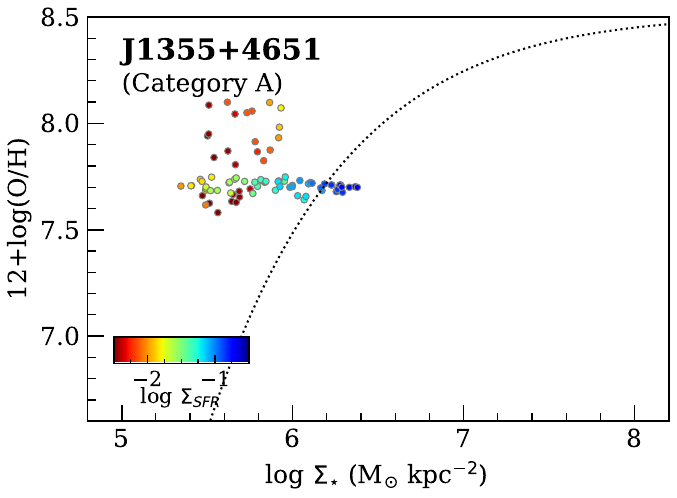}
	   \put(61,12){\includegraphics[bb=0 0 399 285, scale=0.23]{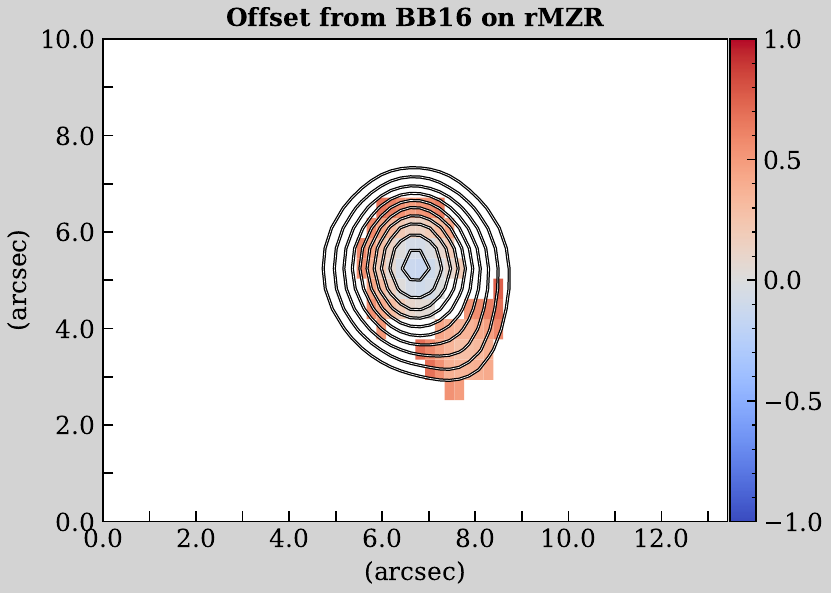}}
         \end{overpic}
        \end{center}
      \end{minipage}
      \begin{minipage}{0.48\hsize}
        \begin{center}
	 \begin{overpic}[bb=0 0 325 239, width=1.0\textwidth]{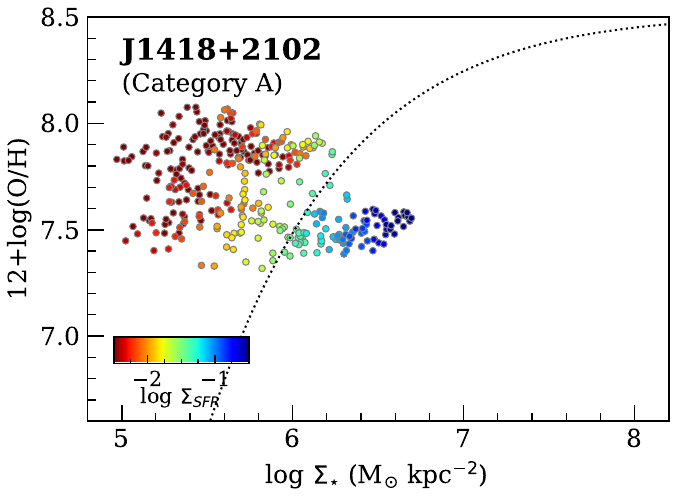}
	   \put(61,12){\includegraphics[bb=0 0 399 285, scale=0.23]{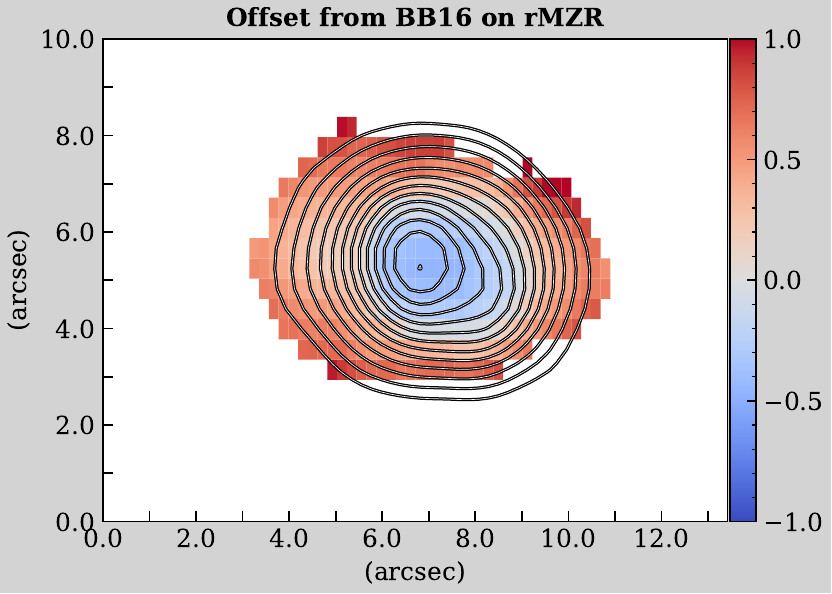}}
         \end{overpic}
        \end{center}
      \end{minipage}
      \\
      \begin{minipage}{0.48\hsize}
        \begin{center}
	 \begin{overpic}[bb=0 0 325 239, width=1.0\textwidth]{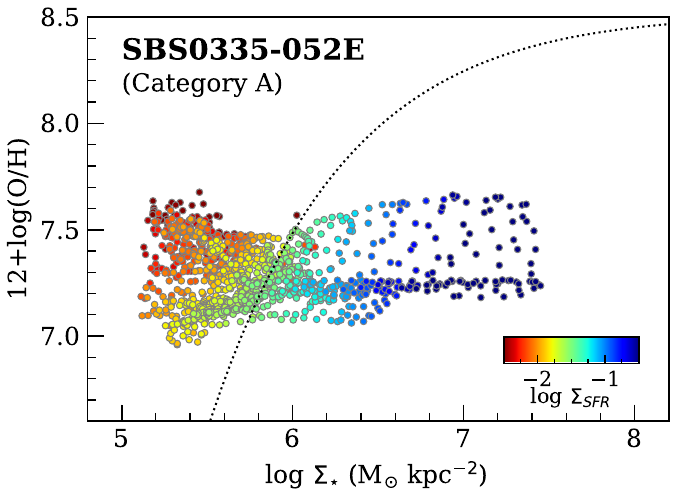}
           \put(63.5,45.8){\includegraphics[bb=0 0 399 285, scale=0.215]{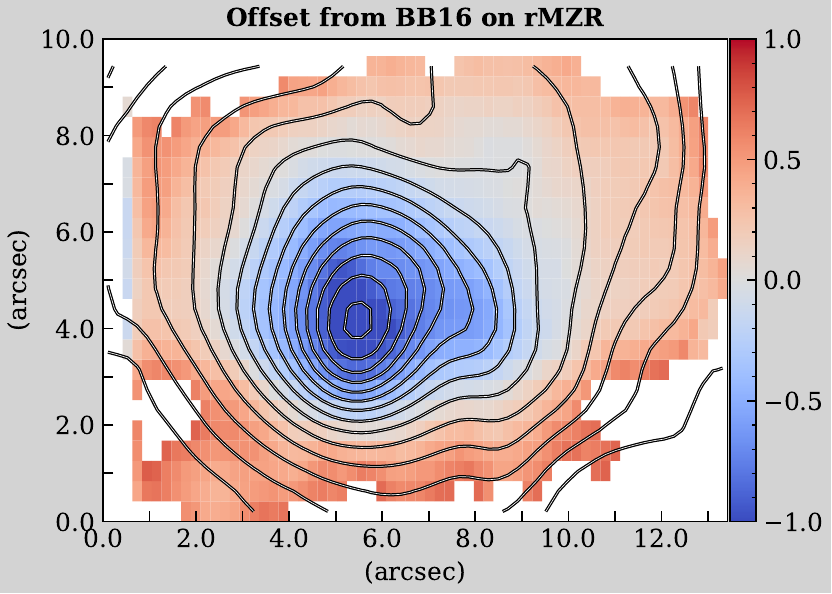}}  
         \end{overpic}
        \end{center}
      \end{minipage}      %
      \begin{minipage}{0.48\hsize}
        \begin{center}
	 \begin{overpic}[bb=0 0 325 239, width=1.0\textwidth]{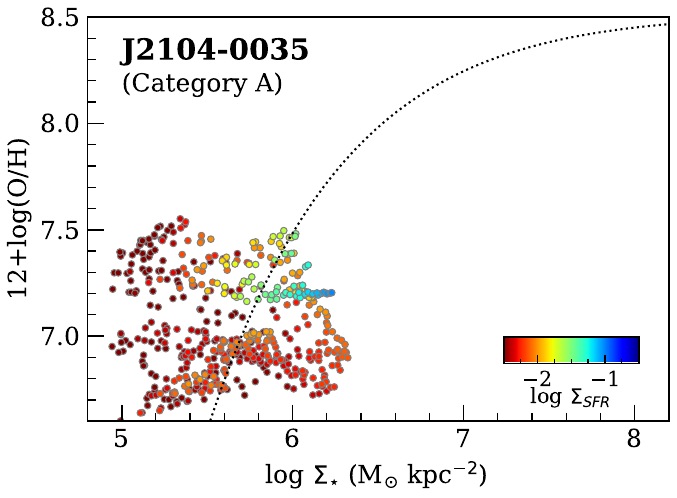}
           \put(61,44){\includegraphics[bb=0 0 399 285, scale=0.23]{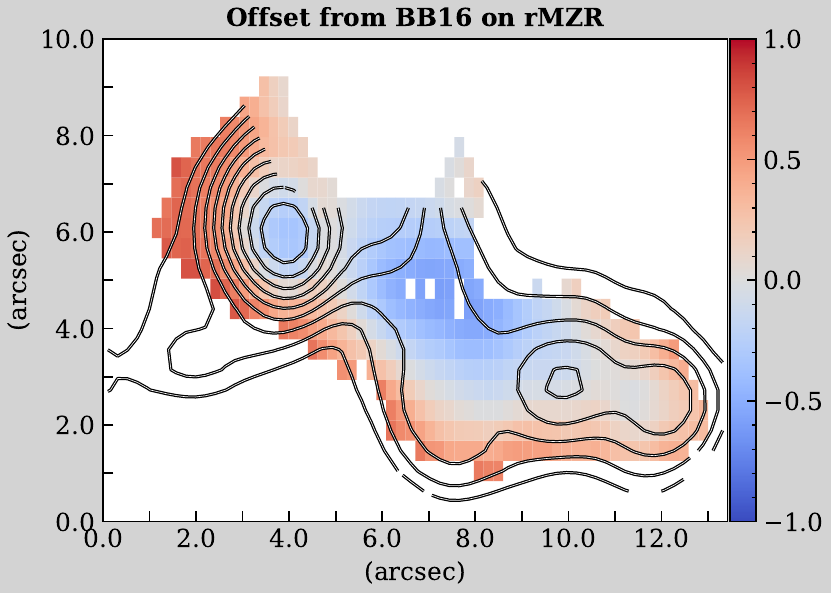}}  
         \end{overpic}
        \end{center}
      \end{minipage}      \\
    \end{tabular}
    \caption{%
      Same as Figure \ref{fig:rMZR_individual_a}, but for the other individual EMPGs in Catefory A.
    }
    \label{fig:rMZR_individual_a_entire}
\end{figure*}


\begin{figure*}[!t]
  \addtocounter{figure}{-1}
  \centering
    \begin{tabular}{c}      %
      \begin{minipage}{0.48\hsize}
        \begin{center}
	 \begin{overpic}[bb=0 0 325 239, width=1.0\textwidth]{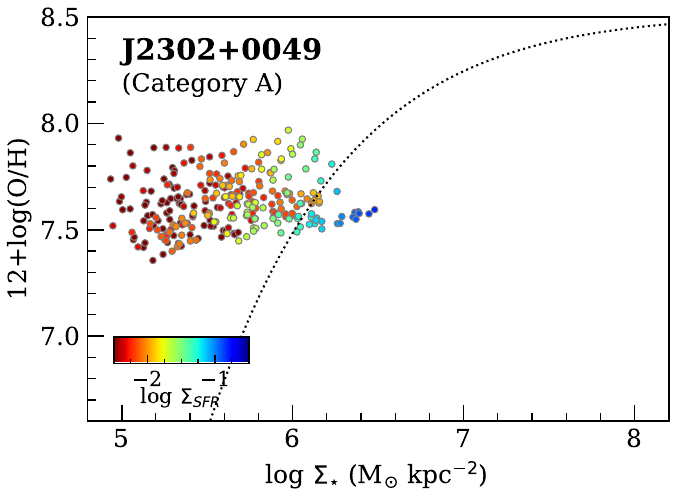}
	   \put(61,12){\includegraphics[bb=0 0 399 285, scale=0.23]{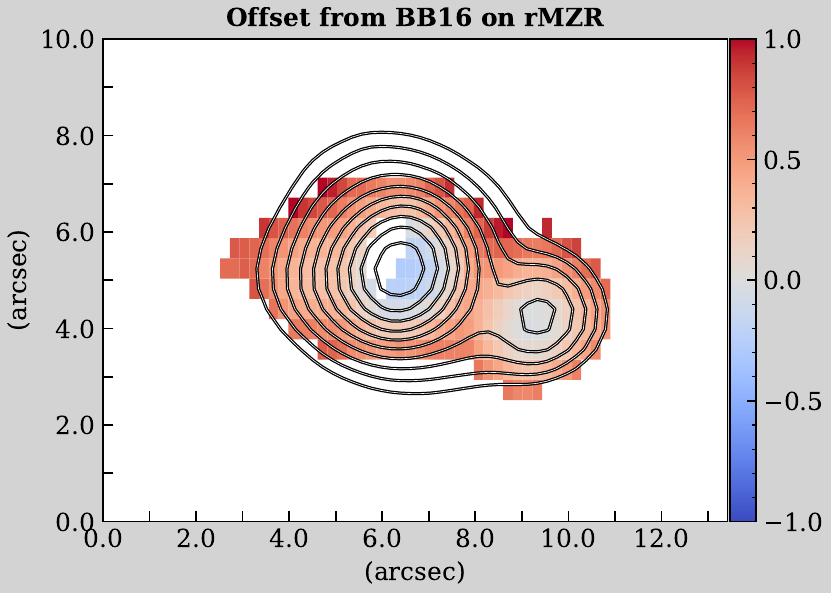}}
         \end{overpic}
        \end{center}
      \end{minipage}
      \begin{minipage}{0.48\hsize}
        \begin{center}
	 \begin{overpic}[bb=0 0 328 234, width=1.0\textwidth]{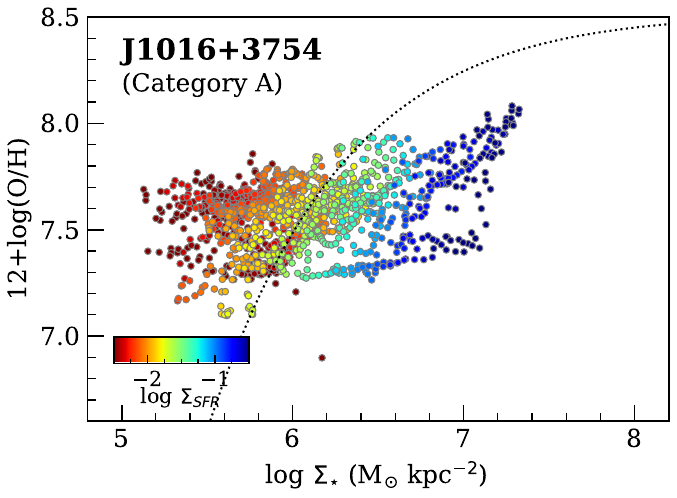}
           \put(64.8,11.6){\includegraphics[bb=0 0 399 284, scale=0.2]{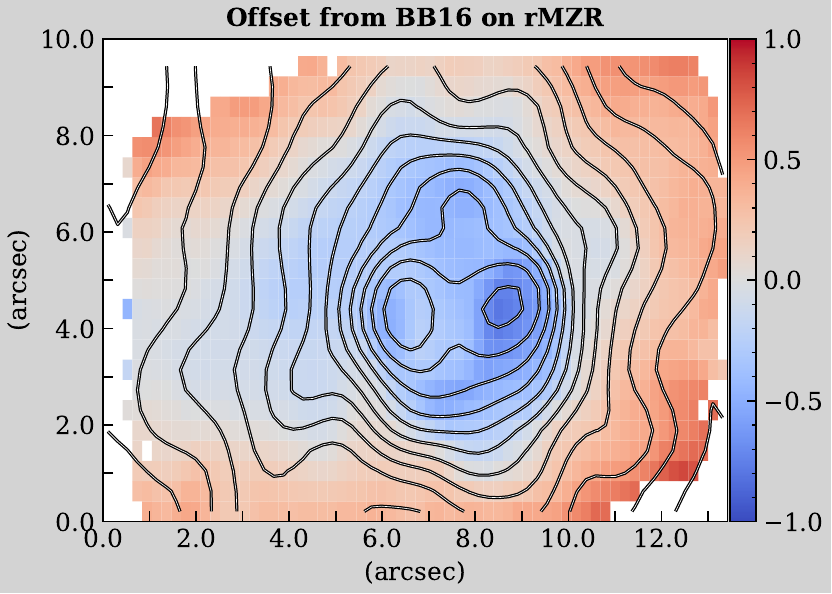}}  
         \end{overpic}
        \end{center}
      \end{minipage}
      \\
      \begin{minipage}{0.48\hsize}
        \begin{center}
	 \begin{overpic}[bb=0 0 328 234, width=1.0\textwidth]{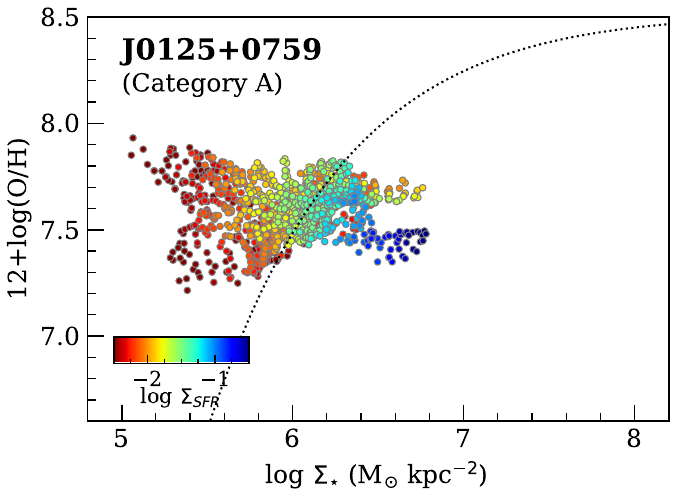}
           \put(64.8,11.6){\includegraphics[bb=0 0 399 284, scale=0.2]{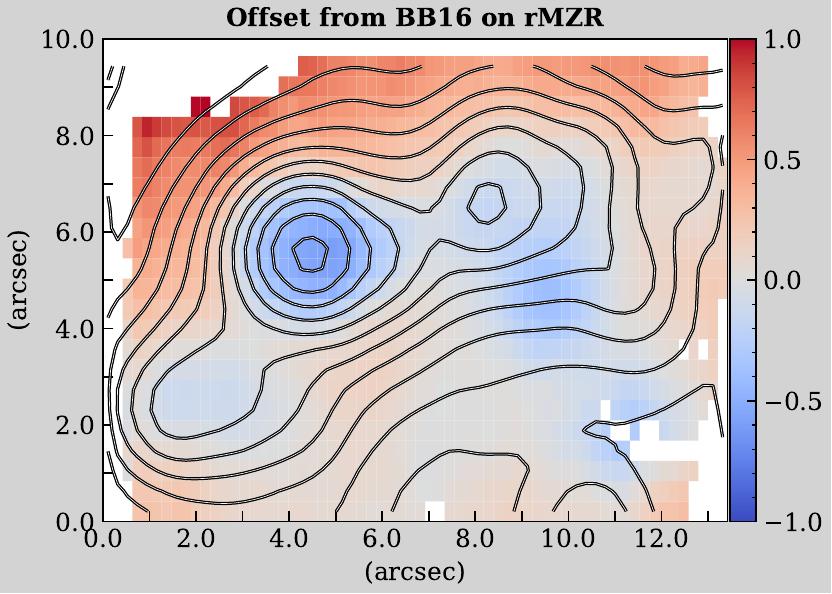}}  
         \end{overpic}
        \end{center}
      \end{minipage}
      \begin{minipage}{0.48\hsize}
        \begin{center}
	 \begin{overpic}[bb=0 0 325 239, width=1.0\textwidth]{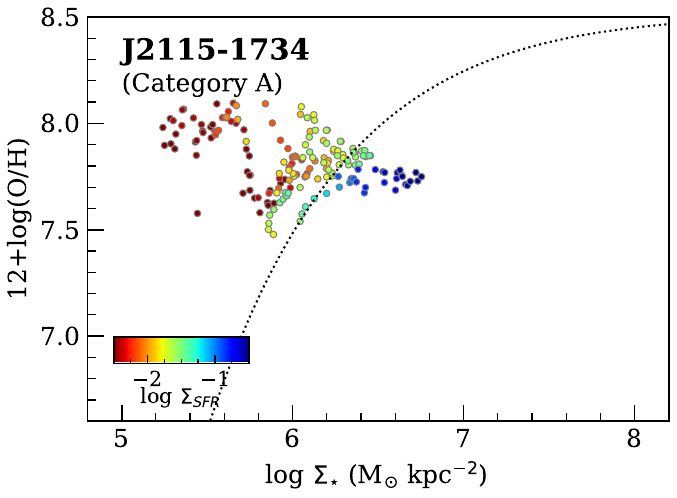}
	   \put(61,12){\includegraphics[bb=0 0 399 285, scale=0.23]{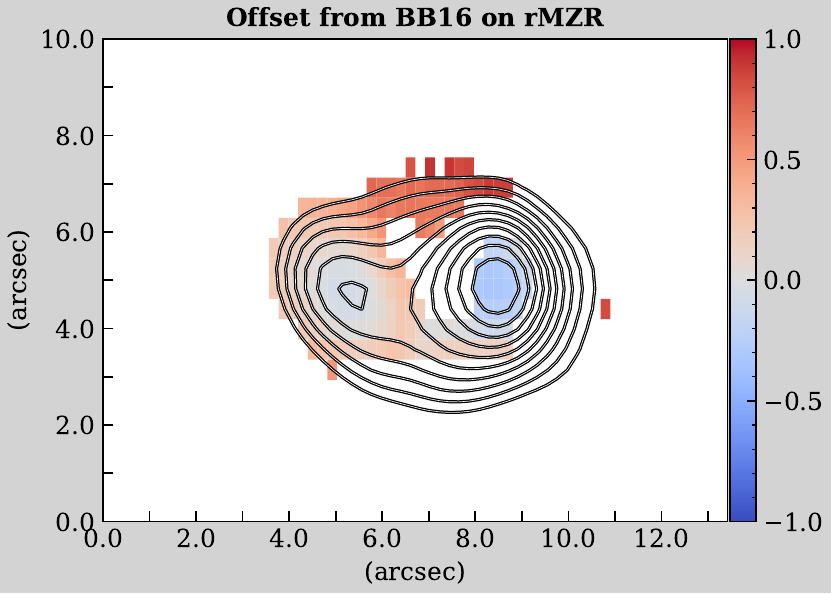}}
         \end{overpic}
        \end{center}
      \end{minipage}
      \\
    \end{tabular}
    \caption{%
      (Continued.)
    }
\end{figure*}



\subsection{Metallicity gradients} \label{ssec:results_gradZ}

The spatial distribution of metallicity across individual galaxies provides valuable insights into the processes driving chemical enrichment. A radial metallicity gradient is a commonly used measure to quantify the metallicity distribution within galaxies. For each target, we extract pixels along the semi-major axis (allowing a vertical width of $\Delta = \pm 1$\,pixel) and calculate the metallicity gradient, centered on the $\Sigma_{\rm SFR}$ peak position. In cases where multiple star-forming clumps are detected, we generate a segmentation map to define their clumps and the border(s) using \verb+SExtractor+ \citep{BA1996} and only use the pixels associated with the main clump.

Figure \ref{fig:gradZ} summarizes the observed metallicity gradients for the 24 EMPGs as a function of distance from the center along the semi-major axis (blue points) and their best-fit linear functions overlaid (gray lines). The fitting assumes a symmetric metallicity distribution for simplicity. To accurately capture the complex metallicity distributions observed in some EMPGs, two data points are shown at certain radii without averaging, representing measurements in both directions along the semi-major axis.

The EMPGs indeed display a wide range of metallicity slopes. Ten of the 24 galaxies exhibit the expected negative gradient, while 7 show an unusual positive gradient, and the remaining 7 present a flat metallicity distribution. Some EMPGs also exhibit clear asymmetric features in their metallicity distributions. 
Interestingly, EMPGs with positive gradients, such as J1418$+$2102, J1044$+$0353, and J2136$+$0414 show different trends in their inner and outer regions. In the inner regions, the gradient becomes flatter or even negative, whereas the outer regions generally display higher metallicities compared to the inner regions, resulting in an overall positive gradient. This finding indicates that a more detailed analysis of the spatial metallicity distribution is needed for EMPGs, as metallicity gradients alone may fail to fully describe the complexity of chemical enrichment in these galaxies. In particular, current methods do not account for metal distributions outside the semi-major axis or those associated with secondary clump(s). Furthermore, as confirmed in Sect.~\ref{ssec:results_rMZR}, the strong dependence of $\Sigma_{\rm SFR}$ on the rMZR highlights the need to investigate how metallicity varies as a function of both $\Sigma_{\star}$ and $\Sigma_{\rm SFR}$ within each EMPG. These aspects will be explored in more detail in the next section.

\subsection{Zooming in individual EMPG on rMZR} \label{ssec:results_rMZR_individual}

In Sect.~\ref{ssec:results_rMZR}, we demonstrate a tight relationship between metallicity, $\Sigma_{\star}$, and $\Sigma_{\rm SFR}$ by assembling the spaxels from the 24 EMPGs. However, several EMPGs exhibit complex star-formation activities, which contribute to the variety of observed metallicity gradients (Sect.~\ref{ssec:results_gradZ}). In this section, we revisit the rMZR for each EMPG to investigate the spatial distribution of metallicity and its dependence on $\Sigma_{\star}$ and $\Sigma_{\rm SFR}$ on an individual basis.

To robustly analyze the spatial properties of individual EMPGs, it is essential to focus on systems that are sufficiently spatially resolved. We define a characteristic radius, $r_{A}$, where $r_{A} = \sqrt{(A/\pi)}$, and $A$ corresponds to the area (i.e., the number of pixels) with reliable metallicity measurements. To ensure spatial resolution, we select EMPGs with $r_{A}$ greater than $1.5\times$ the FWHM of the seeing during observations.
Based on this criterion, we present a detailed analysis of 22 spatially-resolved EMPGs, while the remaining two objects are classified as unresolved and excluded from further spatial analysis.

After analyzing the individual EMPGs on the rMZR, we identify some interesting patterns shared among them. Based on these patterns, we categorize the sample into four groups, beginning with:
\\

\noindent
\textbf{\textsf{Category A. Metal-poor clump at the core, surrounded by more metal-enriched regions:}}\\
Figure \ref{fig:rMZR_individual_a} illustrates an example of a system belonging to this category. The main panel presents the rMZR, color-coded by $\Sigma_{\rm SFR}$ as in Figure \ref{fig:rMZR}(a), but displaying only spaxels from HS0822$+$3542. On the diagram, the lowest metallicity spaxels shape a sequence, which is flat or slightly increasing extending toward the $\Sigma_{\star}$ peak. This feature is referred to as a ``metal-poor horizontal branch'' on the rMZR in this paper. This sequence connects with more metal-enriched spaxels in the low-$\Sigma_{\star}$ regime.

Eleven of the 22 EMPGs exhibit similar features on the rMZR, as summarized in Figure \ref{fig:rMZR_individual_a_entire}, i.e., possessing a characteristic metal-poor horizontal branch accompanied by more metal-enriched regions, with metallicities reaching up to $0.1$--$0.2$\,\Zsun\ in the low-$\Sigma_{\star}$ regime. Notably, half of these galaxies contain multiple star forming clumps, with one or more clumps showing a metal-poor horizontal branch.

For each EMPG in this category, the main rMZR panel, color-coded by $\Sigma_{\rm SFR}$ (as in Figure \ref{fig:rMZR}a), shows that the tip of the metal-poor branch aligns with regions exhibiting the highest $\Sigma_{\rm SFR}$. The inserted sub-panels in Figures \ref{fig:rMZR_individual_a} and \ref{fig:rMZR_individual_a_entire} display the spatial distribution of the offsets from the rMZR of \citet{barrera-ballesteros2016}, confirming that the metal-poor horizontal branches originate from the central core regions where both $\Sigma_{\star}$ and $\Sigma_{\rm SFR}$ peak, while the more metal-enriched, low-$\Sigma_{\rm SFR}$ regions surround the core areas.

This behavior on the rMZR can be interpreted as a consequence of metal-poor gas inflows. These EMPGs likely experienced prior star formation that enriched the surrounding regions to $0.1$--$0.2$\,\Zsun. A subsequent metal-poor gas inflow into the central potential well would lower the central gas-phase metallicity. The fresh gas would also replenish the inner parts and feed local star formation, following the Kennicutt-Schmidt law \citep{schmidt1959,kennicutt1998_KSLaw}, triggering episodic star formation within the system. This scenario aligns with trends observed in nearby tadpole galaxies, whose cores show similarly low metallicities around $0.1$\,\Zsun\ \citep{sanchez-almeida2014}. We will further explore the implications of this mechanism in Sect.~\ref{sec:discussions}.

If this scenario is correct, the EMPGs in this category are classified as such due to a recent infall of metal-poor gas, triggering a burst of star formation and producing intense nebular emission lines. This characteristic is preferentially detected by the EMPG classification criteria, which rely on strong hydrogen lines with comparatively weak metal lines, accompanied by an extremely faint continuum (i.e., high specific SFR; \citealt{kojima2020}, \citealt{nishigaki2023}). However, these EMPGs may not serve as reliable analogs of galaxies in the early stages of evolution.
Moreover, in the absence of such metal-poor gas infall, many of these systems might no longer meet the EMPG classification criteria, as their surrounding regions are already enriched to metallicity levels of $0.1$--$0.2$\,\Zsun. This could also contribute to the significant scatter observed in the global mass--metallicity relation at the low-mass end (e.g., \citealt{kojima2020, nishigaki2023}).

The interpretation also helps explain the different metallicity gradients observed between the inner and outer regions of certain EMPGs (Sect.~\ref{ssec:results_gradZ}). The metal-poor horizontal branch on the rMZR corresponds to the inner metallicity gradient.
Revisiting the inner metallicity gradients of EMPGs in this category reveals that their slopes are nearly flat or mildly negative, indicating that metallicity remains constant or decreases slightly from the root to the tip. These findings indicate that the core regions of these EMPGs typically exhibit flat or mildly negative (``ordinary'') metallicity gradients, with no clear evidence of positive (``extraordinary'') gradients. This supports an inside-out galaxy evolution scenario, where central starburst-driven outflows rapidly mix metals, leading to a lack of positive gradients near the core. The short mixing timescales inferred for these low-mass clumps highlight the dynamic and efficient redistribution of metals within these systems.

Note that sources in this category might be expected to exhibit positive metallicity gradients across their entire systems. However, not all EMPGs in this category display positive gradients when considering the full region (e.g., J1016$+$3754, J2115$-$1734). This is because metallicity contributions from regions outside the semi-major axis or those associated with secondary clumps are not fully captured in the gradient measurements. Analyzing the full distribution of metals on the rMZR on an individual basis provides a more comprehensive interpretation of the evolutionary stages of these sources. This approach is particularly important for EMPGs, which are likely undergoing episodic star formation and showing complex metallicity structures.

\begin{figure}[!t]
    \centering   
     \begin{overpic}[bb=0 0 325 239, width=0.99\columnwidth]{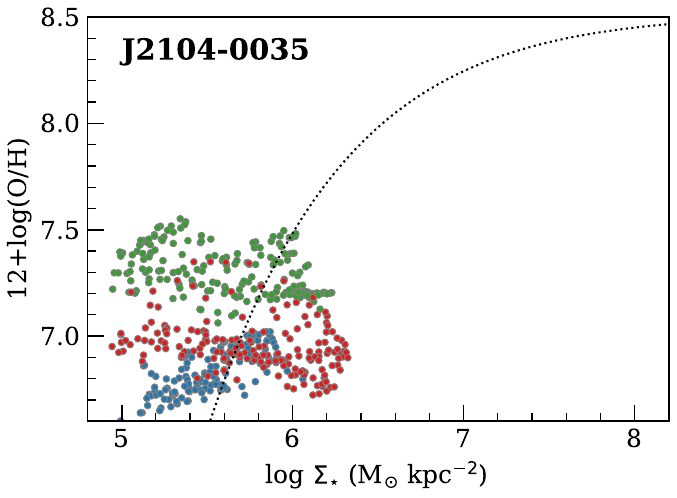}
      \put(63.1,42.5){\includegraphics[bb=0 0 484 386, scale=0.175]{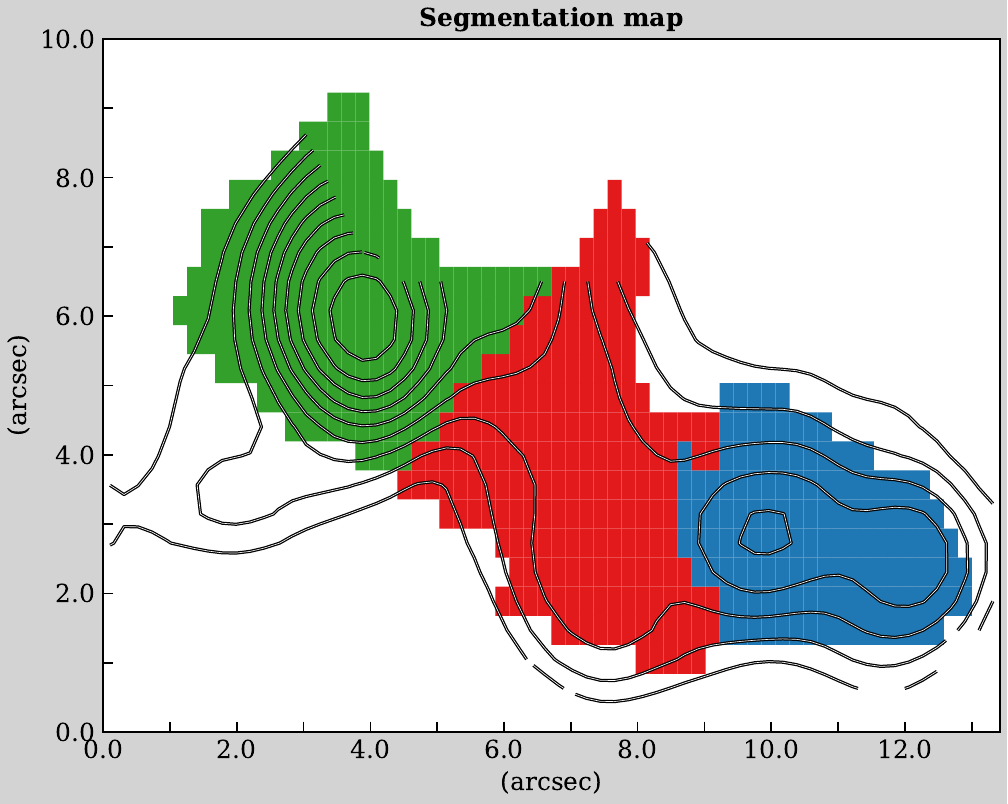}}  
     \end{overpic}
    \caption{%
    		The rMZR for J2104$-$0035, color-coded by three distinct regions as indicated in the inset map. This object is classified as Category A based on the patterns observed in the green-colored clump. In contrast, the blue-colored clump displays characteristics similar to Category B, while the red-colored regions appear to be in a transition phase, resembling Category C.
		}
    \label{fig:rMZR_individual_a_highlight}
\end{figure}

\begin{figure*}
  \centering
    \begin{tabular}{c}      %
      \begin{minipage}{0.48\hsize}
        \begin{center}
	 \begin{overpic}[bb=0 0 325 239, width=1.0\textwidth]{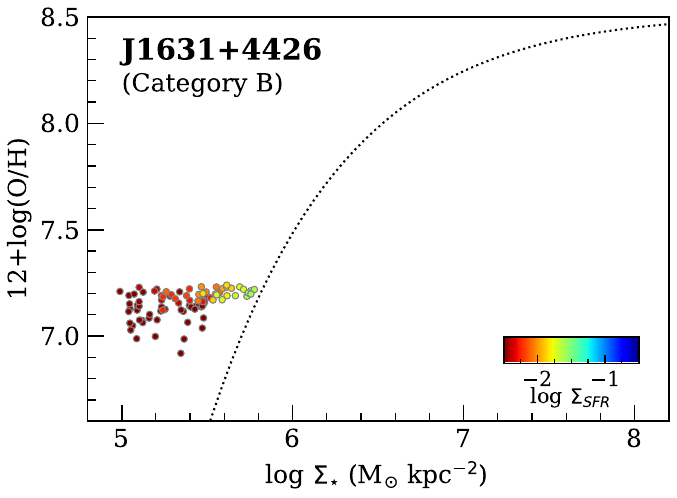}
           \put(61,44){\includegraphics[bb=0 0 399 285, scale=0.23]{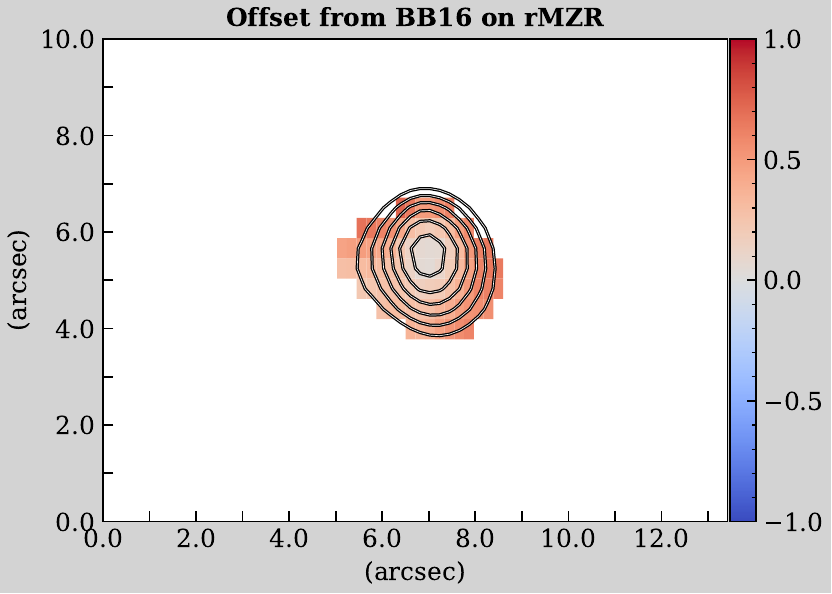}}  
         \end{overpic}
        \end{center}
      \end{minipage}      %
      \begin{minipage}{0.48\hsize}
        \begin{center}
	 \begin{overpic}[bb=0 0 325 239, width=1.0\textwidth]{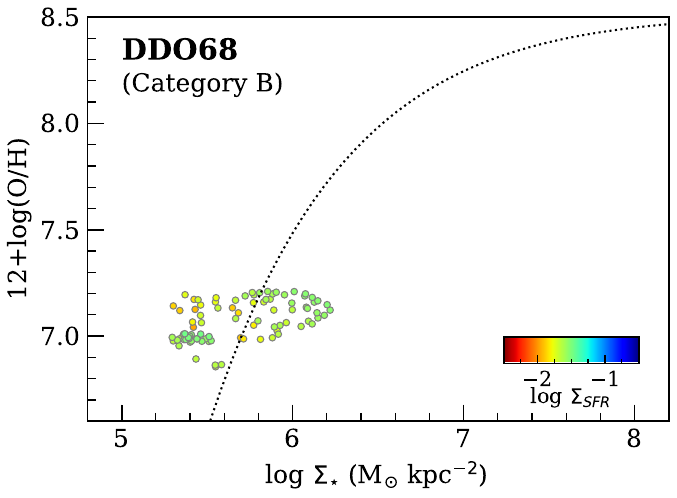}
           \put(61,44){\includegraphics[bb=0 0 399 285, scale=0.23]{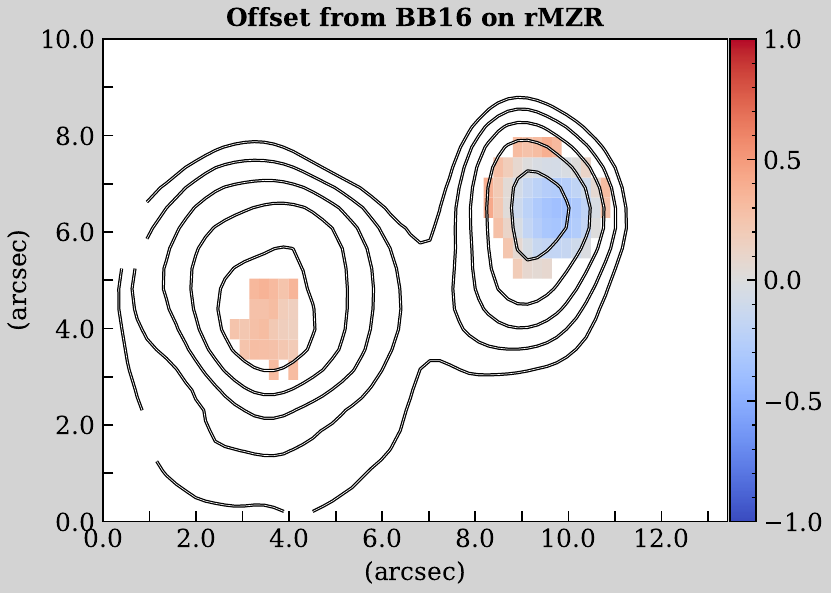}}  
         \end{overpic}
        \end{center}
      \end{minipage}      \\
      \begin{minipage}{0.48\hsize}
        \begin{center}
	 \begin{overpic}[bb=0 0 325 239, width=1.0\textwidth]{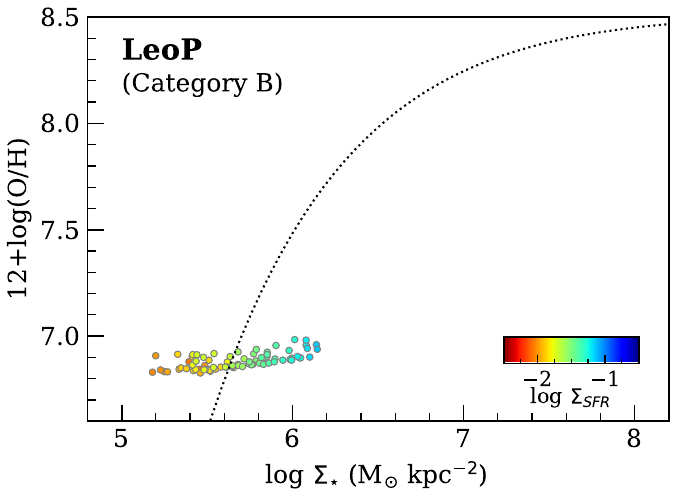}
           \put(61,44){\includegraphics[bb=0 0 399 285, scale=0.23]{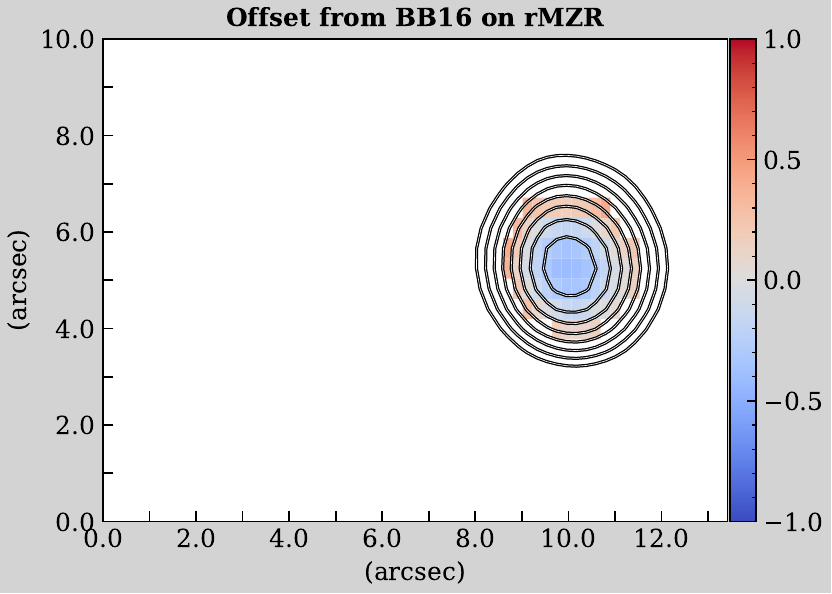}}  
         \end{overpic}
        \end{center}
      \end{minipage}      %
      \begin{minipage}{0.48\hsize}
        \begin{center}
	 \begin{overpic}[bb=0 0 325 239, width=1.0\textwidth]{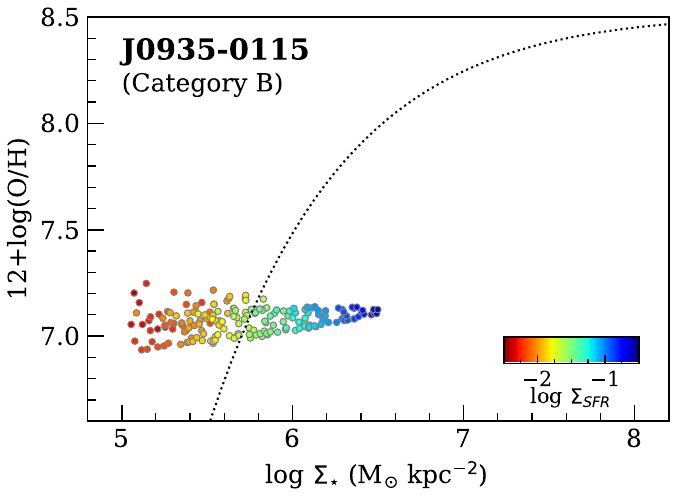}
           \put(61,44){\includegraphics[bb=0 0 399 285, scale=0.23]{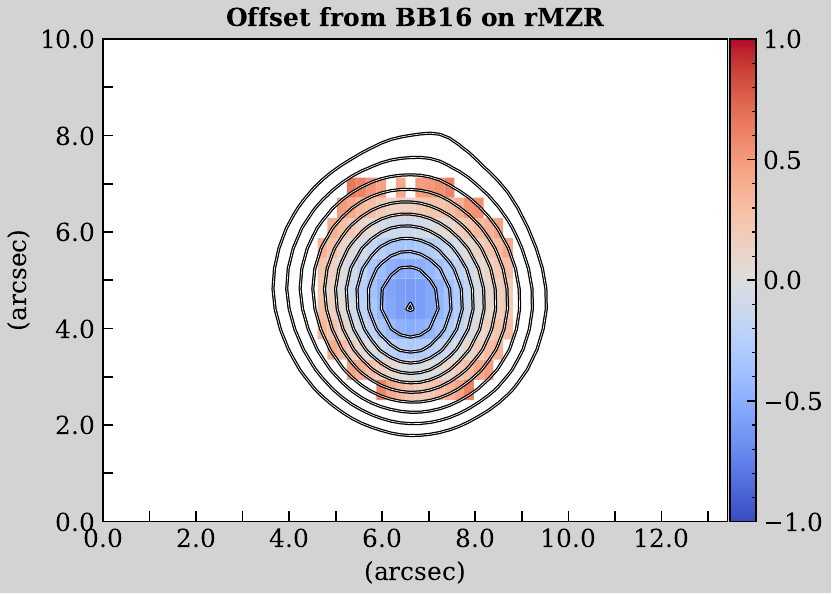}}  
         \end{overpic}
        \end{center}
      \end{minipage}      \\
    \end{tabular}
    \caption{%
      Same as Figure \ref{fig:rMZR_individual_a}, but for individual EMPGs in Category B. These sources likely represent galaxies undergoing their initial phases of star formation and chemical enrichment in the local universe.
    }
    \label{fig:rMZR_individual_b}
\end{figure*}



One particularly noteworthy EMPG in this category is J2104$-$0035, which presents a complex metallicity structure. As a multi-clump system, it exhibits distinct behavior on the rMZR. To highlight the contributions from individual clumps, Figure \ref{fig:rMZR_individual_a_highlight} shows the rMZR of J2104$-$0035 color-coded according to spatial position, with the corresponding clumps marked on the inserted map.
The upper-left clump, shown in green, presents a typical metal-poor horizontal branch at its center, surrounded by a chemically enriched region, similar to other EMPGs in this category. This clump appears to be undergoing active star formation, fueled by fresh gas inflow. However, the chemically enriched region is confined locally and does not extend further into the outer regions.
In contrast, the lower-right clump, shown in blue, exhibits an extremely low metallicity, as low as $0.01$--$0.02$\,\Zsun, with a decreasing metallicity gradient toward its outskirts. Meanwhile, the bridging region (in red) between these two clumps shows low metallicity, despite containing a relatively high stellar mass, further emphasizing the complex nature of these systems.
We speculate that the blue-colored clump is in the early phase of chemical evolution (Category B), while the red-colored region represents a transition phase (Category C). 
Alternatively, this could be a merging phase between systems in Categories A and B, where metals have yet to fully mix.
These different categories will be further described in the followings. 
\\

\noindent
\textbf{\textsf{Category B. Just metal-poor clump(s): Experiencing the first phase of chemical evolution?}}\\
Four of the 22 EMPGs exhibit a metal-poor horizontal branch on the rMZR, similar to those in Category A. However, unlike Category A, these EMPGs do not show any associated metal-rich surrounding regions. Figure \ref{fig:rMZR_individual_b} presents the four EMPGs in this category: J1631$+$4426, DDO68, LeoP, and J0935$-$0115. These galaxies are likely in the early stages of chemical evolution, experiencing their first episodes of enrichment, and may represent an early phase of galaxy formation.
Interestingly, these four EMPGs are the most metal-poor objects in the sample, with compact morphologies. A tantalizingly similar trend to those observed in this category is hinted at in J1234$+$3901, despite the spatial limitations.
With global stellar masses typically below $10^6$\,\Msun, where high-redshift galaxies are rarely sampled (cf. \citealt{vanzella2023_metalpoor}), these objects offer a unique opportunity to probe the earliest phases of galaxy evolution in the universe.

A possible caveat is that the absence of metal-rich surrounding regions may reflect observational limitations. The detection limit could prevent us from probing the faint outskirts of these systems, leaving only the brighter core regions accessible for analysis. This is particularly evident in the case of DDO68, which shows diffuse, extended \Ha\ emission around two clumps. The half-light radii of these clumps exceed $r_A$, indicating that the metallicity distribution likely extends beyond what is currently observed.
Deeper observations will be required to confirm the classification of these galaxies and further explore their extended metallicity distributions.
\\

\begin{figure*}
  \centering
    \begin{tabular}{c}      %
      \begin{minipage}{0.48\hsize}
        \begin{center}
	 \begin{overpic}[bb=0 0 325 239, width=1.0\textwidth]{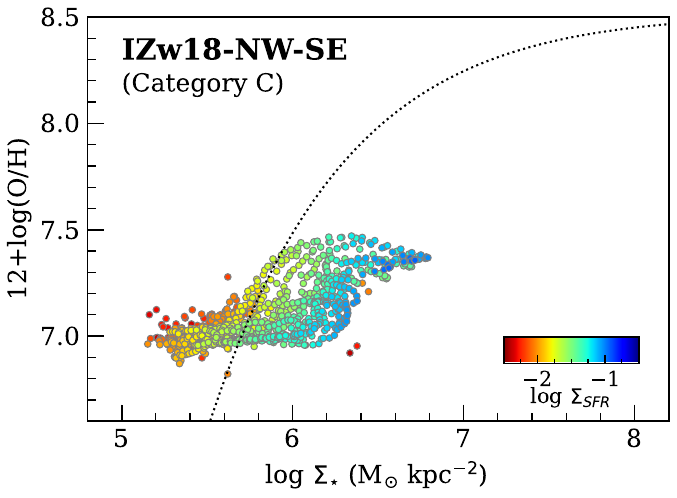}
           \put(61,44){\includegraphics[bb=0 0 399 285, scale=0.23]{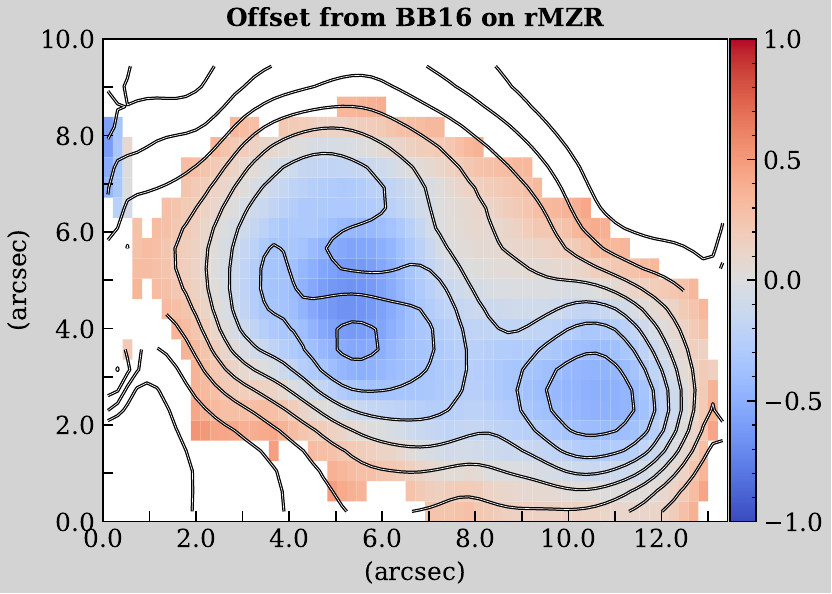}}  
         \end{overpic}
        \end{center}
      \end{minipage}      %
      \begin{minipage}{0.48\hsize}
        \begin{center}
	 \begin{overpic}[bb=0 0 325 239, width=1.0\textwidth]{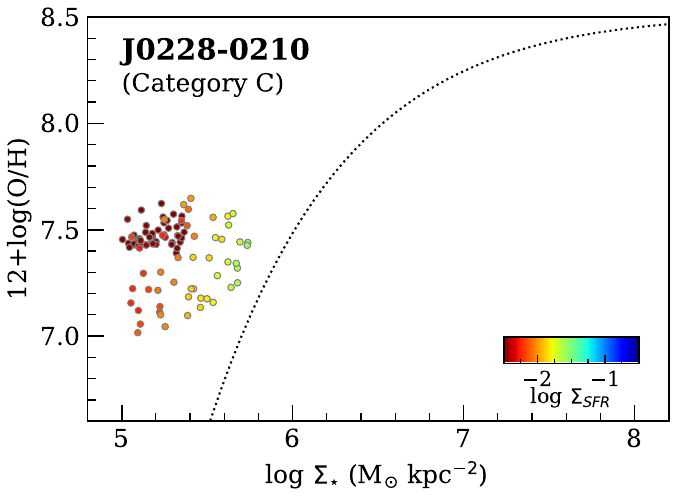}
           \put(61,44){\includegraphics[bb=0 0 399 285, scale=0.23]{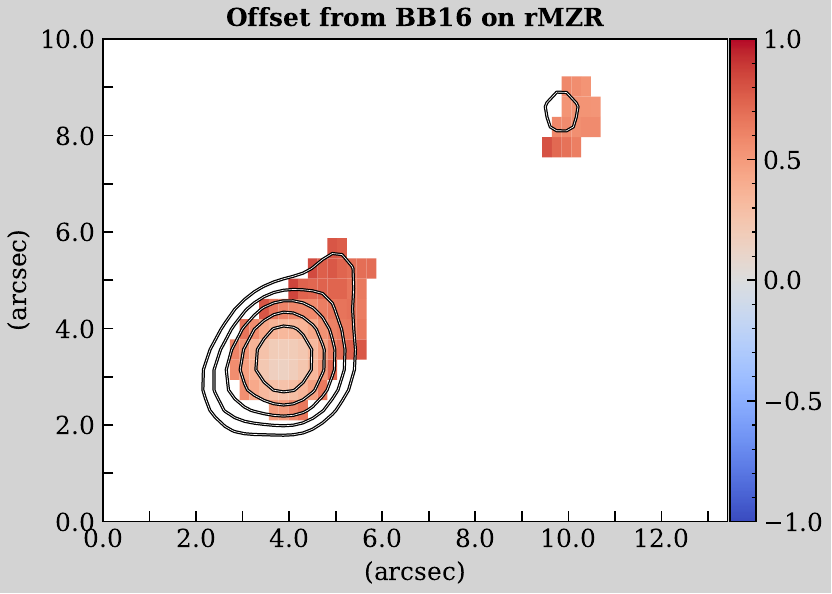}}  
         \end{overpic}
        \end{center}
      \end{minipage}      \\
      \begin{minipage}{0.48\hsize}
        \begin{center}
	 \begin{overpic}[bb=0 0 325 239, width=1.0\textwidth]{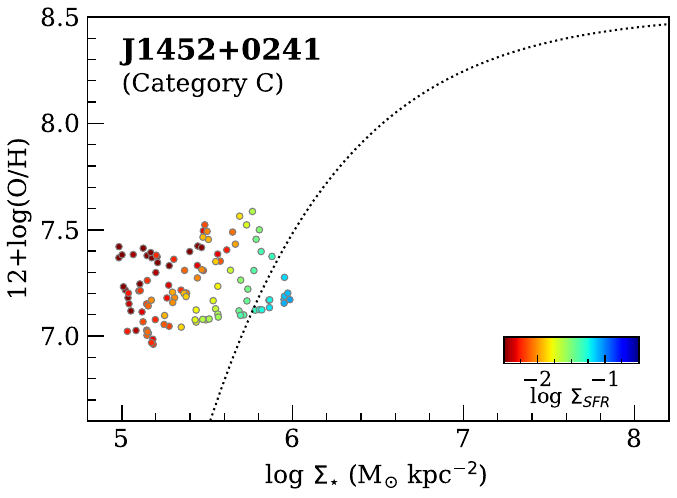}
           \put(61,44){\includegraphics[bb=0 0 399 285, scale=0.23]{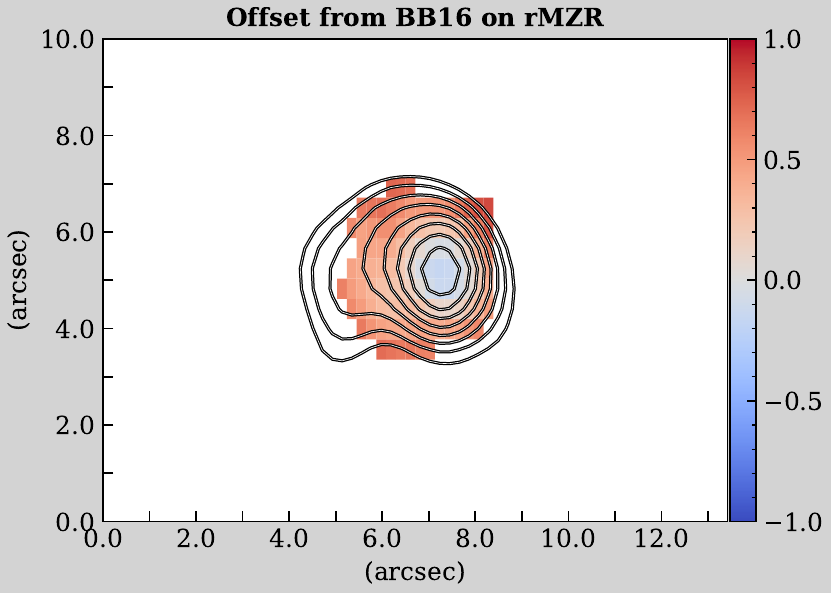}}  
         \end{overpic}
        \end{center}
      \end{minipage}      %
      \begin{minipage}{0.48\hsize}
        \begin{center}
	 \begin{overpic}[bb=0 0 325 239, width=1.0\textwidth]{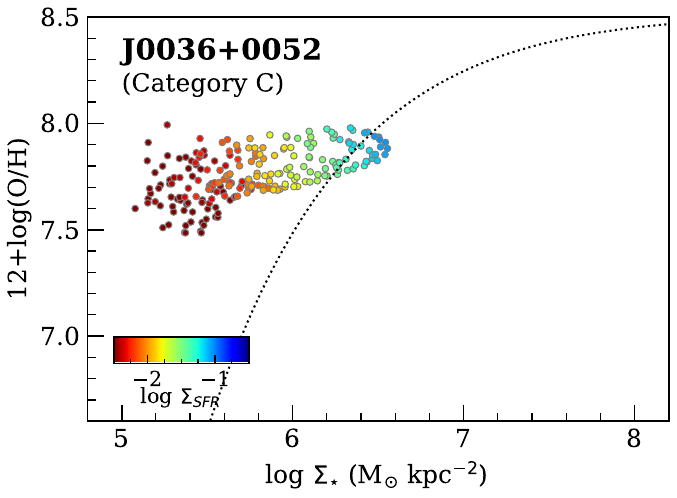}
	   \put(61,12){\includegraphics[bb=0 0 399 285, scale=0.23]{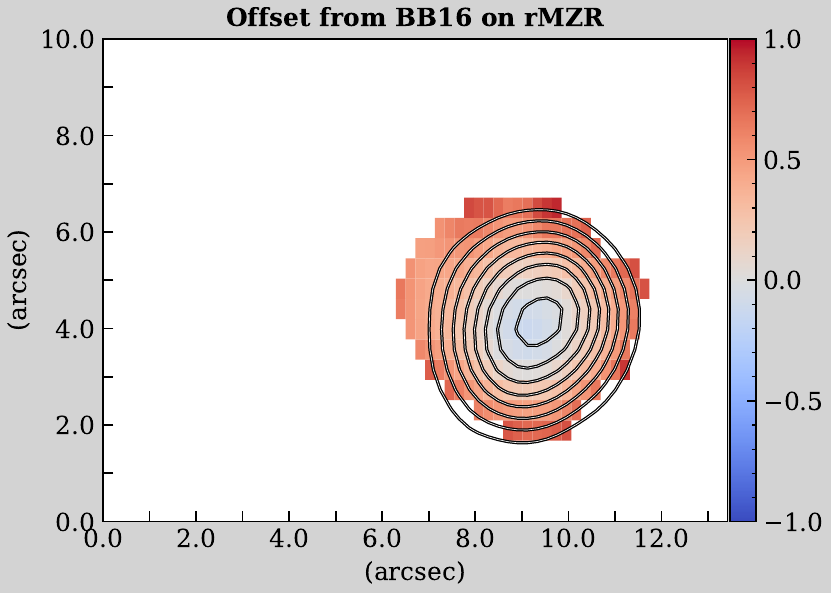}}
         \end{overpic}
        \end{center}
      \end{minipage}
      \\
    \end{tabular}
    \caption{%
      Same as Figure \ref{fig:rMZR_individual_a}, but for individual EMPGs in Category C, likely in a transition phase.
    }
    \label{fig:rMZR_individual_c}
\end{figure*}


\begin{figure*}
  \addtocounter{figure}{-1}
  \centering
    \begin{tabular}{c}      %
      \begin{minipage}{0.48\hsize}
        \begin{center}
	 \begin{overpic}[bb=0 0 325 239, width=1.0\textwidth]{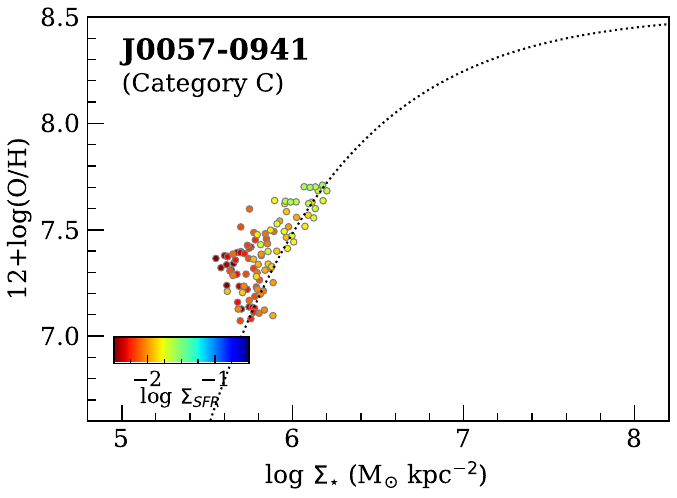}
	   \put(61,12){\includegraphics[bb=0 0 399 285, scale=0.23]{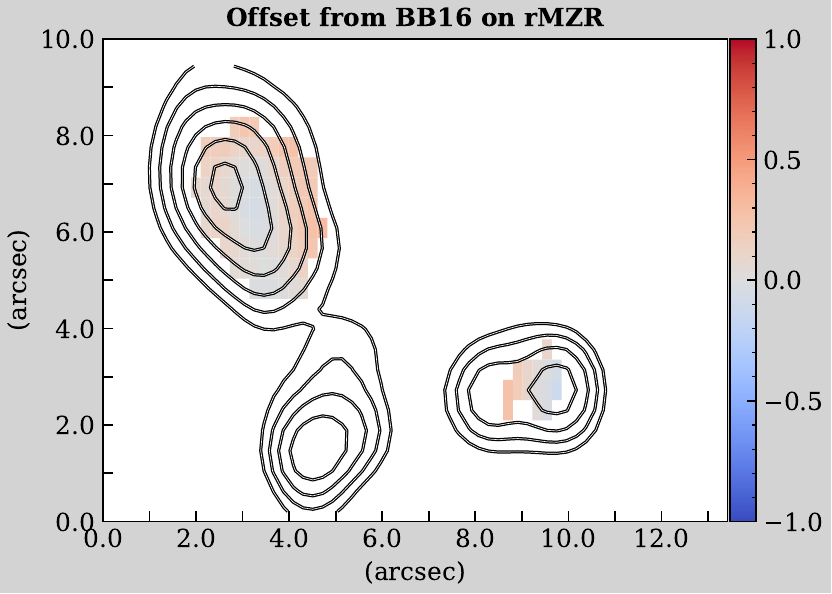}}
         \end{overpic}
        \end{center}
      \end{minipage}
      \begin{minipage}{0.48\hsize}
        \begin{center}
	 \begin{overpic}[bb=0 0 325 239, width=1.0\textwidth]{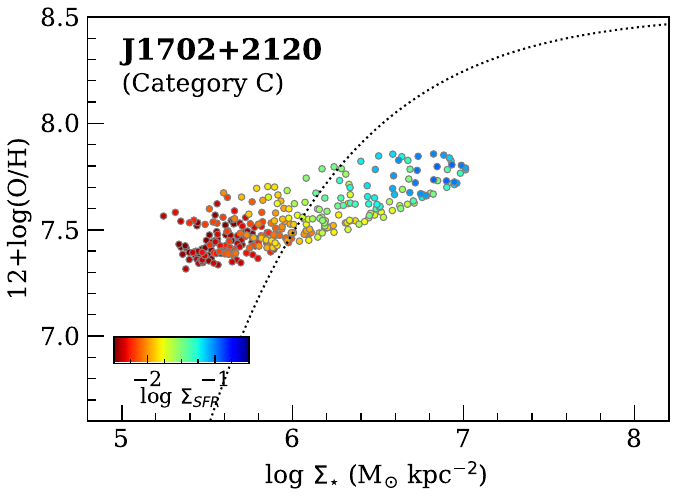}
	   \put(61,12){\includegraphics[bb=0 0 399 285, scale=0.23]{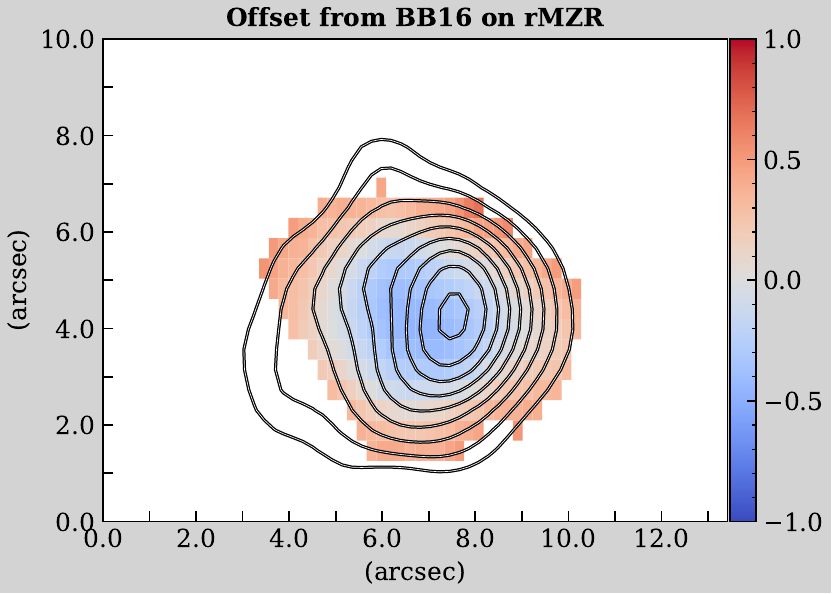}}
         \end{overpic}
        \end{center}
      \end{minipage}
      \\
      \begin{minipage}{0.48\hsize}
        \begin{center}
	 \begin{overpic}[bb=0 0 325 239, width=1.0\textwidth]{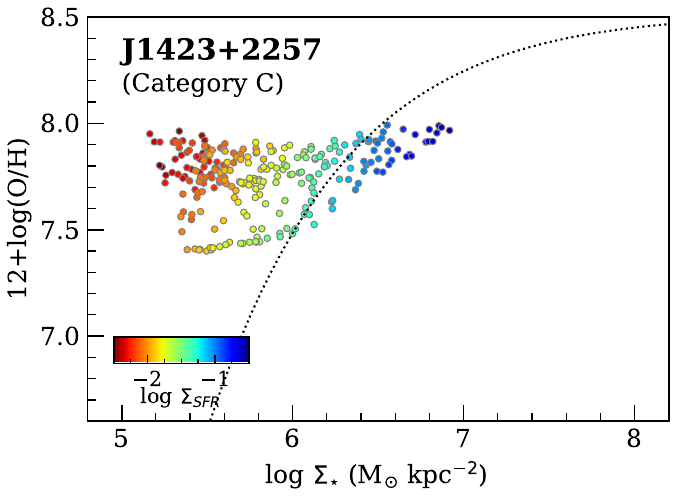}
	   \put(61,12){\includegraphics[bb=0 0 399 285, scale=0.23]{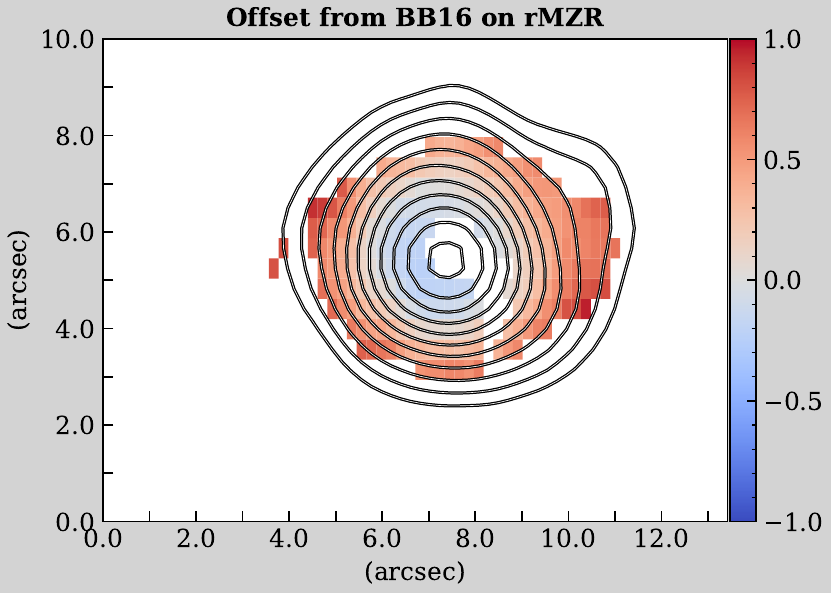}}
         \end{overpic}
        \end{center}
      \end{minipage}
    \end{tabular}
    \caption{%
      (Continued.)
    }
\end{figure*}



\begin{figure*}
  \centering
    \begin{tabular}{c}      %
      \begin{minipage}{0.48\hsize}
        \begin{center}
	 \begin{overpic}[bb=0 0 325 239, width=1.0\textwidth]{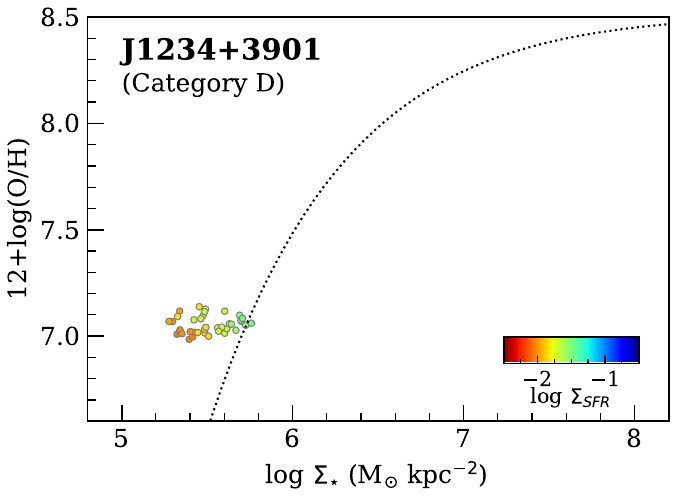}
           \put(61,44){\includegraphics[bb=0 0 399 285, scale=0.23]{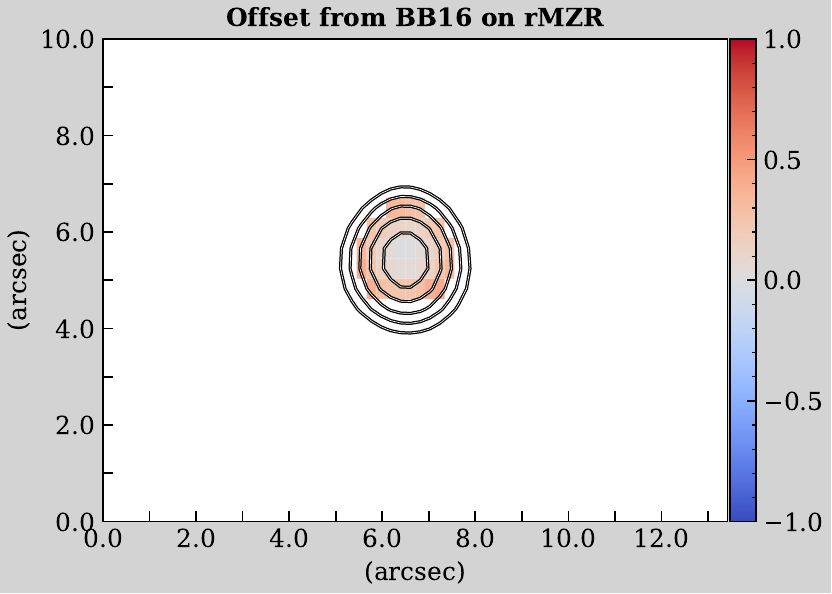}}  
         \end{overpic}
        \end{center}
      \end{minipage}      %
      \begin{minipage}{0.48\hsize}
        \begin{center}
	 \begin{overpic}[bb=0 0 325 239, width=1.0\textwidth]{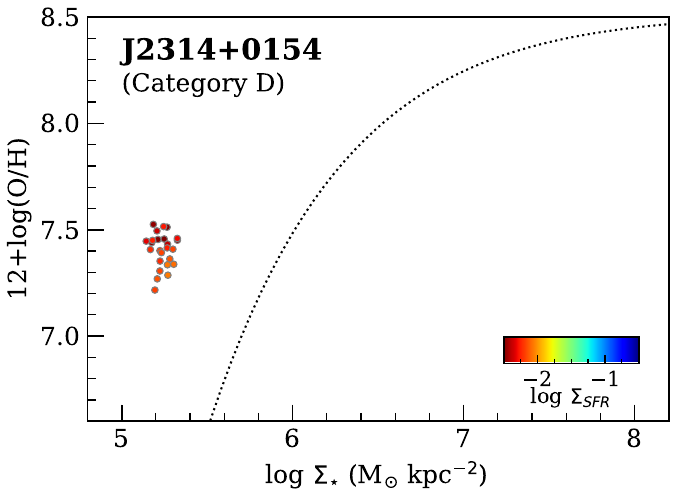}
           \put(61,44){\includegraphics[bb=0 0 399 285, scale=0.23]{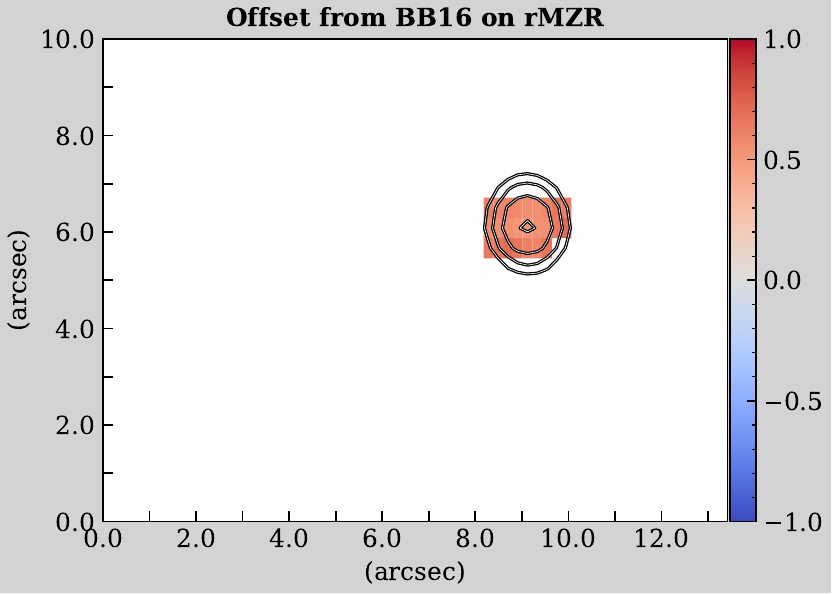}}  
         \end{overpic}
        \end{center}
      \end{minipage}      \\
    \end{tabular}
    \caption{%
      Same as Figure \ref{fig:rMZR_individual_a}, but for individual EMPGs in Category D, Unresolved. 
    }
    \label{fig:rMZR_individual_d}
\end{figure*}



\noindent
\textbf{\textsf{Category C. No clear metal-poor horizontal branch: In a transition phase?}}\\
Figure \ref{fig:rMZR_individual_c} summarizes 7 galaxies in the third category: galaxies where no clear metal-poor horizontal branch is observed on the rMZR. Instead, the central $\Sigma_{\star}$ peak regions exhibit chemical enrichment. In these systems, the metallicity in the outer regions is comparable to or slightly lower than that in the central $\Sigma_{\star}$ peak regions. 

We speculate that these galaxies are in a transition phase, moving away from a recent burst of star formation. The chemically enriched central regions are likely consuming their fresh gas, resulting in higher metallicity as star formation proceeds. Several objects, such as I\,Zw\,18, J1452$+$0241, J0036$+$0052, J1702$+$2120, and J1423$+$2257, still exhibit a visible branch on the rMZR, though no longer in a horizontal form. In contrast, J0228$-$0210 and J0057$-$0941 follow the general rMZR trend, suggesting they may have had sufficient time for internal mixing of metals within the system.

Notably, I\,Zw\,18, J1452$+$0241, and J0057$-$0941 are associated with outer regions that remain extremely metal-poor, with metallicities around $Z\sim 0.02$\,\Zsun. These galaxies might be in a transitional state or have just transitioned from a phase similar to those in Category B. This could indicate a recent cessation of fresh gas inflows, marking the shift from gas accretion-driven star formation to more stabilized, internal evolution.
\\

\noindent
\textbf{\textsf{Category D. Unresolved:}}
Figure \ref{fig:rMZR_individual_d} lists the remaining two EMPGs that are too compact to be conclusively classified. As mentioned earlier, J1234$+$3901 tentatively exhibits a metal-poor horizontal branch with $Z\sim 0.02$\,\Zsun\ and could potentially belong to Category B. 
J2314$+$0154 does not show any discernible branch on the rMZR and lacks signs of active star formation, suggesting it may share characteristics with the galaxies in Category C, which are likely in a transition phase. 
J1323$-$0132, although spatially well-resolved, presents spaxels with \OIII/\Hb\ ratios that lie outside the valid range of the R3-index \citep{nakajima2022_empressV}, likely due to the galaxy's relatively high metallicity. As a result, a precise classification remains uncertain for this object.

The categories are summarized in Table \ref{tbl:objects}.

\section{Discussions} \label{sec:discussions}

Based on the spatially-resolved properties of 24 EMPGs with $Z\lesssim 0.1$\,\Zsun, our results reveal significant metallicity variations driven by local stellar mass and star formation activity (Sect.~\ref{ssec:results_rMZR}), as well as by distance from the center of the galaxies (Sect.~\ref{ssec:results_gradZ}). Through an individual examination of each EMPG, we identified three distinct categories based on their spatial metallicity distribution (Sect.~\ref{ssec:results_rMZR_individual}). In this section, we further examine these findings and explore their implications for understanding the nature and evolution of EMPGs in the local universe.

\begin{figure}[t]
    \begin{center}
     \includegraphics[bb=0 0 350 436, width=0.99\columnwidth]{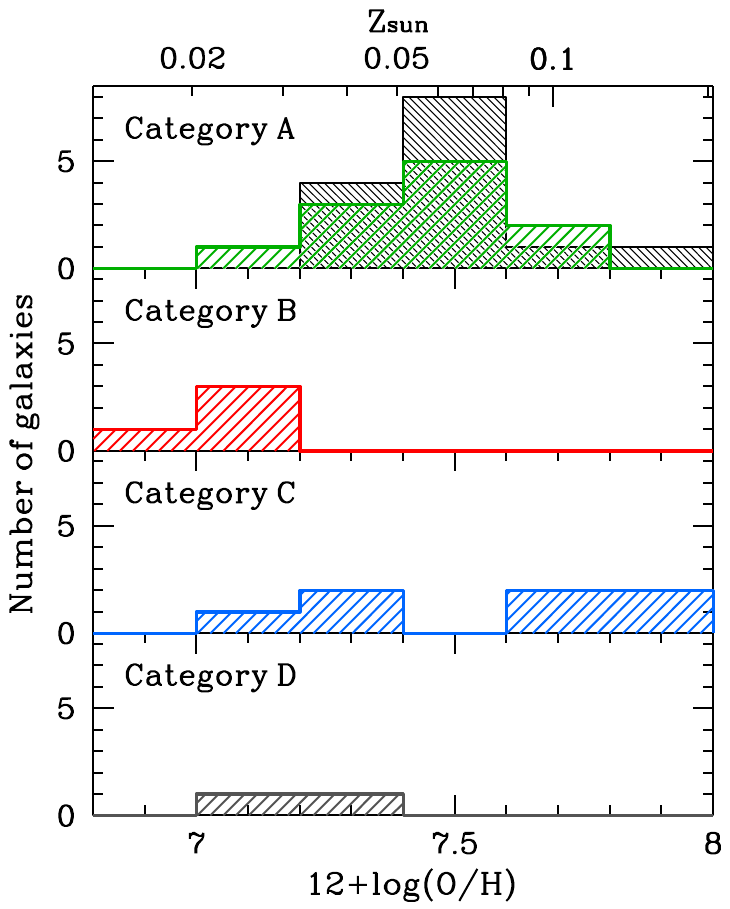}
    \caption{%
    		Metallicity distributions of galaxies in the four categories (Sect. \ref{ssec:results_rMZR_individual}). The metallicities correspond to literature values (Table \ref{tbl:objects}, derived from observations that integrate the central regions of the systems. In the histogram of Category A, the green shade presents the distribution of EMPGs studies in this paper. In addition, the black shade shows the distribution of tadpole galaxies exhibiting metallicity drops around the peak of $\Sigma_{\rm SFR}$ from the literature \citep{sanchez-almeida2013, sanchez-almeida2014, sanchez-almeida2015}.
		}
    \label{fig:histo_category}
    \end{center} 
\end{figure}

\begin{figure*}
  \centering
    \begin{tabular}{c}      %
      \begin{minipage}{0.48\hsize}
        \begin{center}
	 \begin{overpic}[bb=0 0 325 239, width=1.0\textwidth]{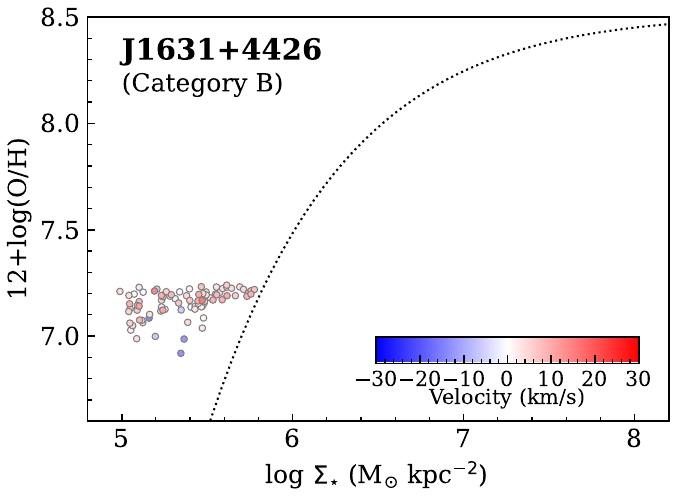}
           \put(64,44){\includegraphics[bb=0 0 484 375, scale=0.175]{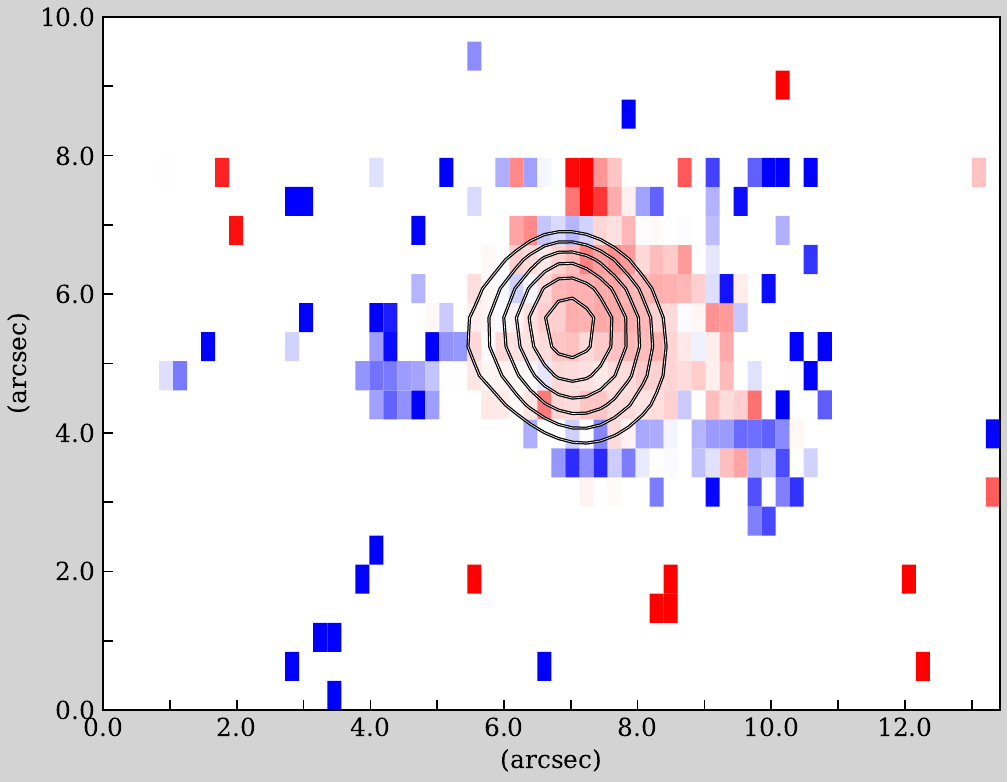}}  
         \end{overpic}
        \end{center}
      \end{minipage}      %
      \begin{minipage}{0.48\hsize}
        \begin{center}
	 \begin{overpic}[bb=0 0 325 239, width=1.0\textwidth]{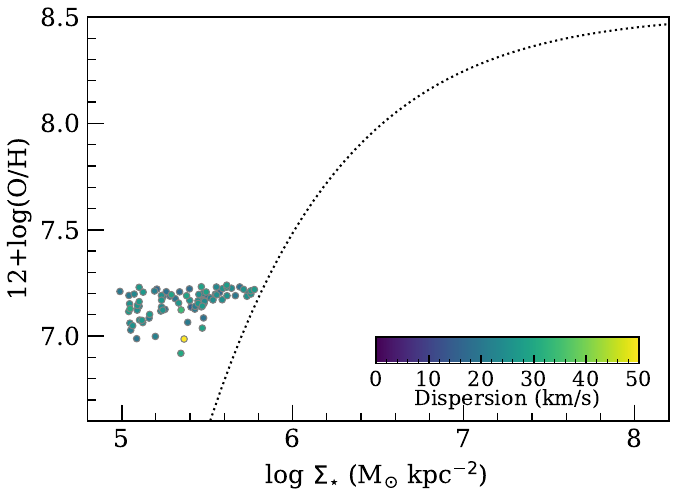}
           \put(64,44){\includegraphics[bb=0 0 484 375, scale=0.175]{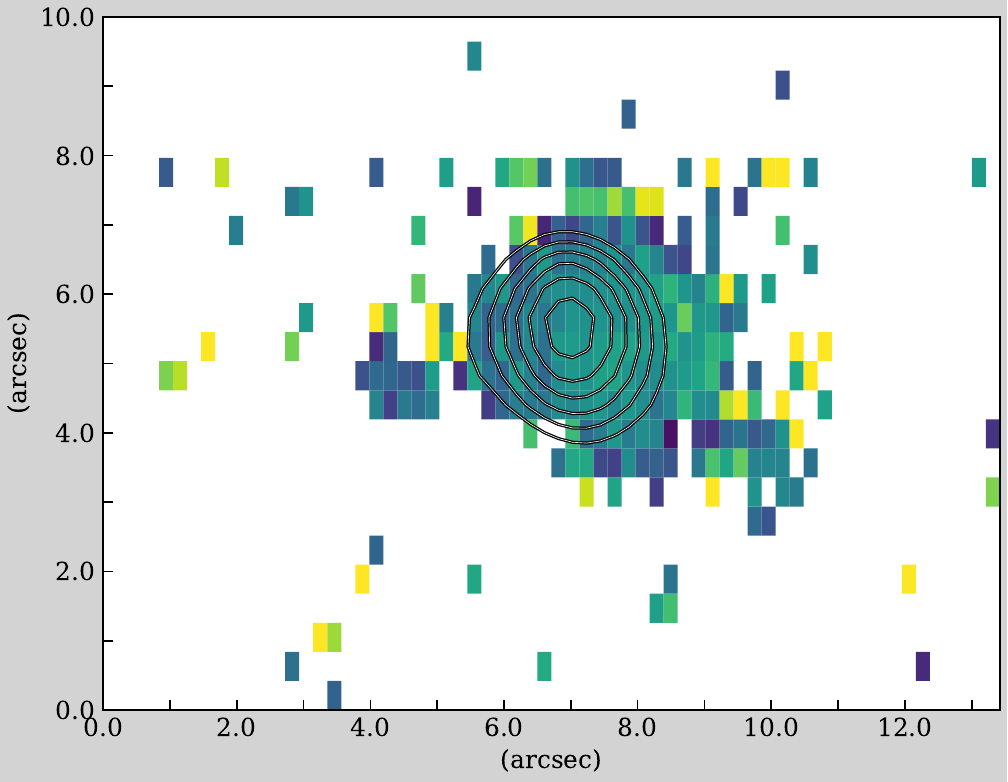}}  
         \end{overpic}
        \end{center}
      \end{minipage}      \\
      \begin{minipage}{0.48\hsize}
        \begin{center}
	 \begin{overpic}[bb=0 0 325 239, width=1.0\textwidth]{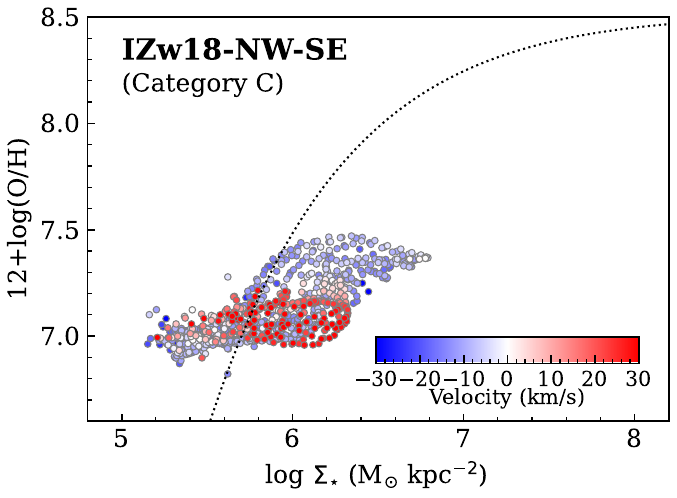}
           \put(64,44){\includegraphics[bb=0 0 484 375, scale=0.175]{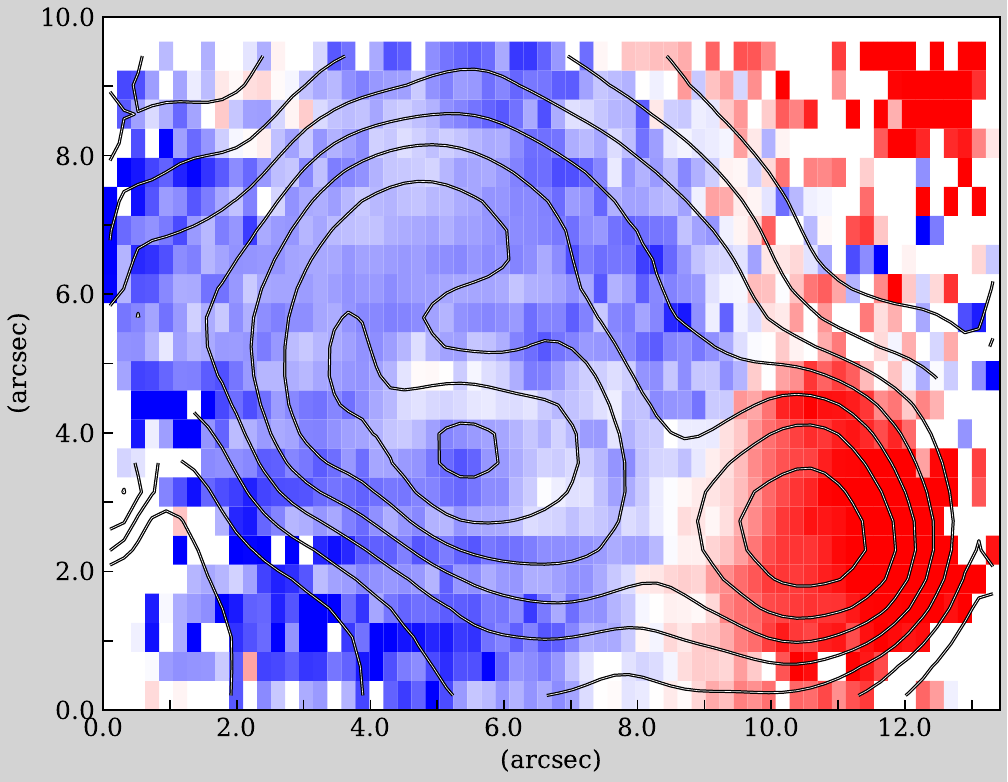}}  
         \end{overpic}
        \end{center}
      \end{minipage}      %
      \begin{minipage}{0.48\hsize}
        \begin{center}
	 \begin{overpic}[bb=0 0 325 239, width=1.0\textwidth]{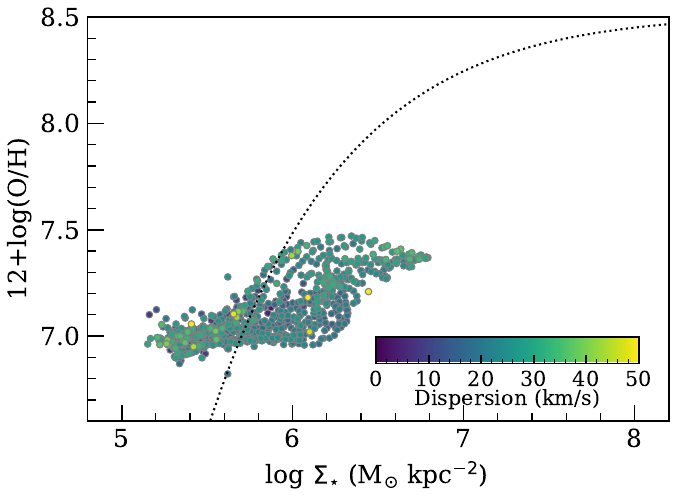}
           \put(64,44){\includegraphics[bb=0 0 484 375, scale=0.175]{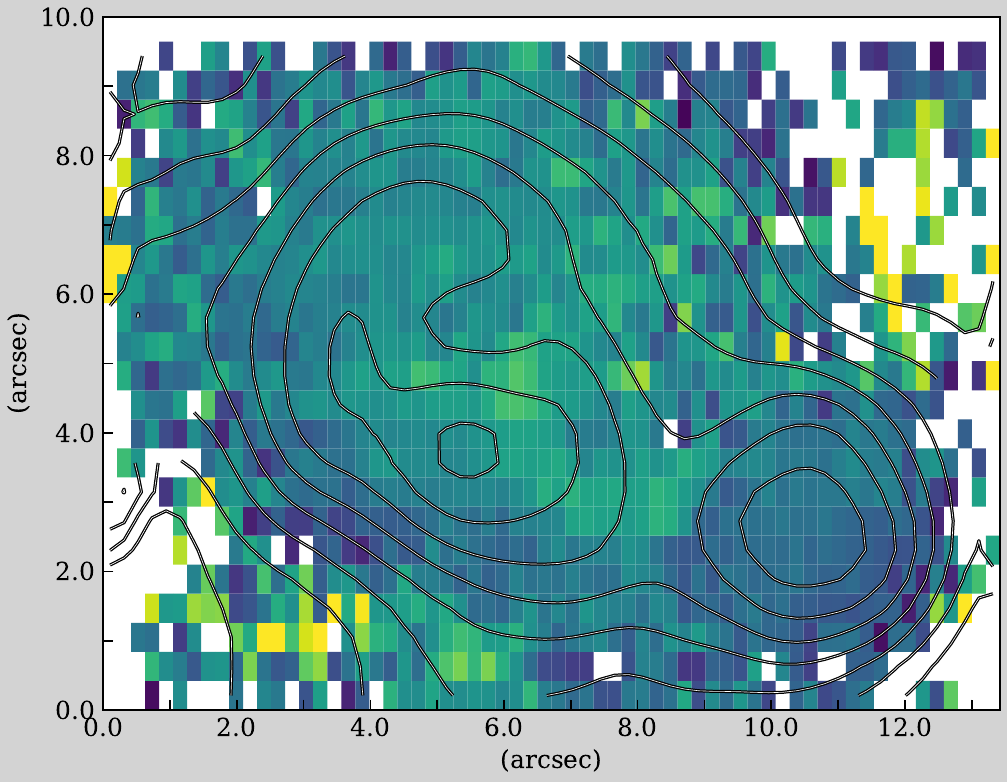}}  
         \end{overpic}
        \end{center}
      \end{minipage}      \\
      \begin{minipage}{0.48\hsize}
        \begin{center}
	 \begin{overpic}[bb=0 0 325 239, width=1.0\textwidth]{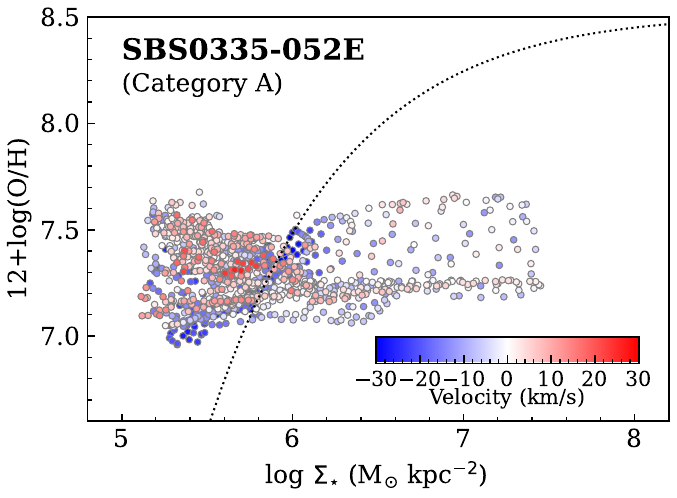}
           \put(66.2,45.7){\includegraphics[bb=0 0 484 375, scale=0.164]{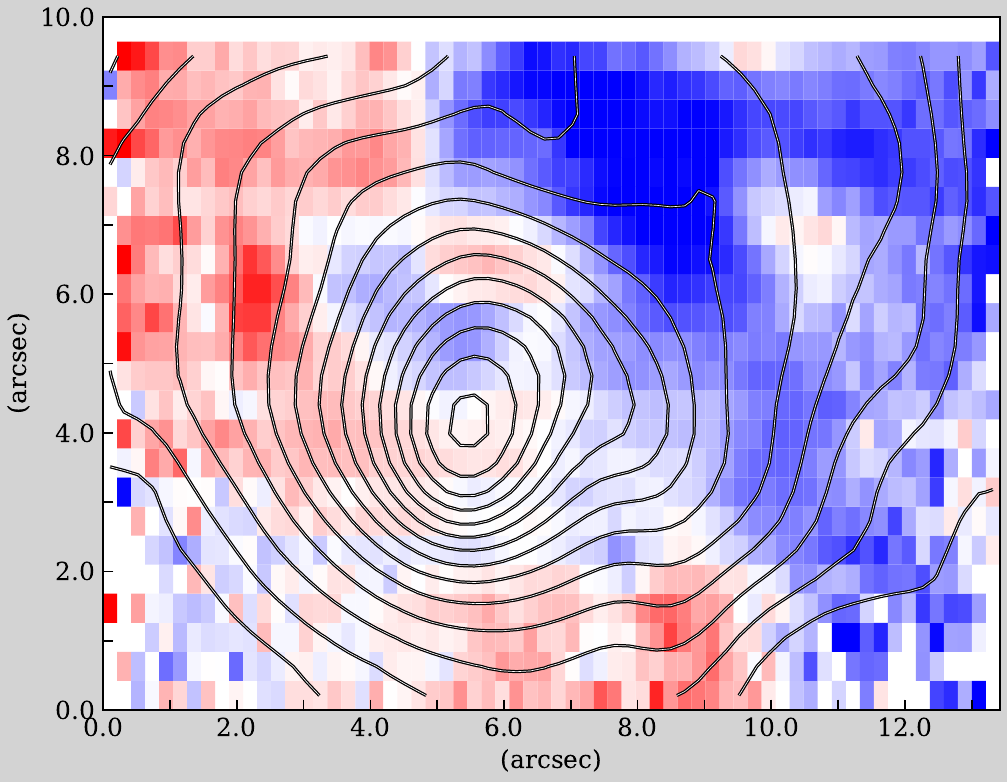}}  
         \end{overpic}
        \end{center}
      \end{minipage}      %
      \begin{minipage}{0.48\hsize}
        \begin{center}
	 \begin{overpic}[bb=0 0 325 239, width=1.0\textwidth]{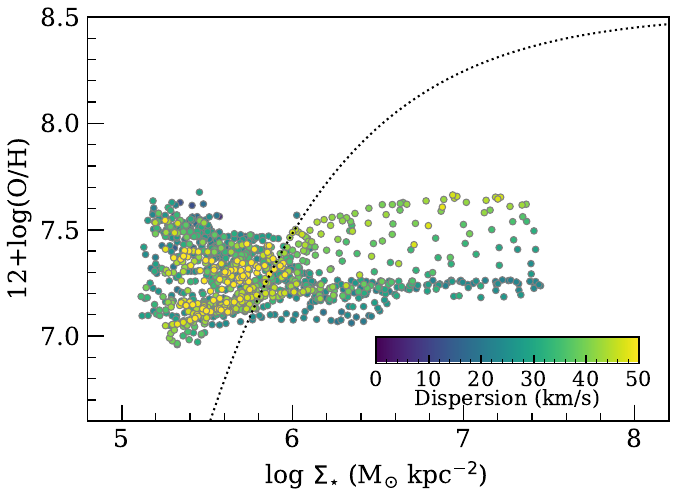}
           \put(66.2,45.7){\includegraphics[bb=0 0 484 375, scale=0.164]{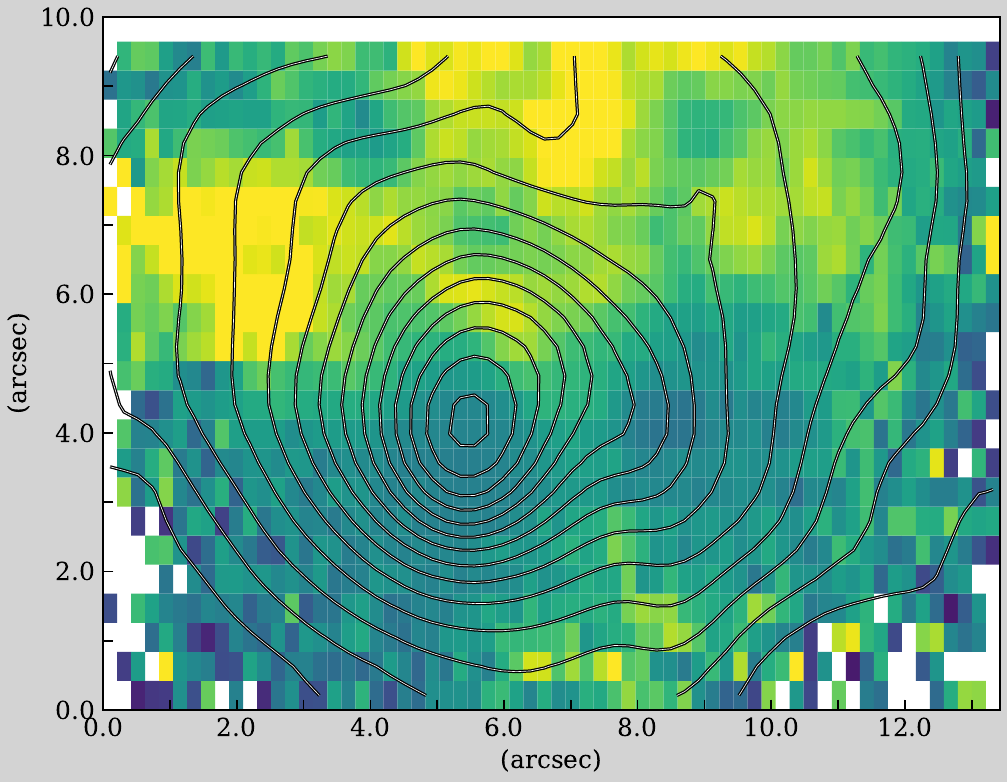}}  
         \end{overpic}
        \end{center}
      \end{minipage}      \\
    \end{tabular}
    \caption{%
       The rMZRs for six EMPGs, J1631$+$4426, IZw18, SBS0355-052E, HS0822$+$3542, J1044$+$0353, and J2115$-$1734 (from top to bottom), with kinematic properties available from high-resolution IFU observations \citep{isobe2023_focasifu}. For each galaxy, the left panel shows the rMZR color-coded by velocity (inset map), while the right panel displays the rMZR color-coded by dispersion (inset map).
    }
    \label{fig:rMZR_kin}
\end{figure*}


\begin{figure*}
  \addtocounter{figure}{-1}
  \centering
    \begin{tabular}{c}      %
      \begin{minipage}{0.48\hsize}
        \begin{center}
	 \begin{overpic}[bb=0 0 325 239, width=1.0\textwidth]{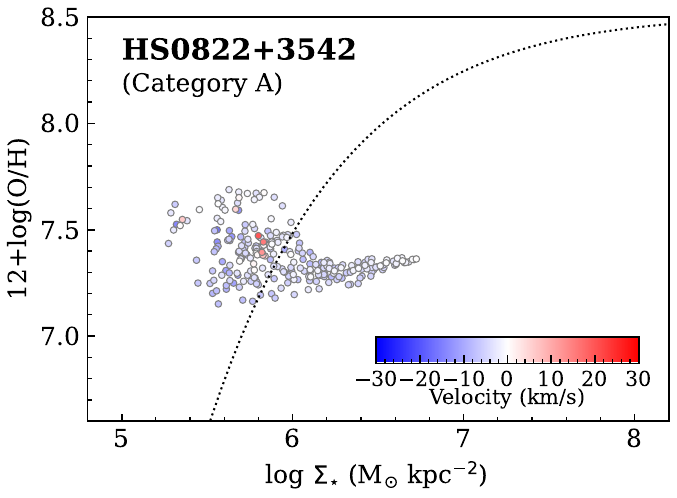}
           \put(64,44){\includegraphics[bb=0 0 484 375, scale=0.175]{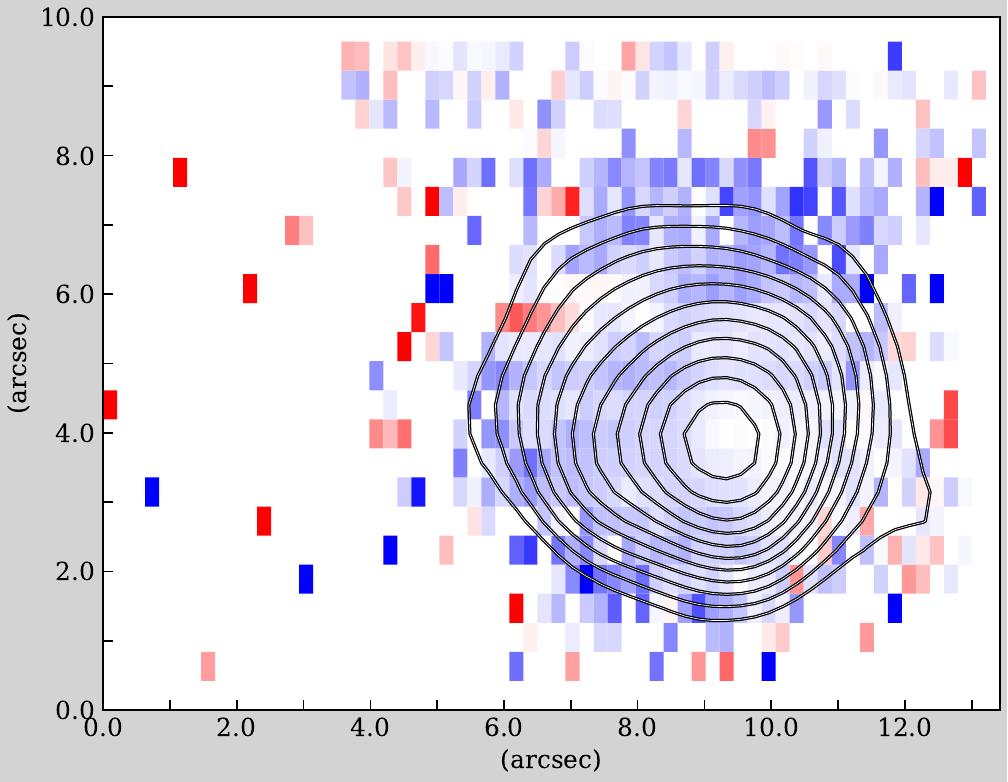}}  
         \end{overpic}
        \end{center}
      \end{minipage}      %
      \begin{minipage}{0.48\hsize}
        \begin{center}
	 \begin{overpic}[bb=0 0 325 239, width=1.0\textwidth]{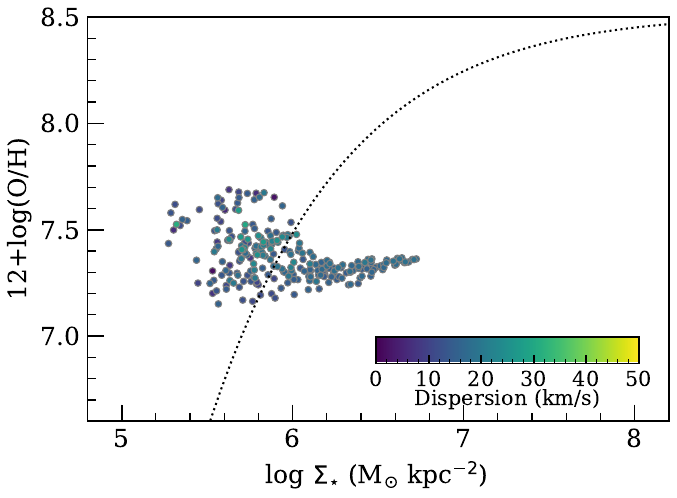}
           \put(64,44){\includegraphics[bb=0 0 484 375, scale=0.175]{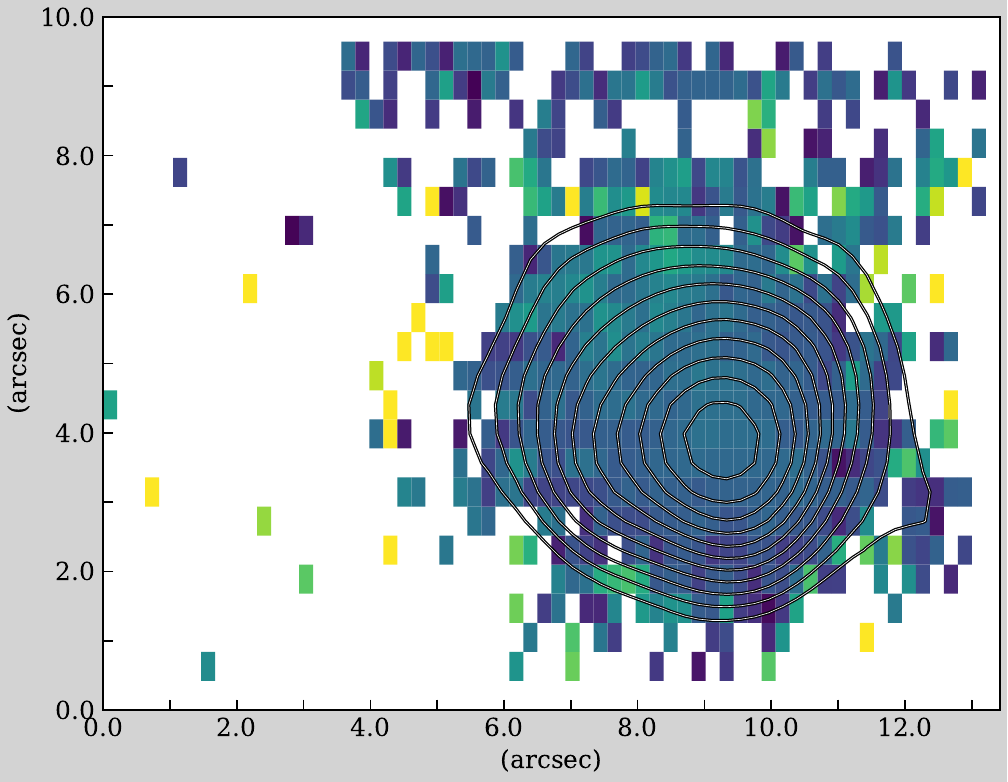}}  
         \end{overpic}
        \end{center}
      \end{minipage}      \\
      \begin{minipage}{0.48\hsize}
        \begin{center}
	 \begin{overpic}[bb=0 0 325 239, width=1.0\textwidth]{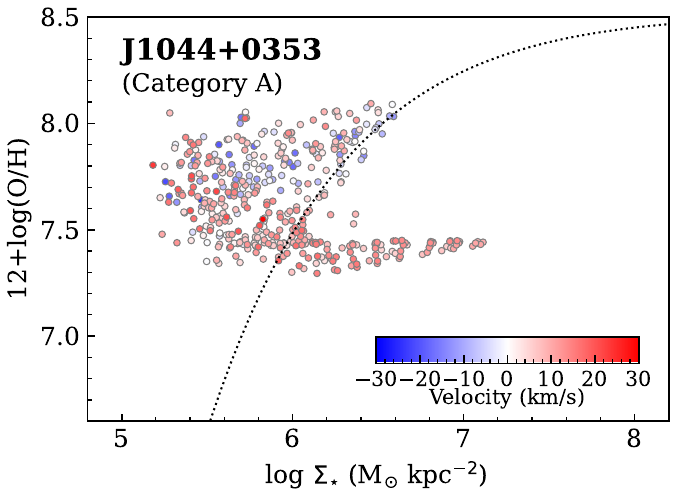}
           \put(64,44){\includegraphics[bb=0 0 484 375, scale=0.175]{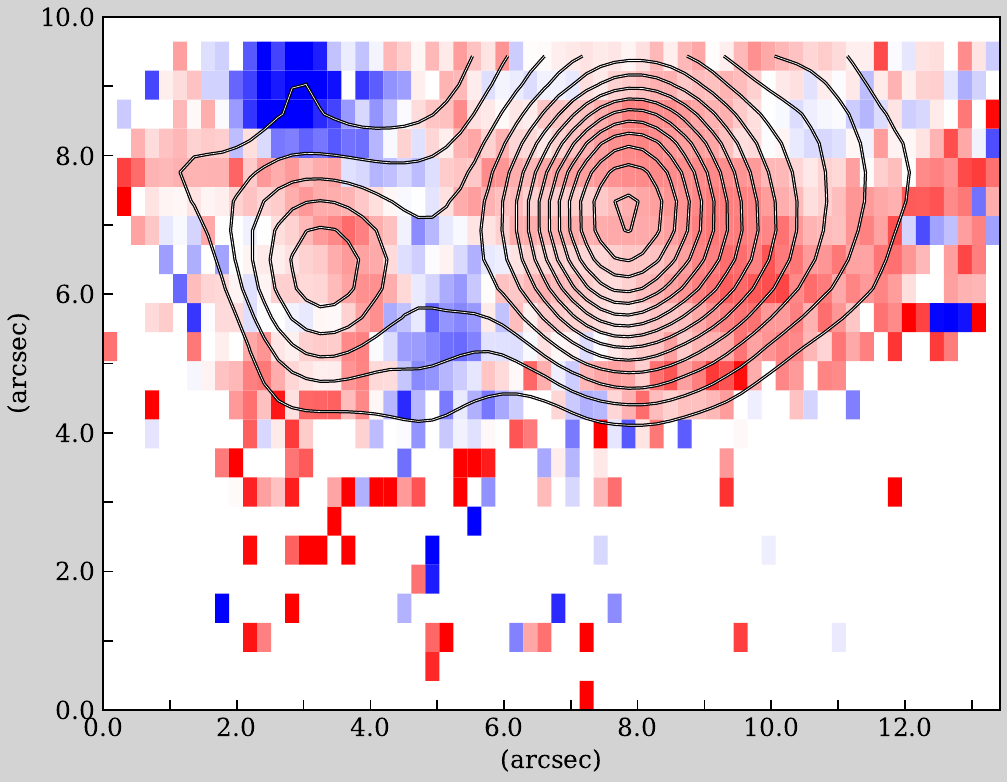}}  
         \end{overpic}
        \end{center}
      \end{minipage}      %
      \begin{minipage}{0.48\hsize}
        \begin{center}
	 \begin{overpic}[bb=0 0 325 239, width=1.0\textwidth]{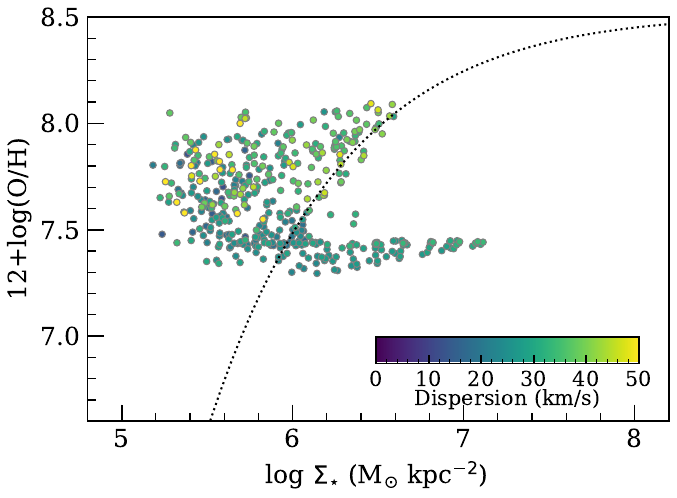}
           \put(64,44){\includegraphics[bb=0 0 484 375, scale=0.175]{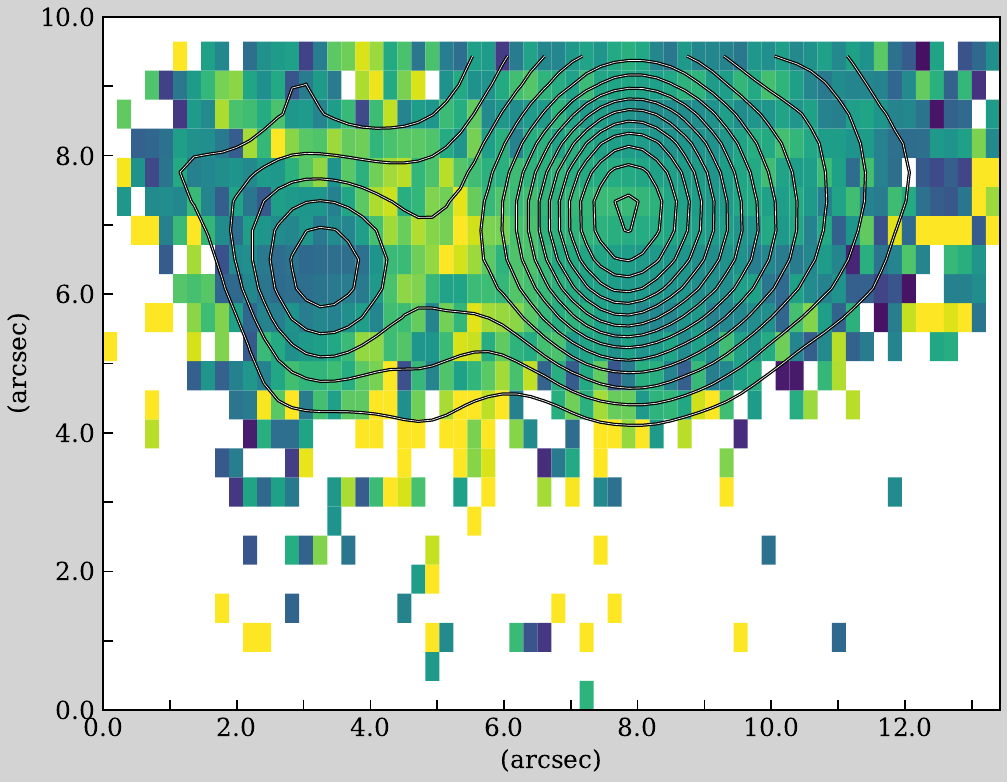}}  
         \end{overpic}
        \end{center}
      \end{minipage}      \\
      \begin{minipage}{0.48\hsize}
        \begin{center}
	 \begin{overpic}[bb=0 0 325 239, width=1.0\textwidth]{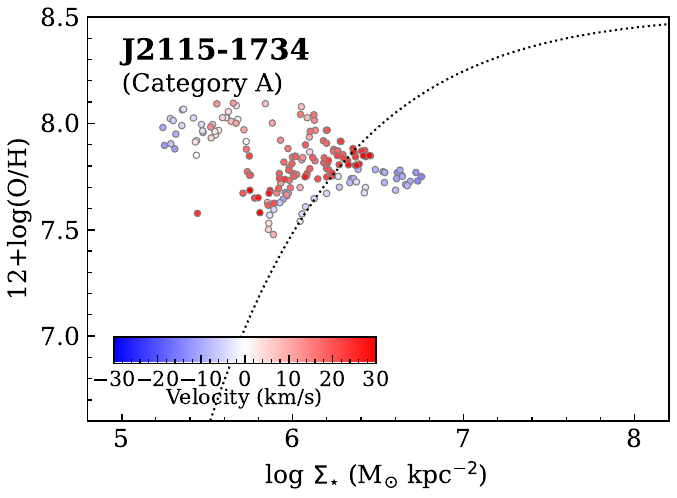}
           \put(64,12.5){\includegraphics[bb=0 0 484 375, scale=0.175]{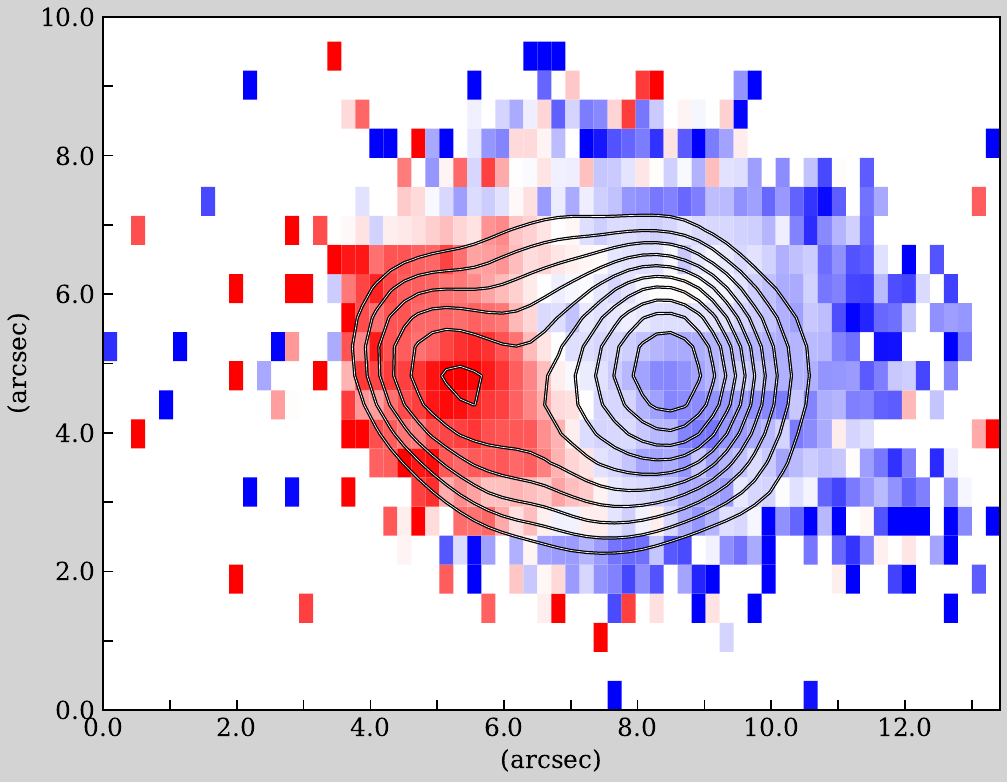}}  
         \end{overpic}
        \end{center}
      \end{minipage}
      \begin{minipage}{0.48\hsize}
        \begin{center}
	 \begin{overpic}[bb=0 0 325 239, width=1.0\textwidth]{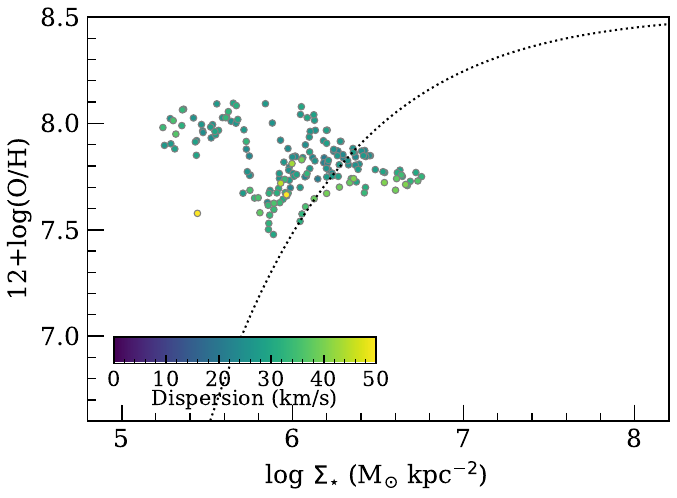}
           \put(64,12.5){\includegraphics[bb=0 0 484 375, scale=0.175]{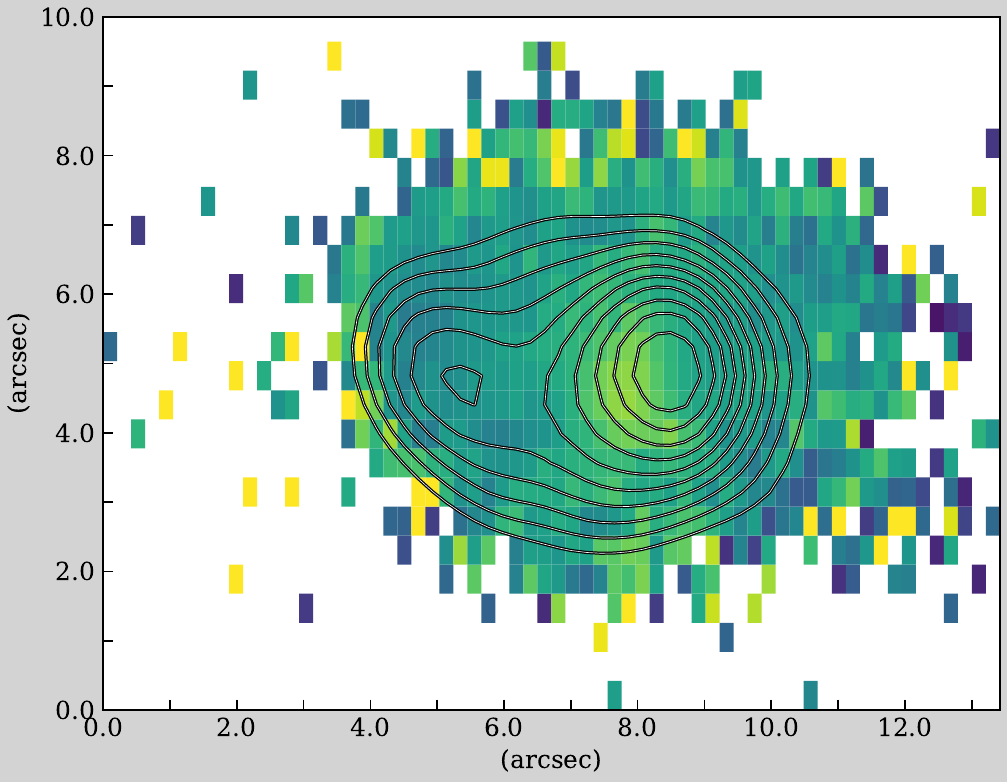}}  
         \end{overpic}
        \end{center}
      \end{minipage}
      \\
    \end{tabular}
    \caption{%
      (Continued.)
    }
\end{figure*}



\subsection{EMPGs in the early phase of galaxy evolution?} \label{ssec:discussions_EMPGs}

One of our initial motivations for studying EMPGs in the local universe was the assumption that these galaxies represent young systems in the early stages of galaxy evolution, making them valuable analogs of high-redshift sources. We now revisit this assumption by evaluating the chemical enrichment processes occurring within EMPGs.

As demonstrated in Sect.~\ref{ssec:results_rMZR_individual}, not all EMPGs in the local universe appear to be undergoing their first burst of star formation; instead, many seem to experience episodic star formation. Nearly half of the sample displays a distinct feature on the spatially-resolved mass-metallicity relation, characterized by a ``metal-poor horizontal branch.'' This feature typically corresponds to regions with locally enhanced stellar mass and star formation surface densities, while the surrounding regions exhibit moderate chemical enrichment. These EMPGs, classified as Category A, are likely influenced by inflows of metal-poor gas from the cosmic web, as predicted by numerical simulations of galaxy evolution (e.g., \citealt{keres2005, dekel2009_nature, schaye2010, fumagalli2011, vandeVoort2012, ceverino2016, mandelker2018}). Such inflows can reduce the gas-phase metallicity (O/H) and trigger starbursts by compressing gas as it interacts with the galactic disk, or by increasing the gas mass within the disk, leading to instabilities and subsequent star formation. If this scenario is correct, these EMPGs may not serve as ideal analogs for galaxies in the earliest stages of evolution. Instead, they may represent systems undergoing fresh gas replenishment and fueling at moderately high redshifts (e.g., at $z=3-6$; \citealt{cresci2010,marconcini2024}). Crucially, these low-metallicity systems can still retain their unique chemical signatures if the gas accretion and starburst triggering occur on timescales similar to or shorter than the metal-mixing timescale. EMPGs in Category A, identified efficiently through their distinct selection criteria (see Sect.~\ref{ssec:results_rMZR_individual}), exemplify such systems, providing insights into gas accretion and episodic star formation in low-mass galaxies.

A similar phenomenon has been observed in local dwarf galaxies, particularly in systems with off-center starbursts, such as ``tadpole galaxies'' \citep{elmegreen2012, sanchez-almeida2013}. \citet{sanchez-almeida2013} and subsequent studies report that some tadpole galaxies, especially those with a minimum metallicity below $0.1$\,\Zsun, exhibit localized metallicity drops accompanied by enhanced star-formation activity at their starburst regions (the ``head''). Further studies suggest that a metallicity threshold around $Z = 0.1$\,\Zsun\ is required for such drops in metallicity to become apparent (see also \citealt{sanchez-almeida2014,sanchez-almeida2014_review,sanchez-almeida2015}).

By targeting EMPGs with metallicities as low as $Z = 0.01$--$0.02$\,\Zsun, our observations further probe the metallicity distribution in extremely metal-poor environments. We identify a distinct category (B) of EMPGs that consists solely of metal-poor clumps without any surrounding metal-rich regions. Figure \ref{fig:histo_category} presents the metallicity distributions (local minima) across different categories. For reference, the histogram for Category A includes the tadpole galaxies exhibiting metallicity drops reported by \citet{sanchez-almeida2013, sanchez-almeida2014, sanchez-almeida2015}.

Our results confirm the existence of Category A galaxies just below one-tenth of the solar metallicity. However, at lower metallicities of $Z \lesssim 0.03$\,\Zsun (\Oabundance\ $\lesssim 7.2$), Category A galaxies become sparse, and Category B candidates begin to emerge. This trend suggests that EMPGs with $Z \lesssim 0.03$\,\Zsun\ provide a more suitable sample of galaxies in the earliest stages of evolution.

This finding aligns with predictions about the metallicity of the cosmic web and the intergalactic medium (IGM), which is expected to reach $Z \sim 0.01$\,\Zsun\ at $z \sim 0$ (e.g., \citealt{fumagalli2011, oppenheimer2012, sanchez-almeida2014_review}). If such pristine gas from the cosmic web accretes into these systems, it could trigger the first episodes of star formation.

One noteworthy case is J2104$-$0035, a Category A galaxy with a metallicity close to the critical threshold of \Oabundance\ $\sim 7.2$, which suggests it may be an exceptional example. This galaxy exhibits multiple clumps, with some regions potentially falling into Category B (Sect.~\ref{ssec:results_rMZR_individual} and Figure \ref{fig:rMZR_individual_a_highlight}).

Interestingly, EMPGs with \Oabundance\ $\lesssim 7.2$ also exhibit blue UV continuum slopes ($\beta < -2$), close to the intrinsic value predicted for primordial star-forming galaxies \citep{nakajima2022_empressV}. This further supports our conclusion that the metallicity threshold of $Z \lesssim 0.03$\,\Zsun\ is a useful criterion for identifying galaxies undergoing their first star formation and chemical enrichment.

Metal abundance ratios, particularly N/O as a function of O/H, are widely employed to explore the presence of metal-poor gas inflow. These ratios primarily reflect the cumulative star-formation history rather than being directly influenced by the influx of metal-poor gas (e.g., \citealt{AM2013}). \citet{isobe2022_fe} argue that extremely metal-poor galaxies in their sample are likely experiencing their first episodes of star formation and chemical enrichment, based on their low N/O ratios. Notably, one of the EMPGs from their study, J1631$+$4426, is classified as a Category B object in this work.

However, we find that not all EMPGs exhibit low N/O ratios, especially those with metallicities close to the EMPG threshold of $Z=0.1$\,\Zsun\ (e.g., \citealt{sanchez-almeida2016}). This observation aligns with our identification of Category A galaxies, further supporting the idea that these EMPGs may represent systems with more complex star-formation histories or ongoing gas accretion.
A more detailed discussion of the EMPRESS 3D objects, incorporating other abundance ratios such as N/O, will be presented in a forthcoming work.

\subsection{Comparisons with Kinematic properties} \label{ssec:discussions_kinematics}

The presence of gas inflow triggering star formation is expected to produce some sort of minor kinematic disturbance (e.g., \citealt{glazebrook2013, ceverino2016, el-badry2016, el-badry2018}). Out of the 22 spatially resolved EMPGs, six were also observed with the high-resolution mode of FOCAS IFU, enabling a detailed kinematic analysis by \citet{isobe2023_focasifu}, including velocity and velocity dispersion maps. Four of these six EMPGs belong to Category A, with one each classified in Categories B and C.

Figure \ref{fig:rMZR_kin} presents the rMZR for the six EMPGs, color-coded by their local velocity (left) and velocity dispersion (right), as shown in the accompanying kinematic maps. Most of these EMPGs, except for HS0822$+$3542, exhibit complex kinematic structures. Notably, distinct velocity shifts are observed along the metal-poor horizontal branch relative to the surrounding metal-enriched regions, suggesting discontinuities in kinematics at the peaks of local star formation activity.
While a clear correlation between velocity dispersion and metallicity is not evident across the sample, intriguing trends emerge in specific cases. For instance, SBS0335$-$052E and J1044$+$0353 shows disturbed regions with enhanced metallicity and high velocity dispersion near star-forming clumps, which could indicate chemical enrichment in the outskirts driven by outflows. These results hint at the diversity of physical processes, such as gas inflows, mergers, and outflows, that may contribute to the complex chemical and kinematic properties of EMPGs. Indeed, I\,Zw\,18 show kinematic features that may represent gas-rich mergers; multiple metal-poor star-forming clumps are evident, with one associated primarily with the metal-poor branch (redshifted component) and another showing more metal-enriched regions (blueshifted component). This highlights the interplay between mergers and localized chemical enrichment in low-metallicity systems.

Finally, we highlight key findings from the companion studies by \citet{isobe2023_focasifu} and \citet{xu2024_focasifu}, which investigate the kinematic properties of EMPGs. A notable result is the low ratio of rotation velocity to velocity dispersion ($v_{\rm rot}/\sigma_0 = 0.29$--$0.80 < 1.0$) in the star-forming clump regions, indicating that turbulence or random motions dominate over ordered rotation. Moreover, the exceptionally high gas mass fractions in these systems ($f_{\rm gas} = 0.9$--$1.0$) suggest that these galaxies are gas-rich and highly primed for star formation. These findings align well with our proposed scenario of metal-poor gas inflow triggering starbursts within the clump regions of EMPGs and enhancing the velocity dispersion as the gas releases its potential energy. This mechanism likely plays a central role in the evolution of the EMPGs.

Interestingly, \citet{xu2024_focasifu} identify a tentative trend suggesting that EMPGs in Category B exhibit smaller $v_{\rm rot}/\sigma_0$ values compared to those in Category A ($0.34\pm 0.08$ vs.~$0.68\pm 0.10$). This supports our hypothesis that EMPGs in Category A contain older, metal-enriched populations that may have already formed a rotational disk, which is now disturbed by recent gas inflow. In contrast, Category B EMPGs appear to be undergoing their first significant chemical enrichment event, potentially driven by gas accretion, and present a more dispersion-dominated kinematic structure due to the lack of a stable rotational disk.

\section{Summary} \label{sec:summary}

We have presented the initial 3D spectroscopic analysis of 24 local extremely metal-poor galaxies (EMPGs) as part of the EMPRESS 3D collaboration. This paper focuses on the spatially-resolved optical nebular properties of EMPGs, with particular emphasis on their metallicity distributions in the extremely metal-poor regime, i.e., below one-tenth of the solar value. Our key findings are summarized as follows:

\begin{itemize}
    \item Compiling data from 9,177 spatial pixels across the 24 EMPGs, all with metallicity measurements based on the latest calibrations and no clear AGN signatures, we confirm a steep relationship between stellar mass surface density ($\Sigma_{\star}$) and metallicity in the low-metallicity regime ($Z < 0.1$\,\Zsun) on the spatially-resolved mass-metallicity diagram (rMZR). The scatter in this relation is significantly reduced when accounting for the dependence on star-formation surface density ($\Sigma_{\rm SFR}$), consistent with trends observed in global scaling relations.

    \item We analyze the metallicity gradients along the semi-major axis of each EMPG. Approximately half of the galaxies exhibit negative or flat gradients, consistent with inside-out chemical enrichment processes, while a few show complex distributions, such as positive gradients, indicating diverse mixing processes and evolutionary pathways.

    \item We categorize the resolved EMPGs into four groups based on their rMZR profiles: (A) Metal-poor cores with enriched surroundings, (B) Isolated metal-poor clumps, (C) Systems in transition with no clear metal-poor branches, and (D) Unresolved compact systems. This classification reveals distinct evolutionary stages among EMPGs and helps us understand their diverse properties.

    \item Nearly half of the well-resolved EMPGs (11 out of 22) display a characteristic feature on the rMZR: a distinct metal-poor horizontal branch that appears at the peaks of $\Sigma_{\star}$ and $\Sigma_{\rm SFR}$, surrounded by a higher metallicity region (Category A). This pattern likely indicates recent or ongoing infall of pristine gas into the central potential well, triggering localized bursts of star formation and diluting the local gas metallicity (i.e., lowering O/H). If this scenario holds, quite a few EMPGs in the local universe are undergoing episodic star-formation bursts following prior star formation and chemical enrichment. These EMPGs, particularly those with central metallicities just below the $Z = 0.1$\,\Zsun\ threshold, may not serve as ideal analogs for galaxies in the early universe.

    \item The kinematic analysis of EMPGs reveals discontinuities in kinematics, particularly at the peaks of local star formation activity. This is accompanied by exceptionally high gas mass fractions ($f_{\rm gas} = 0.9$--$1.0$) and low rotation-to-dispersion ratios ($v_{\rm rot}/\sigma_0 < 1$), indicating turbulence-dominated systems. These findings are consistent with a scenario where inflows of metal-poor gas trigger localized starbursts, enhancing velocity dispersion and fueling star formation.

    \item The four most metal-poor galaxies in our sample (with $Z < 0.03$\,\Zsun), show only a metal-poor horizontal branch on the rMZR without any associated higher metallicity regions (Category B). This stricter metallicity threshold isolates extreme EMPGs, which likely represent an early phase of galaxy evolution in the present-day universe.

\end{itemize}


\section*{Acknowledgements}

This work is based on observations made with the Subaru Telescope, which is operated by the National Astronomical Observatory of Japan (NAOJ). We are honored and grateful for the opportunity to observe the Universe from Maunakea, which has cultural, historical, and natural significance in Hawaii. We thank the staff at the Subaru Telescope for their help with the observations. 
This paper is supported by World Premier International Research Center Initiative (WPI Initiative), MEXT, Japan, as well as the joint research program of the Institute of Cosmic Ray Research (ICRR), the University of Tokyo. 
In addition, we acknowledge support from JSPS KAKENHI Grant: JP20K22373 and JP24K07102 (K.\,Nakajima), JP20H00180 and JP21H04467 (M. Ouchi), JP21J20785 (Y.\,Isobe), JP21K13953 and JP24H00245 (Y.\,Harikane), JP22KJ0157, JP21H04499, JP21K03614, and JP22H01259 (Y.\,Hirai), JP21K03622 (M.\,Onodera), JP20H01895, JP21K13909, and JP21H05447 (K.\,Hayashi), JP21H04489 (H.\,Yajima), and JP21H01128 and JP24H00247 (T.\,T.\,Takeuchi).
We also acknowledge support from JST; H.\,Yajima is supported by FOREST Program: JPMJFR202Z, and M.\,Nishigaki is supported by the establishment of university fellowships towards the creation of science technology innovation: JPMJFS2136.
J.\,H.\,Kim acknowledges the support from the National Research Foundation of Korea (NRF) under grant No.\,2021M3F7A1084525 and No.\,2020R1A2C3011091 and the Institute of Information \& Communications Technology Planning \& Evaluation (IITP) grant, No. RS-2021-II212068 funded by the Korean government (MSIT).
S.\,Fujimoto acknowledges support from the European Research Council (ERC) Consolidator Grant funding scheme (project ConTExt, grant No. 648179). 
This work has also been supported in part by the Sumitomo Foundation Fiscal 2018 Grant for Basic Science Research Projects (180923), and the Collaboration Funds of the Institute of Statistical Mathematics, ``Machine Learning Approaches to Cosmic Structures: From Structure Formation to Galaxy Evolution'' (2024-ISMCRP-2025).
The Cosmic Dawn Center is funded by the Danish National Research Foundation under grant No. 140. 
This project has received funding from the European Union's Horizon 2020 research and innovation program under the Marie Sklodowska-Curie grant agreement No. 847523 ``INTERACTIONS''.
This research was supported by a grant from the Hayakawa Satio Fund awarded by the Astronomical Society of Japan.


\bibliographystyle{aasjournal}{}
\bibliography{Refs_paper.bib}{}



\end{document}